\documentclass[twocolumn,amsmath,aps,prd,preprintnumbers,amssymb,nofootinbib,showpacs,floatfix]{revtex4}
\usepackage{amssymb}
\usepackage{amsmath}
\usepackage{epsfig}
\usepackage{subfigure}
\usepackage{mathrsfs}
\usepackage{longtable}
\usepackage[usenames,dvipsnames]{xcolor}
\usepackage{mathtools}
\usepackage{relsize}
\begin{document}
%Dan's definitions
 \renewcommand{\thefigure}{\arabic{figure}}
\newcommand{\noj}{}

\newcommand{\apjl}{Astrophys. J. Lett.}
\newcommand{\aap}{Astron. Astrophys.}
\newcommand{\apjs}{Astrophys. J. Suppl. Ser.}
\newcommand{\sa}{Sov. Astron. Lett.}.
\newcommand{\jpb}{J. Phys. B.}
\newcommand{\natu}{Nature (London)}
\newcommand{\aaps}{Astron. Astrophys. Supp. Ser.}
\newcommand{\aj}{Astron. J.}
\newcommand{\aas}{Bull. Am. Astron. Soc.}
\newcommand{\mnras}{Mon. Not. R. Astron. Soc.}
\newcommand{\pasp}{Publ. Astron. Soc. Pac.}
\newcommand{\jcap}{JCAP.}
\newcommand{\jmat}{J. Math. Phys.}
\newcommand{\prep}{Phys. Rep.}
\newcommand{\jtep}{Sov. Phys. JETP.}
\newcommand{\plb}{Phys. Lett. B.}
\newcommand{\pla}{Phys. Lett. A.}
\newcommand{\jhep}{Journal of High Energy Physics}

%Doddy's definitions
\newcommand{\be}{\begin{equation}}
\newcommand{\ee}{\end{equation}}
\newcommand{\bea}{\begin{align}}
\newcommand{\eea}{\end{align}}
\newcommand{\lsim}{\mathrel{\hbox{\rlap{\lower.55ex\hbox{$\sim$}} \kern-.3em \raise.4ex \hbox{$<$}}}}
\newcommand{\gsim}{\mathrel{\hbox{\rlap{\lower.55ex\hbox{$\sim$}} \kern-.3em \raise.4ex \hbox{$>$}}}}
\newcommand{\grad}{\ensuremath{\vec{\nabla}}}
\newcommand{\adotoa}{\ensuremath{{\cal H}}} 
\newcommand{\Uc}{\ensuremath{{\cal U}}}
\newcommand{\Vc}{\ensuremath{{\cal V}}}
\newcommand{\Jc}{\ensuremath{{\cal J}}}
\newcommand{\Mc}{\ensuremath{{\cal M}}}

\newcommand{\unit}[1]{\ensuremath{\, \mathrm{#1}}}

%Ewan's definitions
\newcommand{\gb}{\gamma_{\rm b}}
\newcommand{\dx}{\delta x}
\newcommand{\dy}{\delta y}
\newcommand{\dz}{\delta z}
\newcommand{\dr}{\delta r}
\newcommand{\ds}{\delta s}
\newcommand{\dt}{\delta t}
\newcommand{\uns}{\rmunderscore}
\newcommand{\chimin}{\langle \chi \rangle}

%Scott's definitions
\newcommand{\bi}{\begin{itemize}}
\newcommand{\ei}{\end{itemize}}
\newcommand{\ben}{\begin{enumerate}[itemsep=0.1in,parsep=0.1in]}
\newcommand{\een}{\end{enumerate}}
\newcommand{\tben}{\begin{enumerate}[itemsep=0.0in,parsep=0.0in]}
\newcommand{\teen}{\end{enumerate}}
\newcommand{\ud}{{\rm d}}
\newcommand{\drm}{\mathrm{d}}
\newcommand{\rhos}{\rho_\phi}
\newcommand{\rhor}{\rho_r}
\newcommand{\pr}{p_r}
\newcommand{\rhom}{\rho_\mathrm{dm}}
\newcommand{\rhost}{\tilde{\rho}_\phi}
\newcommand{\gam}{\Gamma_\phi}
\newcommand{\cms}{\; {\rm cm^3}/{\rm s}}
\newcommand{\gev}{\; \mbox{GeV}}
\newcommand{\tev}{\; \mbox{TeV}}

%%%%%%%%%%%%% colortext comments %%%%%%%%%%%%%%
\newcommand{\tkDM}[1]{\textcolor{red}{#1}}                     % Doddy
\newcommand{\tkGSW}[1]{\textcolor{blue}{#1}}		  % Scott
%%%%%%%%%%%%%%%%%%%%%%%%%%%%%%%%%%%%%%%%%%

\title{Constraining SUSY with Heavy Scalars -- using the CMB}

\author{Luca Iliesiu$^{1}\footnote{liliesiu@princeton.edu} 
$,~David J. E. Marsh$^{2}$,~Kavilan Moodley$^{3}$, and Scott Watson$^{4}$}
\affiliation{$^{1}$Princeton University Department of Physics, Jadwin Hall, Washington Road, Princeton, NJ  08544, USA}
\affiliation{$^{2}$Perimeter Institute, 31 Caroline St N, Waterloo, ON, N2L 6B9, Canada}
     \affiliation{$^{3}$Astrophysics and Cosmology Research Unit, University of KwaZulu-Natal, Durban, 4041, SA}
 \affiliation{$^{4}$Department of Physics, Syracuse University, Syracuse, NY 13244, USA}
%\author{Luca Iliesiu}
%\email{liliesiu@princeton.edu}
%\affiliation{Princeton University Department of Physics, Jadwin Hall, Washington Road, Princeton, NJ  08544, USA}

%\author{David J. E. Marsh}
%\email{dmarsh@perimeterinstitute.ca}
%\affiliation{Perimeter Institute, 31 Caroline St N, Waterloo, ON, N2L 6B9, Canada}

%\author{Kavilan Moodley}
%\email{moodleyk41@ukzn.ac.za}
%\affiliation{Astrophysics and Cosmology Research Unit, University of KwaZulu-Natal, Durban, 4041, SA}
%\affiliation{Centre for High Performance Computing, CSIR Campus, 15 Lower Hope St., Rosebank, Cape Town, SA}

%\author{Scott Watson}
%\email{gswatson@syr.edu}
%\affiliation{Department of Physics, Syracuse University, Syracuse, NY 13244, USA}

\date{\today}

 %---------------------- ABSTRACT -------------------------
\begin{abstract}

If low-energy SUSY exists, LHC data favors a high mass scale for scalar superpartners (above a TeV), while sfermions and the dark matter can be parametrically lighter -- leading to a so-called split-spectrum. When combining this fact with the motivation from fundamental theory for shift-symmetric scalars (moduli) prior to SUSY breaking, this leads to a non-thermal history for the early universe. Such a history implies different expectations for the microscopic properties of dark matter, as well as the possibility of dark radiation and a cosmic axion background. In this paper we examine how correlated and mixed isocurvature perturbations are generated in such models, as well as the connection to dark radiation. WMAP constraints on multiple correlated isocurvature modes allow up to half of the primordial perturbations to be isocurvature, contrary to the case of a single isocurvature mode where perturbations must be dominantly adiabatic. However, such bounds are strongly prior dependent, and have not been investigated with the latest Planck data. In this paper we use the example of a SUSY non-thermal history to establish theoretical priors on cosmological parameters. Of particular interest, we find that priors on dark radiation are degenerate with those on the total amount of isocurvature -- they are inversely correlated. Dark radiation is tightly constrained in the early universe and has been used recently to place stringent constraints on string-based approaches to beyond the standard model.  Our results suggest such constraints can require more input from theory. Specifically, we find that in many cases constraints on dark radiation are avoidable because the density can be reduced at the expense of predicting an amount of multi-component isocurvature. The latter are poorly constrained by existing probes, and lead to the interesting possibility that such models could have new predictions for the next generation of observations. Our results are not only important for establishing the post-inflationary universe in the presence of SUSY, but also suggest that data from cosmological probes -- such as Planck -- can help guide model building in models of the MSSM, split-SUSY, and beyond. Our model also demonstrates the utility of UV models in constructing cosmological priors.

\end{abstract}

\maketitle

\section{Introduction}
The discovery of a $125$ GeV Higgs at the Large Hadron Collider (LHC) -- and nothing else -- has left 
the relevance of supersymmetry (SUSY) in question, particularly as a mechanism for stabilizing the hierarchy between the Electroweak and Planck scales.  Within the minimal SUSY standard model (MSSM), Natural-SUSY still remains possible, but at the cost of increasing the level of complexity of models (see e.g.  \cite{Randall:2012dm}).  There are a number of alternative ways to reconcile SUSY with the data, including models of Split-SUSY \cite{Wells:2003tf,ArkaniHamed:2004fb,Arvanitaki:2012ps}, or simply by accepting that some fine-tuning may just be an accident of nature\footnote{It is noteworthy that even defining the level of fine-tuning can be an issue, see e.g. \cite{Baer:2013gva}}.  Regardless of one's viewpoint, it seems that if low-energy SUSY will prevail it will require the existence of a new scale at around $10-100$ TeV.  In many models 
this scale is set by SUSY breaking and scalar superpartners masses will be around this range,  whereas fermion superpartners (and dark matter) will be parametrically lighter at around $100-1000$ GeV -- providing a so-called {\em split spectrum}. 

Any additional light scalars, or moduli, resulting from beyond the standard model physics would also generically receive masses around the $100$ TeV range \cite{Acharya:2009zt}.  Cosmologically this is very interesting, since if these moduli are only gravitationally coupled to other matter (which is typically the case) this mass range is precisely what is necessary to avoid the cosmological moduli problem (CMP) \cite{coughlan1983,Watson:2009hw,Acharya:2008bk}. Moreover, moduli in this mass range will decay early enough to avoid disrupting Big Bang Nucleosynthesis (BBN), and lead to a new {\it non-thermal history} for the early universe.  This leads to new expectations for early universe cosmology including: altered predictions for confronting inflation models with Cosmic Microwave Background (CMB) data \cite{Easther:2013nga}; different expectations for the microscopic properties of dark matter (DM) \cite{Acharya:2008bk,Watson:2009hw,Cohen:2013ama,Fan:2013faa,Allahverdi:2013noa}; enhanced small-scale structure \cite{erickcek2011}; and the possible existence of dark radiation (DR) and a relic cosmic axion background \cite{higaki2013,Conlon:2013isa}.

It has been said that Planck \cite{Ade:2013ktc} cosmological constraints \cite{Ade:2013zuv} make for a `maximally boring universe', described with exquisite precision by the six parameter $\Lambda$CDM standard cosmological model in which all perturbations are gaussian and adiabatically produced. One phenomenological extension of this model is the possible existence of isocurvature modes in the dark or visible sector \cite{bucher2000}. WMAP \cite{Hinshaw:2012aka,savelainen2013} and Planck \cite{Ade:2013uln} place constraints on the level of isocurvature in various models. The constraints on single-mode isocurvature are strong, and limit the isocurvature fraction at the percent or sub-percent level depending on the model, with correlated models being more tightly constrained. However, when multiple isocurvature modes are allowed, degeneracies between the modes can allow the isocurvature fraction to be almost half \cite{bucher2004,moodley2004,Bean:2006qz}. In the two-mode model with DM and neutrino isocurvature that we will consider in this paper, the fraction is lowered to around 30 to 40\%, but is still substantially larger than that allowed for single modes.  In such a scenario the universe is certainly not `maximally boring'. The key question to be answered is whether such a model is theoretically well motivated and can exist within a UV extension of the SM. 

The existence of DR (parameterised by the effective number of neutrino species $N_{\rm eff}$) having departures from its canonical value, $\Delta N_{\rm eff}=N_{\rm eff}-3.04$ is also a generic and well-motivated extension of the six parameter $\Lambda$CDM model \cite{linde1979}. Constraints on $\Delta N_{\rm eff}$ from Planck and other CMB experiments has caused it to receive much attention over the last few years \cite{dunkley2010,hamann2010a,fischler2011,keisler2011,kobayashi2011,marsh2011b,abazajian2012,benetti2013,calabrese2013,Conlon:2013isa,Weinberg:2013kea}. The interpretation of constraints to $N_{\rm eff}$ can depend on the theoretical model underpinning the departure from the canonical value \cite{archidiacono2013,angus2013}, while the constraints themselves can have a dependence on the priors coming from such a model \cite{gonzales-morales2011,wyman2013,Verde:2013cqa}.

Under certain generic assumptions, which we outline below, the moduli of SUSY pick up isocurvature perturbations during inflation. In the subsequent decay of the moduli these perturbations are passed on to the DM, just like in the well-known curvaton scenario \cite{enqvist2001,lyth2002,moroi2001}. Moduli effectively behave as scalar fields, and in SUSY come partnered with pseudo-scalar axion fields to which they are coupled. The shift symmetry of the axions protects their masses, allowing them to be light compared to the moduli, while also suppressing their couplings and making them long lived. As such, heavy moduli can decay into light, relativistic axions, providing a component of DR that also inherits an isocurvature perturbation correlated to the DM isocurvature.
%In this paper we focus on how bounds on the level of isocurvature perturbations from WMAP and Planck can be used to restrict model building in the SUSY frameworks above.
%As we review in the next section, the decay of moduli into dark matter and dark radiation can in some cases lead to the prediction of a large isocurvature component in the CMB.
%WMAP \cite{Hinshaw:2012aka}, and more recently Planck \cite{Ade:2013uln}, place constraints on the level of isocurvature, and given these measurements we identify when these lead to meaningful constraints on models, and when they can be safely be neglected.
Much of our analysis has overlap with existing studies of curvatons and related toy models (see e.g. \cite{Lemoine:2009is,Ade:2013uln} and references within), but with the added twist of DR, and non-thermal DM.  

The cosmological constraints to correlated DM-DR isocurvature are strongly prior dependent \cite{bucher2004,moodley2004}, so that such a situation cries out for a UV model able to fix the priors based on other considerations. We will discuss the importance and implications of this in some detail. In the context of our model there is also a prior, and therefore interpretation, for constraints to $\Delta N_{\rm eff}$ as an axion background produced by decay of a modulus with  certain mass, width, and branching ratios. The goals of this work are to establish the importance of isocurvature constraints for SUSY models, find the cosmological priors implied in a SUSY set up for correlated DM-DR isocurvature, and how the constraints and degeneracies from DR production can complement the isocurvature phenomenology.

The remainder of the paper is organised as follows.  In Section~\ref{sec:susywimps} we review the connection between SUSY-based model building after LHC and a non-thermal history for the post inflationary universe -- including what this implies for the expected microscopic properties of DM, and how DR can be produced. We discuss how in some cases this can lead to a substantial amount of isocurvature in the primordial temperature fluctuations, which would be in conflict with observations.  We then identify which cases are most severely restricted by observations, which we find correspond to cases where the decaying field responsible for the non-thermal history has sub-Hubble mass during inflation, leading to the production of isocurature modes. In Section~\ref{sec:iso_production} we establish the basic formalism for computations in this model, computing cosmological observables and relating them to the CMB spectrum. We present the results of these calculations in Section~\ref{sec:results}, where we use priors on the modulus parameters in Split and Natural-SUSY to compute the priors on cosmological parameters, discussing how constraints to DR and isocurvature can be complementary. In the last section we conclude and discuss what this implies for the current status of SUSY dark matter of non-thermal origin, and outline future directions, in particular how the results of this work can be used to accurately constraint SUSY using Planck data. The details of our numerical calculation are relegated to Appendix~\ref{appendix:computation}, while some details of the power spectrum normalisation and spectral indices are given in Appendix~\ref{appendix:spectra}.

\section{SUSY WIMPs, Non-thermal Histories, Inflation and Reheating}
\label{sec:susywimps}

The motivation from LHC for higher than anticipated superpartner masses implies that the scale of SUSY breaking $\Lambda_{\rm SUSY}$ should be around 
$\Lambda_{\rm SUSY} = (m_{3/2} M_{pl})^{1/2} \sim 10^{12}$ GeV, where the gravitino mass $m_{3/2} \approx 10-100$ TeV sets the mass scale of the scalar superpartners and $M_{pl} = 1/\sqrt{8 \pi G} \approx 2.4 \times 10^{18}$ GeV is the reduced Planck mass.
However, superpartner fermions -- one of which plays the role of WIMP DM -- can be parametrically below this scale due to loop suppression and R-symmetry \cite{Moroi:1999zb,Wells:2003tf,ArkaniHamed:2004fb}.  When this type of framework is required to have a high-energy (UV) completion within supergravity (SUGRA) or string theory (which is necessary for self-consistently), additional scalars with little or no potential (moduli) will naturally appear leading to a non-thermal cosmological history \cite{Watson:2009hw,Acharya:2008bk,Acharya:2009zt}.  We briefly review this in the next section, followed by a discussion of how this can lead to a large generation of isocurvature perturbations in the primordial spectrum, and production of DR.

\subsection{Non-thermal dark matter and cosmological moduli}

\begin{figure}[t]
\begin{center}
\includegraphics*[scale=0.25]{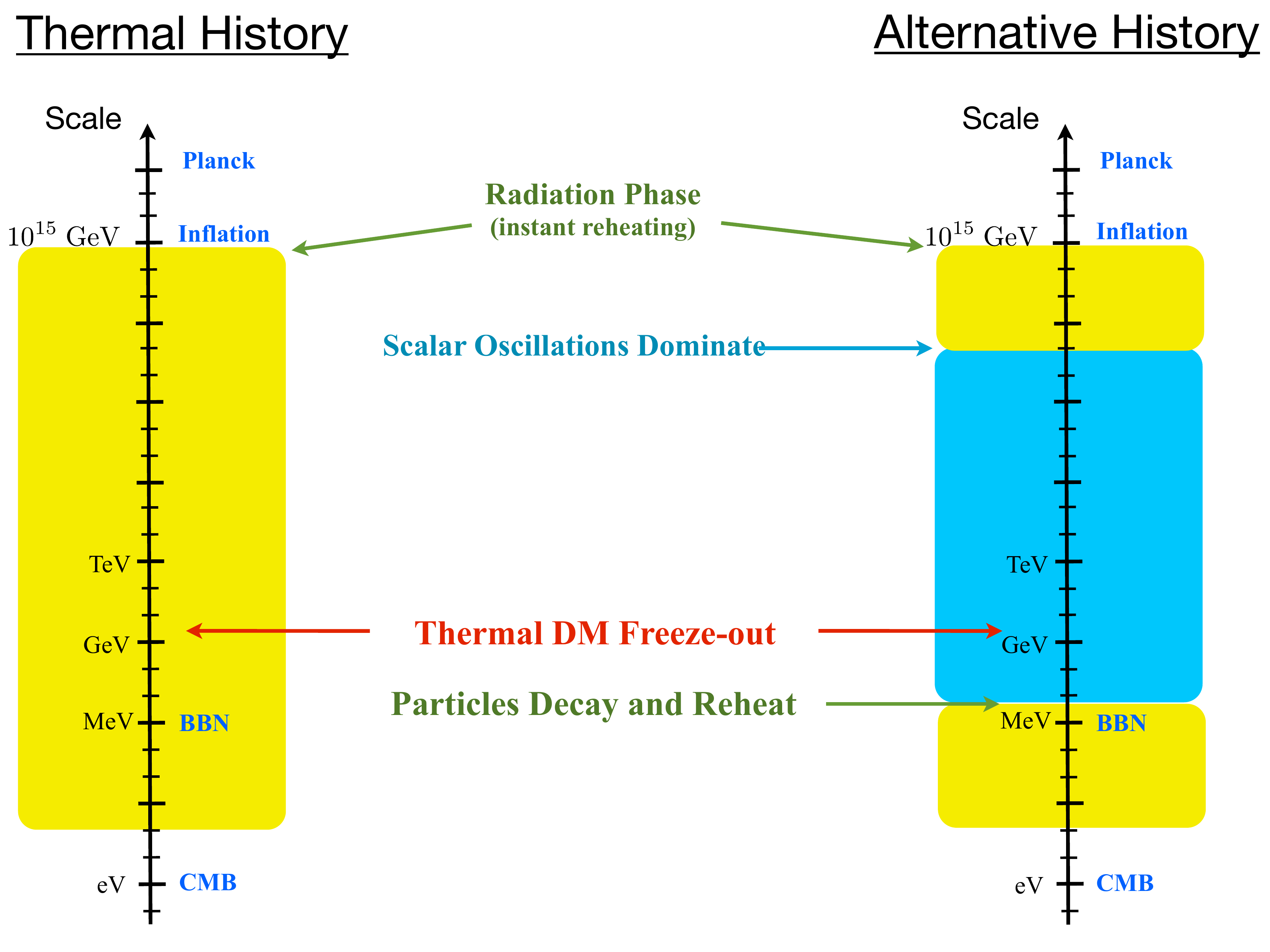}\label{fig1}
\hspace{0.2cm}
\end{center}
\caption{The left timeline represents the standard assumption of a thermal history for the early universe where dark matter is populated in the thermal bath that emerges shortly after inflation.  
The right timeline represents a non-thermal history resulting from SUSY models with moduli and a split-SUSY-like spectrum (scalars heavy, fermions / dark matter light) where dark matter production occurs directly from scalar decay.}
\end{figure}

The non-thermal post-inflationary history of the universe which we outline in this subsection is sketched in Fig.~\ref{fig1}.

Given a Split-SUSY-like spectrum in theories that contain moduli (such as SUGRA and string theories), a non-thermal cosmological history naturally results \cite{Watson:2009hw,Acharya:2008bk,Acharya:2009zt}.  As an example, consider a modulus or scalar field $\sigma$ with a shift symmetry so that naively $V(\sigma) = 0$.  If this remains a good symmetry until SUSY breaking, we expect the field to get a mass $\sim m_{3/2}$, where the split spectrum implies
\be
m_\sigma = c_0 m_{3/2} =c_0 {\frac{\Lambda_{\rm SUSY}^2}{M_{pl}}} \approx 10 - 1000 \; \mbox{TeV},
\label{eqn:mass}
\ee
where the constant $c_0$ is typically not more than $100$ and in string based examples is frequently related to the hierarchy $c_0 \sim \ln \left( M_{pl} / m_{3/2} \right)$ \cite{LoaizaBrito:2005fa,Watson:2009hw}. However, we note that in models of Split-SUSY (where the electroweak hierarchy is addressed anthropically) the gravitino mass can be significantly higher than the TeV scale, and so in those models the moduli mass will take values that can range all the way up to the Planck scale -- we will consider both possible mass ranges in this paper. We will consider `Natural-SUSY' to have masses below $100$ TeV, and `Split-SUSY' to have masses up to $10^4$ TeV.

The non-thermal history arises from the observation that there is no {\em a-priori} reason why the modulus $\sigma$ should initially begin in its low-energy minimum. As an explicit example, we expect on general grounds that the shift symmetry of the modulus should be broken by both the finite energy density of inflation (another source of SUSY breaking) and quantum gravity effects \cite{Dine:1995kz}.  These considerations imply additional contributions to the effective potential in the form of a Hubble scale mass and a tower of non-renormalizable operators,
\be \label{potentialex}
\Delta V_1 = -c_1 H_{I}^2 \sigma^2 + \frac{c_{n}}{M^{2n}} \sigma^{4+2n} + \ldots,
\ee
where in the absence of special symmetries $c_1$ and $c_{n}$ are expected to be order one constants (with $n>1$), $H_I$ is the Hubble rate during inflation, and $M$ is the scale of new physics, e.g. quantum gravity. As this will be important later, we note that the dimension of the leading irrelevant operator that lifts the flat direction is model dependent as well as the scale of new physics\footnote{This is a generic expectation in string theories where new thresholds before the Planck scale are common place.  Examples include both the compactification (Kaluza-Klein) scale $M_{kk}$ and string scale $M_s$ where $M \ll M_{pl}$ is required for consistency of the effective theory \cite{Polchinski:1998rq}.}.
Because the contributions (\ref{potentialex}) are the dominant terms in the potential during high scale ($H_I>m_\sigma$) inflation, this implies the minimum of the field at that time will be 
\be
\langle \sigma \rangle \sim M \left( \frac{H_I}{M} \right)^{\frac{1}{n+1}}\, .
\label{eqn:sig_vev_inf}
\ee
The mass at this high energy minimum, $m_\sigma(\langle\sigma\rangle)\propto H_I$, plays a key role in the generation of isocurvature perturbations. The constant of proportionality is set by the model dependent values of $\{c_1,c_n,n\}$ and can be greater or less than unity. Much later, at the time that the low-energy SUSY breaking gives the dominant contributions to the potential, the minimum is near $\langle \sigma \rangle \sim 0$, while the mass is given by Eq.~(\ref{eqn:mass}).  More complicated potentials and contributions are possible, but in this simple case where the mass term $\sim m_{3/2}$ dominates at low energy, this displacement from the low energy minimum leads to energy stored in coherent oscillations of $\sigma$ forming a scalar condensate.
The amplitude of the oscillations is determined by the initial displacement $\sigma_\star \sim \langle \sigma \rangle$, where for the example potential (\ref{potentialex}) we have $\sigma_\star \sim M \left( {H_I}/{M} \right)^{1 /(n+1)}$.

Hubble friction ceases and oscillations will set in when the expansion rate satisfies $H \approx m_\sigma$.  Because the oscillations scale like matter, they dilute more slowly than the primordial radiation (produced during inflationary reheating).  Depending on the initial value $\sigma_\star$, the energy stored in the moduli may quickly come to dominate the energy density of the universe (see e.g. \cite{Acharya:2008bk}).  At the time oscillations begin $t_{\mbox{\tiny osc}} \approx H ^{-1} \approx m^{-1}_\sigma$ the initial abundance is given by
\be
\rho_\sigma(t_{\mbox{\tiny osc}}) = \frac{1}{2} m_\sigma^2  \sigma_\star^2,
\ee
and once the oscillations become coherent (which typically takes less than a Hubble time) they will scale as pressure-less matter \cite{Turner:1983he} with $\rho_\sigma \sim  m_\sigma^2  \sigma_\star^2 /a(t)^3$.  The universe remains matter dominated until the field decays.  Because the field is a modulus we expect it typically to be gravitationally coupled to other particles and so its decay rate is 
\be \label{decayrate} 
\Gamma_\sigma = c_3 \frac{m_\sigma^3}{\Lambda^2},
\ee
where we expect $\Lambda \sim M_{pl}$ and $c_3$ depends on the precise coupling in the fundamental Lagrangian, but typically takes values in the range $1/(4 \pi) \lesssim c_3 \lesssim 100$. 
Most of the field decays\footnote{In most of the literature the moduli decay is treated as instantaneous at $t_{\mbox{\tiny decay}}$.  However, this approximation can be misleading.  For example, in the case where the radiation energy density significantly drops below the moduli energy density ($\rho_\sigma \gg \rho_r$), the continuous (even though small) amounts of particle decay before reheating can lead to changes in the scale factor - temperature relation.  This is because during the decays the entropy is {\it not} conserved, but changes as $\dot{S} = B_{r} \Gamma \rho_\sigma a^4 / T$, which follows from the equation of motion for the radiation (\ref{eqn:densities_source}) with $w_i=1/3$, the definition of the entropy density $S=sa^3=2\pi^2/45 g_s T^3$, and the first law of thermodynamics $\Delta S = \Delta E_r / T$. Using this relation one can show that the scale factor is related to the temperature of the radiation as $a \sim T^{-8/3}$ and so if the modulus dominates the universe so that $H^2 M_{pl}^2 \sim \rho_\sigma \sim 1/ a^3 \sim T^8$ instead of the entropy preserving values (in a matter dominated universe) $H \sim T^{3/2}$ and $a \sim 1/T$.  We refer the reader to \cite{Giudice:2000ex} for further discussion.} at the time $t_{\rm decay} \sim H^{-1} \sim \Gamma_\sigma^{-1}$ and we expect it to decay democratically to Standard Model particles and their super-partners.  
Any heavy super-partners produced will typically decay rapidly into the lightest SUSY partner (LSP), which is stable and will provide the dark matter candidate (which we will denote as $X$). 

In addition to dark matter, light standard model particles that are produced will thermalize and `reheat' the universe for a second time (with inflationary reheating occurring early and at high temperature).  The corresponding reheat temperature is given by 
\be \label{reheattemp}
T_r \approx \sqrt{\Gamma_\sigma \, M_{pl}} \approx c_3^{1/2} \sqrt{\frac{m_\sigma^3}{M_{pl}}}
\ee
and this temperature must be larger than around $3$ MeV to be in agreement with BBN light element abundances \cite{Kawasaki:1999na}. 

If the number density of dark matter particles produced in the decay
 is larger than the critical value approximately given by
\be \label{nc}
n_c = \frac{H}{\langle \sigma v \rangle}\, ,
\ee
then rapid self annihilations will take place until the dark matter abundance reduces to this value, which acts as an attractor value\footnote{In cases where the dark matter abundance is subcritical, then no annihilations take place.  This case is model-dependent, but usually occurs when the moduli do not come to dominate the energy density and/or if decays to super-partners are highly suppressed.  In the former case, this can lead to a large generation of isocurvature as we discuss in the next section.}.   In the equation above, $\langle \sigma v \rangle$ is the averaged annihilation rate and velocity at the time of decay.
Once the fixed point $n_c$ is reached, the resulting abundance of {\it non-thermally produced} dark matter is found to be
\begin{eqnarray}
\Omega_{\rm LSP}h^2  \approx 0.12 \times \left( \frac{10^{-26} \cms}{\langle \sigma v \rangle } \right) \left( \frac{T_f}{T_r} \right),
\label{eqn:lsp_annihilation_fixedpoint}
\end{eqnarray}
where $T_f$ is the freeze-out temperature of thermal dark matter (around a few GeV), and $T_r$ is the reheat temperature following the modulus decay and can be as small as a few MeV.  

A second possibility is that the yield of dark matter from scalar decay is sub-critical $n<n_c$.  In this case, the amount of dark matter depends on the initial scalar density.
If $B_\sigma$ is the branching ratio for decay to superpartners then the amount of dark matter after decay is $\rho_{DM} \sim B_\sigma (m_{DM} / m_\sigma ) \rho_\sigma(t_d)$, where $\rho_\sigma(t_d)$ is the scalar energy density before decay.
As an example, if the modulus dominates the energy density before decay, the comoving amount of dark matter will be independent of its cross-section and will depend primarily on the masses $\rho_{DM} / s(t_r) \sim m_{DM} (m_\sigma / M_{pl})^{1/2}$. Whether one has the sub-critical or super-critical case, depends in practice on the initial displacement of the modulus as this determines the amplitude of oscillations and the amount of energy stored in the oscillations \cite{Moroi:1999zb}.

If the LSP constitutes all of the dark matter we must require $\Omega_{\rm LSP}h^2 = \Omega_c h^2 = 0.1199 \pm 0.0027$ \cite{Ade:2013zuv}. In a standard cosmology where dark matter has a thermal origin this implies $\langle \sigma v \rangle = \langle \sigma v \rangle_{\rm std} \approx 10^{-26} \cms$. However, when dark matter has a non-thermal origin this number can be larger.  

As an example, if we consider a non-thermal history where the modulus decay reheats the universe to a temperature $T_r \approx 10$ MeV, which is significantly below the freeze-out temperature of a typical $250$ GeV WIMP  $T_f \approx m_X / 25 \approx 10$ GeV we find $\langle \sigma v \rangle = 1000 \, \langle \sigma v \rangle_{\rm std}$. For a given candidate (like the MSSM neutralino) this leads to new and interesting predictions for experiments probing the microscopic properties of dark matter such as indirect detection, direct detection, and LHC searches. 

\subsection{Dark Radiation Production \label{DRsection}}
In addition to moduli decays to standard model particles and their superpartners, there may also be decays to hidden sector fields. 
Indeed, this is a common expectation in string-based models that give rise to the non-thermal history for dark matter discussed above
\cite{Higaki:2012ar,Cicoli:2012aq,Higaki:2012ba}. If the particles resulting from decay are light (meaning relativistic)
at the time of BBN and/or recombination, and non-interacting with MSSM particles, this leads to additional radiation coming from the hidden sector.
If these particles contribute substantially to the energy density they will affect 
the expansion rate changing predictions for both the abundances of primordial elements \cite{Steigman:2012ve} and the physics of the CMB~\cite{Abazajian:2013oma}.
Thus, using precision cosmological measurements one can establish constraints 
on the amount of dark radiation that is permitted within a particular class of models -- see \cite{Conlon:2013isa} and references within.

During radiation domination after the decay of the lightest modulus the effect of the hidden sector radiation on the Hubble expansion can be understood through the Hubble equation
$3 H^2 M_{pl}^2 = \rho_r$, where the total relativistic contribution
to the energy density is 
\be \label{basicr} 
\rho_r = \frac{\pi^2}{30}  g_\ast T^4,
\ee
with
\be \label{gdef}
g_\ast = g_{\mbox{\tiny MSSM}} + \sum_{i=bosons} g_i^h \left( \frac{T^h_i}{T} \right)^4 +\frac{7}{8} \sum_{i=fermions} g^h_i \left( \frac{T^h_i}{T} \right)^4
\ee
where the sums are over relativistic hidden sector particles with $g^h_i$ degrees of freedom, the factor of $7/8$ results from Fermi-Dirac statistics of fermions, $T^h$
is the temperature of the hidden sector particles (which importantly need not be equilibrated with standard model radiation), and $g_{\mbox{\tiny MSSM}} $ 
is the visible sector relativistic degrees of freedom, which in the early universe would be at least $g_{\mbox{\tiny MSSM}}=228.75$ in the MSSM, but near the MeV scale only  
the photons and neutrinos contribute with  
\be
g_{\mbox{\tiny MSSM}}(T) = g_\gamma + \frac{7}{8} g_\nu N_{\rm eff} \left( \frac{T_\nu}{T} \right)^4,
\ee
where $g_\gamma =2 $ for the photon, $g_\nu = 2$ for neutrinos, and $N_{\rm eff}$ is the effective number of neutrino species at temperature $T_\nu$.

At the time of BBN, the neutrino temperature tracks the photons so that $T_\nu = T$ and so with three relativistic neutrinos, $N_{\rm eff} =3$, the standard model prediction is $g_{\mbox{\tiny MSSM}}=7.25$.  However, because the neutrinos are weakly interacting $\langle \sigma_\nu v \rangle \sim G_F^2 T^2$
with $G_F \sim 10^{-5}$ GeV$^{-2}$ Fermi's constant, they decouple below the temperature of BBN ($\sim$ MeV) and the entropy in photons increases (so that the total entropy is conserved).
At the time of recombination we have
\begin{eqnarray}
g_{\mbox{\tiny MSSM}}(T_{rec}) &=& g_\gamma + \frac{7}{8} g_\nu N_{\rm eff} \left. \left( \frac{T_\nu}{T} \right)^4 \right\vert_{T=T_{rec}} \nonumber \\
  &=& 2 + \frac{7}{8} \cdot 2 \cdot \left(3.046 \right) \left( \frac{4}{11} \right)^{4/3} \nonumber \\
  &=&3.385,
\end{eqnarray}
where $N_{\rm eff}=3.046 \neq 3$ accounts for a small injection of entropy into neutrinos coming from electron/positron annihilations prior to recombination, as well as energy dependent distortions and and small finite temperature corrections from the plasma \cite{Abazajian:2013oma}.  The increase in the photon entropy following neutrino decoupling leads to an increase in the temperature of $T_\nu / T=(4/11)^{1/3}$ \cite{Abazajian:2013oma}. Given the standard model (MSSM) predictions, any new dark radiation would lead to an additional contribution to $g_\ast$ or 
$N_{\rm eff}$.  For historic reasons, constraints on new hidden sector radiation are typically expressed through $N_{\rm eff}$.

Constraints from BBN result from requiring agreement with both the abundances of $^4He$ and $D$ \cite{Cyburt:2004yc}, which implies $N_{\rm eff}=3.24 \pm 1.2$ at the time of BBN.
At the time of recombination, the Planck satellite \cite{Ade:2013zuv} provides constraints with the current results implying 
$N_{\rm eff}=3.30 \pm 0.27$.  If we make the additional assumption that all radiation was initially in equilibrium with standard model photons (so that the temperature of all relativistic species is the same)
then this corresponds to the bound $g_\ast = 3.50 \pm 0.12$, with a final projected sensitivity for $g_\ast$ of $\pm 0.09$ \cite{Galli:2010it}.  However, as pointed out in \cite{Feng:2008mu} these bounds can be significantly relaxed if the hidden sector radiation does not share the photon temperature. 

For example, if hidden radiation couples different to decaying moduli (or the inflaton during reheating) than standard model particles this 
can lead to different temperature for each species and this will be preserved in the absence of interactions between the systems of particles.
As we will see in the next section, the situation where $T^h \neq T$ and thermal equilibrium with photons is not reached is interesting for the case of isocurvature perturbations.
In such a case, the Planck constraint on $N_{\rm eff}$ from recombination ( and using (\ref{gdef}) ) implies the bound
\be \label{planckg}
g^h_* \left( \frac{T^h_{rec}}{T_{rec}} \right)^4= \frac{7}{8} \cdot 2 \cdot \left( N_{\rm eff} - 3.046 \right) \left. \left( \frac{T_\nu}{T} \right)^4  \right\vert_{T=T_{rec}}\le  0.24
\ee
where (following convention) we have treated the extra radiation as a neutrino species and subtracted the contribution from standard model neutrinos.
Instead we can express this constraint in terms of $\Delta N_{\rm eff}$ where
\be
\Delta N_{\rm eff} = \frac{8}{14} \Delta g_* \left( \frac{T^h_{rec}}{T_\nu} \right)^4 \le 0.42
\ee
with $\Delta g_* \equiv g^h_*$ and if
the hidden sector radiation shares a common temperature with photons then $T^h_{rec}/T_\nu = (11/4)^{1/3}$.
Thus, the $1\sigma$ upper value from Planck of $N_{\rm eff} \le 3.57$ implies an upper bound on the combination of $g^h_*$ and the departure in temperature from the standard model thermal bath.

We close our brief review of  hidden sector radiation by considering the example of axions, which
represent a well-motivated and simple example of dark radiation (c.f. \cite{Conlon:2013isa} and references within).
We can revisit the non-thermal history resulting from moduli decay above, but this time allowing for decay to axions as well.
If we denote by $B_\sigma$ and $B_a$ the branching fraction to dark matter and radiation, respectively, then the remaining fraction to standard model particles is simply $1-B_\sigma-B_a$.
If we consider for simplicity the case that the moduli dominate before decay (and with no dark matter annihilations), then the density in axions will be $\rho_a \sim B_a H_d^2 M_{pl}^2$,
where $H_d \sim \Gamma_\sigma$ is the expansion rate at decay given by (\ref{decayrate}).
Comparing this to the energy density in standard model radiation $\rho_{sm} \sim (1-B_a-B_\sigma) H_d^2 M_{pl}^2$, we can find a constraint on the branching ratios through constraints on $\Delta N_{\rm eff}$.

Following convention and treating the axions as an effective neutrino species, we can use 
(\ref{basicr}), (\ref{gdef}) and (\ref{planckg}) to write the total radiation density as
\begin{eqnarray}
\rho_r &=& \rho_{\rm MSSM} + \rho_a, \nonumber \\
&=& \rho_{\rm MSSM} \left( 1 + \frac{7}{8} \frac{g_{\nu}}{g_{\rm MSSM}} \Delta N_{\rm eff} \left( \frac{T_\nu}{T} \right)^4 \right),
\end{eqnarray}
where $\rho_{\rm MSSM} = \pi^2  g_{\rm MSSM} T^4/ 30$ and $g_\nu=g_\gamma=2$. Identify the axion density with the second term above and inverting 
the expression we have
\be \label{anN}
\Delta N_{\rm eff} = \frac{8}{7} \frac{\rho_a(T)}{\rho_{\rm MSSM}(T)} \left( \frac{T}{T_\nu} \right)^4 \left( \frac{g_{\rm MSSM}(T)}{g_\gamma(T)} \right)
\ee
Following their production the axion's entropy remains fixed and so they simply scale with the expansion as
\be
\rho_a(T) = \rho_a(T_r) \left( \frac{a(T_r)}{a(T)} \right)^4,
\ee
where $T_r$ is the reheat temperature (\ref{reheattemp}). The entropy in photons (MSSM sector) will change following neutrino decoupling, since positrons and electrons will freeze-out
so that $g_{\rm MSSM}$ decreases, while the temperature must increase so that the comoving entropy $S \sim a^3 T^3 g_{\rm MSSM}$ remains constant\footnote{Electron and positron freeze-out
occurs on microscopic time scales so that the cosmic expansion during this event is negligible, i.e. we can take $a=1$ in the expression for the entropy}.  It follows that 
\be
\rho_{\rm MSSM}(T) = \rho_{\rm MSSM}(T_r) \left( \frac{g_{\rm MSSM}(T_r)}{g_{\rm MSSM}(T)} \right)^{1/3} \left( \frac{a(T_r)}{a(T)} \right)^4.
\ee
Using these expressions in (\ref{anN}), and the expression for the energy density in axions and MSSM radiation at the time of reheating, we have
a constraint at the time of recombination 
\begin{eqnarray}
\Delta N_{\rm eff} &=& \frac{8}{7} \left( \frac{11}{4} \right)^{4/3} \left(  \frac{B_a}{1-B_a-B_\sigma} \right) \left( \frac{g_{\rm MSSM}(T_r)}{g_{\rm MSSM}(T_{rec}) }\right)^{1/3} \nonumber \\
&\le& 0.42,
\label{eqn:delta_neff_scott}
\end{eqnarray}
where again we have used $T/T_\nu = (11/4)^{1/3}$ at recombination $T=T_{rec}$, and assumed that the modulus dominates the energy density.

%%%%%%%%%%%%%%%%%%%%%%%%%%%%%%%%%%
%%%%%%%%%%%%%%%%%%%%%%%%%%%%%%%%%%
\subsection{Curvature and Isocurvature Perturbations in Non-thermal Histories \label{pertsection}}
After a brief review of isocurvature perturbations, in this subsection we consider the non-thermal histories discussed above to establish 
how well existing isocurvature constraints restrict SUSY model building in the presence of moduli and identify 
the corresponding observationally interesting cases. 

One can assign a curvature perturbation to each species, $i$, which is defined such that it is exactly conserved on super-horizon scales in the adiabatic limit, when the expansion is dominated by a single species (i.e. once the universe is radiation dominated after modulus decay):
\be
\zeta_i = -\Psi-H\frac{\delta\rho_i}{\dot{\rho}_i} \, .
\ee
where $\Psi$ is the Newtonian potential and dots denote derivatives with respect to cosmic time $t$ (see Appendix~\ref{appendix:cosmo_pert} for conventions used). From this we find the total conserved curvature perturbation
\be
\zeta=\frac{\sum_i(\rho_i+P_i)\zeta_i}{\sum_i(\rho_i+P_i)} \, .
\ee
Then, a gauge invariant definition of an isocurvature perturbation between two fluids $\rho_i$ and $\rho_j$ is given by (e.g. \cite{Malik:2008im})
\be
S_{ij} = 3 \left( \zeta_i - \zeta_j \right).
\ee
In connecting with observations it is convenient to instead define the isocurvature contribution of a particular fluid relative to the total curvature, which in the radiation dominated, post modulus decay universe is approximately given by that in radiation $\zeta_R\approx \zeta$, so that
\be \label{entropyr} 
S_{i} = 3 \left( \zeta_i - \zeta \right)\approx 3 \left( \zeta_i - \zeta_R \right)=S_{iR} \, ,
\ee
where $\zeta_R$ is the spatial curvature on surfaces of constant standard model (MSSM) radiation density.

During inflation, if the mass of the modulus is lighter than the Hubble scale 
\be
m^2_\sigma(\langle \sigma \rangle) \lesssim H_I^2 \, ,
\ee
 then the quasi-deSitter period will result in long-wavelength fluctuations of the field with an average amplitude (e.g. \cite{Linde:2005ht}) 
 \be
 \delta \sigma \sim {H_I}/{2\pi} \, .
 \ee 
This leads to an additional source of cosmological perturbations different from that sourced by the inflaton. Following inflationary reheating -- where typically all of the energy and matter of the universe is assumed to be created -- the modulus can decay leading to an additional source of radiation and matter. Thus, whereas radiation and matter created during inflationary reheating will inherit the inflaton's fluctuation $\zeta_I$, those produced from moduli decay will instead be set by $\delta\sigma$ that initially carries no curvature, implying the existence of isocurvature modes. 

Although we have seen that the moduli decay is an essential part of a non-thermal history, there are still many ways in which isocurvature perturbations may be observationally irrelevant and lead to no new constraints on model building. Firstly, if the mass of the modulus is above the Hubble scale during inflation $m_\sigma (\langle \sigma \rangle) > H_I$, then in a single Hubble time the amplitude of its fluctuations will be exponentially suppressed on large scales by a factor $\exp(-m_\sigma^2 / (3 H_I^2) )$ and so the inflaton will be the only relevant source of cosmological fluctuations \cite{Linde:2005ht}. Another important observation was made by Weinberg, who demonstrated that even if an isocurvature mode is generated initially, if local thermal equilibrium is reached these modes will become adiabatic  \cite{Weinberg:2004kf}. And finally, if the modulus comes to dominate the energy density of the universe (determining the cosmic expansion rate) this also has the effect of washing out any existing isocurvature perturbations. 

To make some of these ideas more precise and establish the cases of observational interest for the rest of the paper we closely follow the formalism of \cite{Langlois:2011zz}.
We will be interested in the decay of moduli into MSSM and hidden sector particles. We define the branching ratio of the decay from moduli to species $\rho_i$ as $B_{i} \equiv \Gamma_i / \Gamma$. Thus, if before the decay the abundance of a species is $\Omega_i^{(0)}$ then the fraction of particles created by the decay is 
\be 
f_i \equiv \frac{ B_i \Omega^{(0)}_\sigma}{\Omega_i^{(0)} +B_i \Omega^{(0)}_\sigma},
\ee
where $\Omega^{(0)} = \rho^{(0)}_\sigma / \rho$ is the initial abundance in moduli compared to the total energy density $\rho$. Thus, if no $\rho_i$ particles are produced in the decay we have $f_i=0$, whereas if all of them are produced then we have $f_i=1$. 

Assuming the moduli scale as pressureless matter ($P_\sigma=w_\sigma \rho_\sigma$ with $w_\sigma=0$) we can express the curvature perturbation for a fluid $\rho_i$ following moduli decay compared to its value before $ \zeta_i^{(0)}$ as
\be
\zeta_i = \sum_j T^{\; j}_i \zeta_j^{(0)},
\ee
where the matrix elements are given by (no trace)
\begin{eqnarray} \label{matrix_elements}
T_{i}^{\;i} &=& 1-f_i+f_i \frac{w_i \Omega^{(0)}_i}{\sum\limits_l (1+w_l) \Omega^{(0)}_l}, \nonumber \\
T_{i}^{\; \sigma}  &=& \frac{f_i}{1+w_i} + f_i \left(  \frac{w_i}{1+w_i} \right) \frac{\Omega^{(0)}_\sigma}{\sum\limits_l (1+w_l) \Omega^{(0)}_l},  \nonumber \\
T_{i}^{\; j}  &=& f_i \left( \frac{w_i (1+w_j)}{1+w_i}\right)  \frac{\Omega^{(0)}_j}{\sum\limits_l (1+w_l) \Omega^{(0)}_l}  \; \; \left( j \neq i,\sigma\right), \;\; \;\;\;\;\;\;
\end{eqnarray}
where the sum is over all significant contributions to the energy density prior to decay.
Using that $\sum_i T^{i}_j=1$ for any $j$, we can then express (\ref{entropyr}) in terms of the matrix elements as
\be  \label{entpert}
S_{ir}=\sum\limits_{j} \left( T^{\; j}_i - T_r^{j} \right) S_j^{(0)},
\ee
where $S_j^{(0)}$ is the entropy perturbation before the decay. 

To see the utility of this approach, consider a three fluid system similar to that expected by the non-thermal history discussed above.
Again treating the scalar oscillations as pressure-less matter, we will have three fluids $\rho_\sigma$, $\rho_{DM}$, and $\rho_R$, which are energy densities of modulus, DM, and radiation, respectively.
The matrix elements (\ref{matrix_elements}) then become
\be \label{thematrix} 
\left(\begin{array}{ccc}
1-f_R+\frac{1}{3}f_R \left( \frac{\Omega^{(0)}_R}{\Omega^{(0)}_T} \right) & \frac{1}{4}f_R \left( \frac{\Omega^{(0)}_{DM}}{\Omega^{(0)}_T}\right) & \frac{3}{4} f_r + \frac{1}{4}f_R \left(\frac{\Omega^{(0)}_\sigma}{ \Omega^{(0)}_T}\right) \\
0 & 1-f_{DM} & f_{DM}  \\
0 & 0 & 0
\end{array}\right)
\ee
where the matrix $T$ above is written in the basis $\zeta = (\zeta_R, \zeta_{DM}, \zeta_\sigma)$ so that $\zeta = T \zeta^{(0)}$ gives the curvatures after the transition and
\be
\Omega^{(0)}_T= \sum\limits_l (1+w_l) \Omega^{(0)}_l=4\Omega_R^{(0)}/3 + \Omega_{DM}^{(0)}+\Omega_\sigma^{(0)},
\ee
is the total weighted relic abundance prior to decay.

The general dark matter isocurvature perturbation following moduli decay is then given by substituting (\ref{thematrix}) into (\ref{entpert})
and we have 
\begin{eqnarray}
S_{DM,R}&= -&\left[ 1-f_R \left( 1-\frac{\Omega^{(0)}_R}{3\Omega^{(0)}_T} \right) \right] \zeta_R^{(0)} \nonumber \\
&+&\left[ 1-f_{DM} -\frac{f_R}{4} \left(\frac{\Omega_{DM}^{(0)}}{\Omega^{(0)}_T} \right)\right] \zeta_{DM}^{(0)}\nonumber \\
&+& \left[ f_{DM} - \frac{1}{4}f_R \left(  3+\frac{\Omega_\sigma^{(0)}}{\Omega_T^{(0)}}  \right)  \right] \zeta_\sigma^{(0)}\, .
\label{eqn:sdmr_scott}
\end{eqnarray}
From this expression we can immediately see the statement earlier that if the modulus comes to dominate the energy density, with $\Omega_T^{(0)}=\Omega_\sigma^{(0)}$, and all of the dark matter and radiation are produced in the decay, $f_{DM}=f_R=1$, then the isocurvature mode above vanishes, $S_{DM,R}=0$. In the other limit, where radiation and dark matter are not produced in the decay $f_R=f_{DM}=0$ we have
\be
S_{DM,R}=\zeta_{DM}^{(0)}-\zeta_R^{(0)},
\ee
and so the existence of an isocurvature mode depends on whether one was initially imprinted.  Thus, if these (prior to decay) sources were produced during inflationary reheating and where thermal equilibrium was established then the isocurvature perturbation vanishes as $\zeta_{DM}^{(0)}=\zeta_R^{(0)}=\zeta_I$ at the time of inflationary reheating where $\zeta_I$ is the curvature fluctuation of the inflaton. When modulus decay from the mode with $\delta\sigma\neq 0$ is included multiple sources of curvature perturbations are present and can generate a non-zero $S_{DM,R}$.

Thus, the cases we will be interested in here correspond to when the modulus does not completely dominate, and/or when the MSSM and any dark sector particles come from multiple sources.  As an example of the latter, some dark matter will be produced thermally in the early universe and moduli decay will lead to an additional source of dark matter.  If the modulus does not come to dominate, we will see this can generate a substantial isocurvature perturbation. Isocurvature requires the modulus to be subdominant prior to decay and this will require us to consider moduli fields with sub-Planckian displacements $\Delta \sigma \ll M_{pl}$ -- since otherwise complete moduli domination will be inevitable. This corresponds in (\ref{potentialex}) to the case where the flat direction is lifted by a low dimension operator and/or the scale of new physics is taken significantly below the Planck scale.

%In this case the isocurvature is said to be {\em uncorrelated} because it comes from different sources.  Whereas, if the modulus decay provides a common source, but never-the-less leads to isocurvature, this is termed {\em correlated}.  In practice, correlated i
\section{Modulus Decay and Correlated Isocurvature}
\label{sec:iso_production}

Having reviewed the instances where isocurvature and dark radiation can be generated in non-thermal cosmologies, we now examine these cases in more detail with emphasis on how these two phenomena can provide complementary constraints. We begin by presenting the background equations and then show how we compute cosmological parameters. The background and perturbations define a system of coupled O.D.E.s that we solve numerically. Details of our numerical procedure to treat the evolution of the perturbations, including equations of motion and initial conditions, are relegated to Appendix~\ref{appendix:computation}.

\subsection{Background Evolution and Parameters}
\label{sec:background}

The background is given by unperturbed flat FRW space, in physical time 
\be
ds^2 = -dt^2 + a(t)^2 \delta_{ij} dx^i dx^j \, .
\ee
The evolution of the scale factor, $a(t)$, is fixed by the Friedmann equation
\be
H^2=\left( \frac{\dot{a}}{a}\right)^2=\frac{1}{3 M_{pl}^2} \sum_i \rho_i \, ,
\ee
where $i$ runs over all species of energy density: $R=$ all standard model and MSSM (visible sector, VS) radiation, including $N_{\rm eff}=3$ massless neutrinos \footnote{Using $N_{\rm eff}=3$ rather than $N_{\rm eff}=3.04$ will have only a minor effect on our value of $\Delta N_{\rm eff}$ for dark radiation.}; $DM =$ dark matter; $DR =$ dark radiation; $\sigma=$ the lightest modulus field \footnote{Our treatment assumes that the heavier moduli decayed producing no isocurvature. Fixing the amplitude of scalar perturbations using inflationary parameters assumes in addition that they did not alter the curvature spectrum, i.e. that they decayed before or during inflation.}. In practice, rather than using time $t$ to evolve our equations, we will use the number of e-folds since the initial time, $N=\ln (a/a_i)$, which absorbs the scale factor normalisation.

We ignore the effects of the baryons. In the background evolution they are sub-dominant, while in the perturbations they are tightly coupled to the photons. Baryon isocurvature modes have the same spectrum as DM isocurvature, and can be accounted for with appropriate scaling \cite{Ade:2013zuv}.

We take the universe to be initially dominated by radiation, which can be assumed to have originated either from the decay of the inflaton, or of the next to lightest modulus, and to contain no DM or DR\footnote{For numerical stability in our code we begin with tiny amounts of DM and DR.}. Following inflationary reheating the modulus field, $\sigma$, has a quadratic effective potential $V(\sigma)=m_\sigma^2 \sigma^2/2$ and begins displaced from the minimum at $\sigma=0$ by some initial value $\sigma(t_i)=\sigma_\star$ and with no initial velocity, $\dot{\sigma}(t_i)=0$.

The energy density and pressure of the modulus field are given by:
\begin{align}
\rho_\sigma &= \frac{1}{2}\dot{\sigma}^2+\frac{1}{2}m_\sigma^2 \sigma^2 \, , \\
P_\sigma &= \frac{1}{2}\dot{\sigma}^2-\frac{1}{2}m_\sigma^2 \sigma^2 \, . 
\end{align}

At early times the modulus field evolves according to the free Klein-Gordon equation
\be \label{KG}
\ddot{\sigma}+3H \dot{\sigma}+m_\sigma^2 \sigma=0 \, ; \quad m\lesssim H .
\ee
The field begins frozen at the initial displacement, $\sigma_\star$. Once the mass overcomes the Hubble friction the modulus begins to roll in its potential and oscillates about the minimum. At this point perturbative decay of the modulus begins, which can be taken into account by introducing an additional friction term given by the modulus decay rate, $\Gamma_\sigma$ \cite{Kofman:1997yn}\footnote{We do not consider the possibility of parametric resonance during this decay, though it may lead to further interesting phenomenology. Given the criterion for the onset of parametric resonance presented in \cite{erickcek2011}, we are safe with our assumption as the majority of our parameter space will satisfy the required bound $\Gamma_\sigma / m_\sigma \ll (m_\sigma / M_{pl})^2$ for negligible parametric instability. }. 
\be
\ddot{\sigma}+(3H +\Gamma_\sigma)\dot{\sigma}+m_\sigma^2 \sigma=0 \, ; \quad t>t_{\rm osc} .
\ee
We define the time when this term is introduced, $t_{\rm osc}$, to be given by the first passage of the modulus field through the minimum of the potential.

Once coherent oscillations begin the average pressure in the modulus field goes to zero, causing it to behave like matter \cite{Turner:1983he}, while the energy density evolves according to the conservation equation
\be \label{conseqn}
\dot{\rho}_\sigma +3H \rho_\sigma = -\Gamma_\sigma \rho_\sigma \, .
\ee 
It is computationally impractical to evolve the two time scales $t_{\rm osc}$ and $\Gamma_\sigma^{-1}$ when $\Gamma_\sigma$ is given by Eq.~(\ref{decayrate}). Therefore after $t_{\rm osc}$ we use Eq.~(\ref{conseqn}) rather than Eq.~(\ref{KG}) to evolve the modulus energy density. We will use a similar approximation for the perturbations, and show in Appendix~\ref{appendix:accuracy_approx} that neither approximation has a substantial effect on our results.

Prior to modulus decay, the other components evolve according to the free conservation equations, while during decay they are sourced by the modulus:
\begin{align}
\dot{\rho_i}+3H (1+w_i)\rho_i &=0 \, ; \quad{t<t_{\rm osc}}, \\
\dot{\rho_i}+3H (1+w_i)\rho_i &=B_i \Gamma_\sigma \rho_\sigma \, ; \quad{t>t_{\rm osc}}, \label{eqn:densities_source}
\end{align}
where $w_i$ is the equation of state for the species and $B_i$ gives the branching ratio of the modulus to species $i$. By conservation of energy $\sum_i B_i=1$. The term $B_i\Gamma_\sigma \rho_\sigma$ accounts for particle production of species $i$ by modulus decay. In addition to decays, dark matter annihilations can also play an important role particularly if the modulus dominates the energy density prior to decay and the branching ratio to dark matter is large.  The effect of annihilations can be captured in an ``effective decay rate", which is how we will deal with them here. As the annihilations happen in less than a Hubble time, their primary effect is to simply reduce the amount of dark matter to that given by (\ref{nc}) and increase the amount of radiation. 

If the modulus comes to dominate the energy density, Eq.~(\ref{eqn:densities_source}) can be solved to give the evolution of $\rho_i (a)$ with two distinct scaling regimes \cite{erickcek2011}:
\be
\rho_i(a) = \rho_{i,\Gamma} a^{-3/2} + \rho_{i,\rm init} a^{-3(1+w_i)} \, .
\label{eqn:densities_source_solution}
\ee
The term proportional to $\rho_{i,\Gamma}$ represents energy density in species $i$ produced by modulus decay, while $\rho_{i,\rm init}$ represents energy density present already (e.g. from inflationary reheating). When the modulus is totally dominant in the energy density, the $a^{-3/2}$ component is universal across species, and dominates their evolution, while as the modulus goes between dominance and sub-dominance the species dependent $a^{-3(1+w_i)}$ term dominates. These scalings will effect the sensitivity of isocurvature observables, which depend on modulus energy density fraction, to the modulus parameters.

The branching fractions to standard model radiation, DM and DR give the reheat temperature and abundances, which we define some number of e-foldings $N_{\rm end}$ after the initial time, when the modulus has decayed completely. After this time the conservation of energy conserves the abundances as in a standard thermal cosmology.  $N_{\rm end}$ is defined by
\be
\frac{\rho_\sigma (N_{\rm end})}{\rho_{DM}(N_{\rm end})}:=10^{-2} \, ,
\ee
when the modulus has decayed such that it is sub-dominant to the DM. Since the modulus is initially dominant over the DM, which to give a standard cosmology with BBN and equality at the correct temperatures is itself substantially sub-dominant to the radiation, this condition guarantees that modulus decay has been completed. The DM abundance, reheat temperature, and amount of dark radiation parameterised by $\Delta N_{\rm eff}$ are all evaluated at $N_{\rm end}$.

The effective value of $B_{DM}$ sets the DM abundance by giving the value of $\rho_{DM}(N_{\rm end})/\rho_{R}(N_{\rm end})$. While this should not vary too much around its central Planck value set by $z_{\rm eq}$, it is strictly a free parameter and we should expect its central value to change in any non-standard cosmology. We choose $B_{DM}$ small to give a reasonable DM abundance, and find that changing its value in any sensible range does not affect the evolution of the perturbations. Firstly, since DM is always sub-dominant prior to $N_{\rm end}$ the actual DM abundance does not affect the expansion rate. Secondly, by assumption we consider only cases where decay of the modulus sources practically all of the DM, either by sub-critical or super-critical production depending on the cross-section. Therefore $f_{DM}$ in Eq.~(\ref{eqn:sdmr_scott}) is always close to unity and $B_{DM}$ does not affect the isocurvature observables (we do not assume the same for radiation: modulus dominance or sub-dominance affects $f_R$). $B_{DM}$ and the DM abundance will play no further role in our analysis, except to stress again that when the modulus dominates the energy density and $T_r<T_f$ the DM is by necessity non-thermal in origin. 

The `reheat temperature'\footnote{Since we allow for decay when the modulus is sub-dominant this is a slightly liberal use of the term.}, $T_r$, is found from the energy density of radiation at $N_{\rm end}$:
\be
\rho_R(N_{\rm end}) = \frac{\pi^2}{30}g_\ast^{VS} (T_r)T_r^4 \, .
\ee
For any finite range of $T$ we assume $g_\ast (T)$ to be a constant and the inversion to find $T_r$ is trivial. Our reheat temperatures are often low, $T_r\lesssim 10$ GeV, and so we take $g_\ast^{VS}(T)=g_\ast^{SM}(T)$ to be given by the Particle Data Group, Ref.~\cite{Beringer:1900zz}. Taking this model for $g_\ast$ only affects $T_r$ and, as we discuss below, $\Delta N_{\rm eff}$, and does not affect the isocurvature observables defined below.

We choose to reject all models where $T_r<T_{BBN}\sim 3$ MeV. This is a hard cut if the modulus is dominating the energy density just prior to decay and is the origin of the Cosmological Moduli Problem (CMP) (see \cite{Watson:2009hw,Acharya:2008bk} and references therein): a non-thermal universe with decaying moduli can ruin the successful predictions of BBN. However, if the modulus is sub-dominant at the time of decay there is a continuous region in parameter space connecting our model to effects in the late universe where decay can occur much later or not at all, such as is the case for (early) dark energy and axion dark matter\footnote{For example, the modulus can decay in a radiation (or DM) dominated universe at very low temperature long after BBN has completed. Such effects should be thought of as decaying Dark Energy, and since cosmological modes will then be entering the horizon these effects should be computed using a Boltzmann code.} (e.g. \cite{marsh2011,marsh2012}). In such a case, the choice taking $T_r>T_{BBN}$ in our model is just a matter of definition separating models of `initial conditions' from models affecting late universe physics.

As discussed in Section \ref{DRsection}, the dark radiation abundance is parameterized by $\Delta N_{\rm eff}=\rho_{DR}/\rho_{1\nu}$ evaluated at neutrino decoupling. This can be obtained from the DR energy density at the time of modulus decay at $N_{\rm end}$, given $g^{VS}_\ast(T)$, using the fact that $\rho_R\propto g_{\ast}^{-1/3} a^{-4}$ (e.g. \cite{Choi:1996vz})
\begin{align}
\Delta N_{\rm eff} &:= \frac{\rho_{DR}(T_\nu)}{\rho_{1\nu}(T_\nu)} \, , \\
&=\frac{g_{\ast}^{VS}(T_\nu)}{g_{\ast,1\nu}(T_\nu)}\left( \frac{g_\ast^{VS}(T_\nu)}{g_\ast^{VS}(T_{r})} \right)^{1/3}\frac{\rho_{DR}(T_{r})}{\rho_R (T_{r})} \, , \\
&=\frac{43}{7}\left( \frac{10.75}{g_\ast^{VS}(T_{r})} \right)^{1/3}\frac{\rho_{DR}(T_{\rm r})}{\rho_R (T_{r})} \, .
\end{align}
Since we allow for the possibility that the modulus does not dominate the energy density at the time of decay we cannot set the ratio of dark to standard model radiation evaluated at $T_{r}$ equal simply to the ratio of branching ratios, as in Eq.~\ref{eqn:delta_neff_scott}. We must therefore evaluate $\Delta N_{\rm eff}$ numerically for each choice of parameters. However, when the modulus is dominant at the time of decay, the amount of DR produced can be computed analytically and approaches the asymptotic value:
\be
\Delta N_{\rm eff} = \frac{43}{7}\left( \frac{10.75}{g_\ast^{VS}(T_{r})} \right)^{1/3}\frac{B_{DR}}{1-B_{DR}} \, ,\quad \text{(dominant decay).}
\label{eqn:delta_neff_dominant}
\ee

The parameters of the background evolution are specified by $\{\sigma_\star, m_\sigma, \Gamma_\sigma, B_{DR}\}$, from which we compute $T_r$ and $\Delta N_{\rm eff}$. Once the initial conditions for the perturbations are fixed by inflationary parameters giving normalisation and spectral indices of the power spectra, the background evolution determines the evolution of the perturbations. Therefore the final amplitudes and correlations between the isocurvature modes are fixed by these same basic parameters of the background evolution.

\begin{figure}[htbp!]
\centering
\includegraphics[scale=0.42]{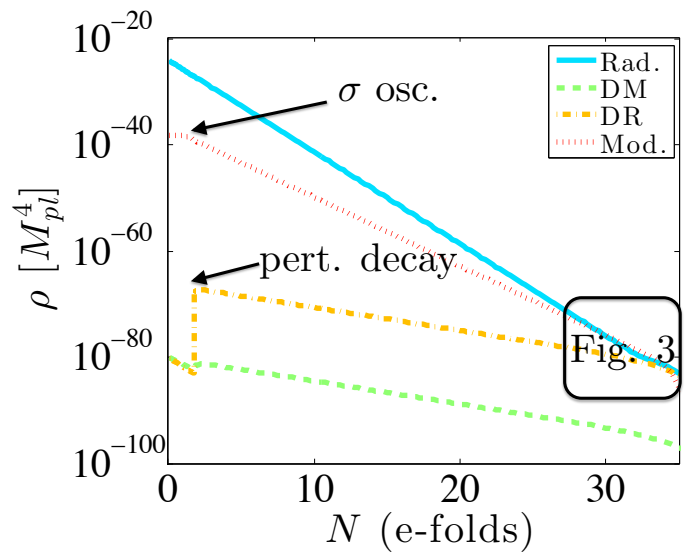}
\caption{Density evolution for dominant modulus decay. The modulus is frozen by Hubble friction very early on, in a universe dominated by radiation from the prior reheating event. Once oscillations begin, perturbative decay sources the DM, DR and a sub-dominant component of radiation. Gravitational coupling leads to low decay rates, so that eventually the modulus comes to dominate the energy density. Its decay sources the final reheating at $T_r\sim T_{BBN}$, shown in Fig.~\ref{fig:density_zoom}, left panel.}\label{fig:density_535_1_all}
\end{figure}

\begin{figure*}[htbp!]
\centering
$\begin{array}{@{\hspace{-0.2in}}l@{\hspace{+0in}}l}
\includegraphics[scale=0.4]{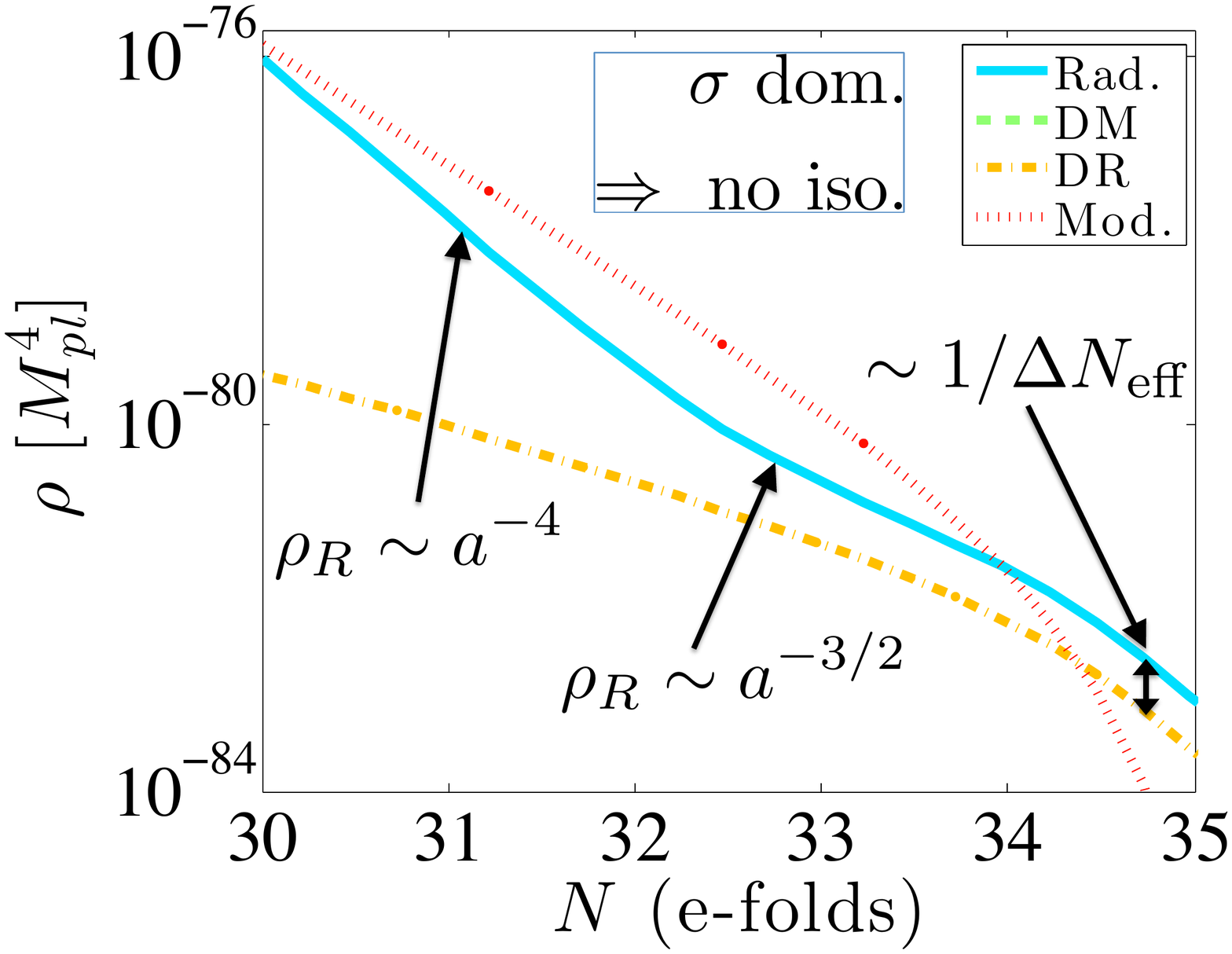}&
\includegraphics[scale=0.4]{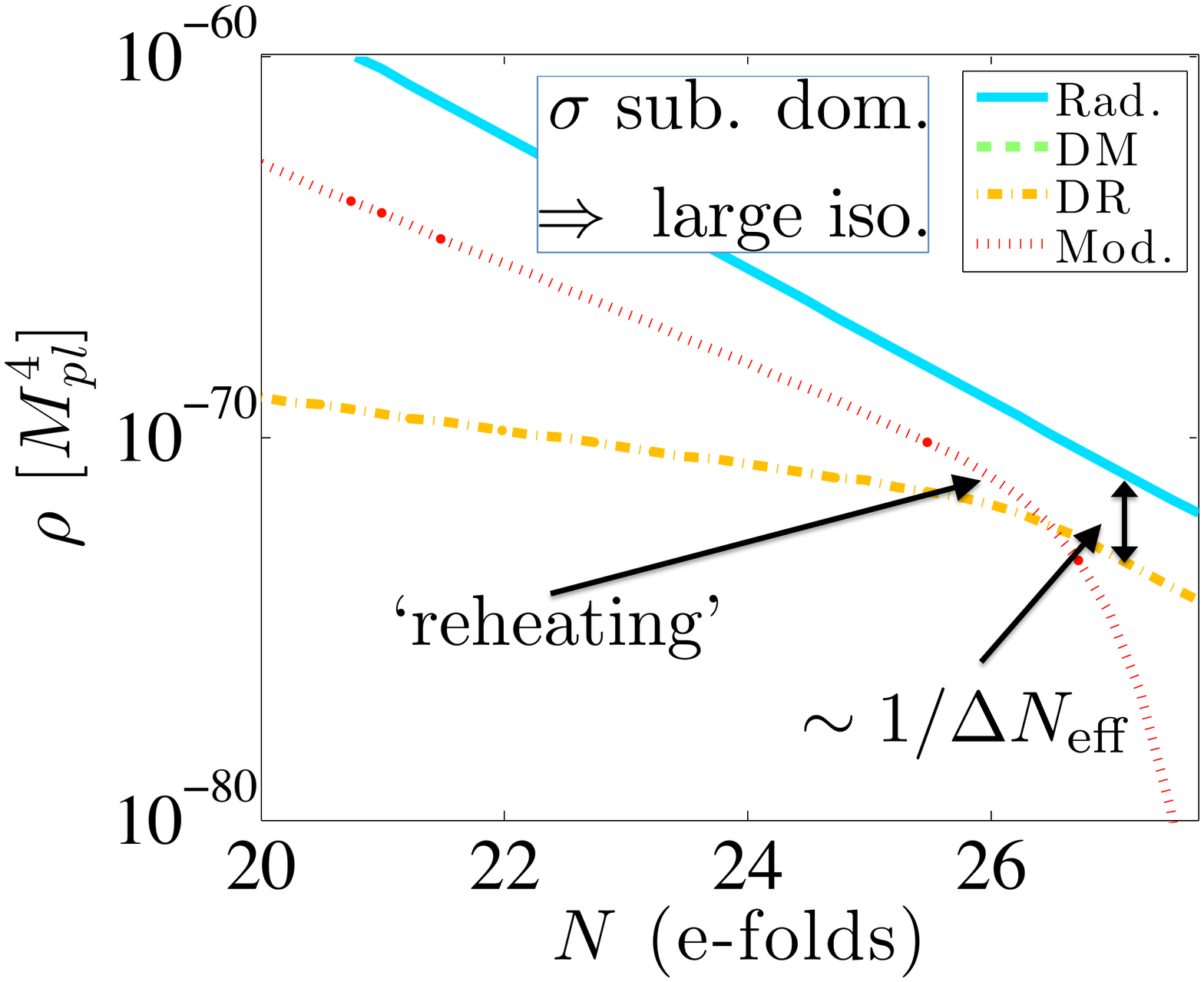}\\[0.0in] 
\end{array}$
\caption{Dominant and sub-dominant decay of a modulus, caused by varying the decay rate parameter $c_3$ with fixed mass, misalignment, and branching to DR. \emph{Left Panel}: (Zoom of Fig.~\ref{fig:density_535_1_all}) Dominant decay, no isocurvature. Small decay rate. When the modulus is dominant all components sourced by it scale as $a^{-3/2}$. The radiation produced by inflaton decay at first reheating scales as $a^{-4}$, dominating at early times. For dominant decay $\Delta N_{\rm eff}$ can be computed analytically from $B_{DR}$ (Eq.~(\ref{eqn:delta_neff_dominant})). \emph{Right Panel}: Sub-dominant decay, large isocurvature. Large decay rate. Only the DR (and DM, which is not visible on this scale) scales as $a^{-3/2}$. Sub-dominant decay produces a smaller amount of DR at fixed branching $B_{DR}$, measured by $\Delta N_{\rm eff}$, and this must be computed numerically. In this case `reheating' is somewhat of a misnomer.}
\label{fig:density_zoom}
\end{figure*}

Fig.~\ref{fig:density_535_1_all} shows the background evolution in an example model where the modulus decays while it dominates the energy density, with $m_\sigma=240$~TeV, $\sigma_\star=10^{-6}$~$M_{pl}$, $c_3=0.028$ (exaggeratedly small for illustration), $B_{DR}=0.21$ \footnote{The value of $B_{DR}$ can be computed in explicit models. For example in \cite{angus2013} it lies in the range 0.3 to 0.5. Our value is chosen semi-arbitrarily. It is $\mathcal{O}(0.1)$ and as we will see later gives variation of $\Delta N_{\rm eff}$ over a range interesting for isocurvature.}. The DR is sub-dominant to the radiation right up until modulus decay has completed, and the value of $\Delta N_{\rm eff}=1.68$ freezes in. Being sourced entirely by modulus decay, both the DM and the DR scale in the same way with e-folding $N$, as $a^{-3/2}$ in terms of scale factor. The SM radiation joins them once the modulus becomes dominant as we see looking more closely in Fig.~\ref{fig:density_zoom} (Left Panel). The two scalings of the radiation during modulus decay, Eq.~(\ref{eqn:densities_source_solution}), $\rho_r\sim a^{-3/2}$ when the modulus is dominant, and $\rho_r\sim a^{-4}$ when it is in transition from sub-dominant to dominant, are clearly visible.

Fig.~\ref{fig:density_zoom} (Right Panel) compares the previous model to another the same except with $c_3\approx 26000$ (exaggeratedly large for illustration), so that the modulus decays while it is sub-dominant. Sub-dominant modulus decay does not affect the scaling of the dominant SM radiation, although the branching ratios in the two models are the same. With sub-dominant decay the amount of DR produced is considerably smaller, $\Delta N_{\rm eff}=0.02$. By analogy with the curvaton and from the results of Section~\ref{pertsection}, the sub-dominant decay produces large amounts of isocurvature, while the dominant decay does not. The detailed understanding of this in our model is the focus of the rest of this paper. The plots shown in Fig.~\ref{fig:density_zoom} thus serve as cartoons to aid in understanding the entire model.

\subsection{CMB Observables}
\label{sec:observables}

Here we build upon the results reviewed in Section~\ref{pertsection}, defining precisely our CMB observables and how we use the curvature perturbation to compute them. Details of the perturbed equations of motion, numerical method, and initial conditions are given in Appendix~\ref{appendix:computation}, while the power spectra are discussed in Appendix~\ref{appendix:spectra}.

In the initially radiation dominated universe the curvature is primarily due to radiation and the adiabatic condition $\delta_i/(1+w_i)=\delta_R/(1+w_R)$ implies for the perturbations `inf' laid down by the inflaton that $S^{\rm inf}_i(0)=0$, with $\zeta^{\rm inf}(0)\neq 0$ and $\delta\sigma^{\rm inf}(0)=0$. With isocurvature initial conditions `mod' seeded by the modulus we have that $\zeta^{\rm mod}(0)=0$, while $\delta\sigma^{\rm mod}(0)\neq 0$ sources non-zero $S_{i} (N_{\rm end})$.

We follow the evolution of all species, keeping track of their $\zeta_i$'s, up until the modulus has decayed and we have entered radiation domination at $N_{\rm end}$, when all $\zeta_i$ freeze-in and set the initial conditions for computation of the CMB power spectrum. Unless otherwise stated, correlators are evaluated at $N_{\rm end}$.

The two point correlations between the total $\zeta_i$ are easy to compute, since by assumption the `inf' and `mod' initial condition modes (see Appendix~\ref{appendix:initial_modes}) are uncorrelated with one another, but totally correlated with themselves. This implies that the correlation matrix is given by
\be
\langle{\zeta_i(\mathbf k) \zeta_j(\mathbf k')}\rangle = (2\pi)^3 \delta^3(\mathbf k - \mathbf k') (\zeta_i^{\rm inf}\zeta_j^{\rm inf} +  \zeta_i^{\rm mod} \zeta_j^{\rm mod}) \, .
\ee 
Using this correlation matrix we can compute any other correllators of total $\zeta$ or $S_i$. Defining the power spectrum by
\be
\langle{X(\mathbf k) Y(\mathbf k')}\rangle = (2\pi)^3 \delta^3(\mathbf k - \mathbf k')P_{XY} \, , 
\ee
then, for example, the total curvature perturbation power spectrum, $P_{\zeta\zeta}=P_{\zeta\zeta}^{\rm mod}+P_{\zeta\zeta}^{\rm inf}$. We construct the observables for the isocurvature fraction, $\alpha_i$, its correlation with the curvature perturbation, $r_i$, and the cross correlation between any two isocurvature modes, $r_{ij}$, all evaluated at the pivot scale, $k_0$:
\begin{align}
\alpha_i&= \frac{P_{S_iS_i}}{\sum_jP_{S_jS_j}+P_{\zeta\zeta}} \, , \\
r_i&=\frac{P_{S_i\zeta}}{\sqrt{P_{S_iS_i}P_{\zeta\zeta}}} \, , \\
r_{ij}&=\frac{P_{S_iS_j}}{\sqrt{P_{S_iS_i}P_{S_jS_j}}} \, ,
\end{align}
where there is no sum implied over repeated indices, unless stated.  The total power from isocurvature is given by considering the total isocurvature, $\mathcal{S}=\sum_i S_i$, and total scalar power, $\mathcal{P}=P_{(\zeta+S)(\zeta+S)}$:
\be
f_{ISO} = \frac{P_{\mathcal{S}\mathcal{S}} + 2 P_{\mathcal{S}\zeta}}{\mathcal{P}}=1-\frac{P_{\zeta\zeta}}{\mathcal{P}} \, .
\ee
Clearly $f_{ISO}$ is not independent of the $\alpha$'s and $r$'s, yet it is useful to compute since it gives an overall measure of the isocurvature power. In the limit of pure isocurvature we have that $f_{ISO}\rightarrow 1$ and $\alpha_{DM}+\alpha_{DR}\rightarrow 1$, with the further contribution to $f_{ISO}$ from the correlation of DM and DR proportional to $r_{DM,DR}\sqrt{\alpha_{DM}\alpha_{DR}}$, as we will see below. That is, in this limit, $f_{ISO}$ and $\sum_i \alpha_i$ are numerically equal to one another.

Our variables are used to construct the total CMB power spectrum, $C_\ell$, as follows
\begin{align}
C_\ell = &A_s \left( (1-\sum_i\alpha_i)\hat{C}_\ell^{\rm ad} +\alpha_{DM} \hat{C}_\ell^{\rm CDI} \right. \nonumber \\
&+\left. \alpha_{DR}\hat{C}_\ell^{\rm DRI}+\alpha_{\rm cor.} \hat{C}_\ell^{\rm cor.}  \right) \, . 
\label{eqn:c_ell_def}
\end{align}
The overall normalisation is given by 
\be
A_s=P_{\zeta\zeta}+\sum_iP_{S_iS_i}\, .
\label{eqn:as_def}
\ee
The unit CMB spectra, $\hat{C}_\ell$, are computed with unit normalisation at the pivot scale, $A_i=1$, from the adiabatic (ad), CDI and DRI initial conditions \cite{bucher2000}, where `CDI' and `DRI' refer to the CDM and DR density isocurvature modes. The DRI mode is related to the more familiar neutrino density isocurvature mode, NDI, by
\be
\hat{C}_\ell^{\rm DRI}=\left(\frac{\Delta N_{\rm eff}}{N_{\rm eff}}\right)^2 \hat{C}_\ell^{\rm NDI} \, .
 \label{eqn:dri_def}
\ee
The factor of $(\Delta N_{\rm eff}/N_{\rm eff})^2$ takes into account that in our model the DRI mode is not sourced by the standard model neutrinos \footnote{Use of NDI and DRI also avoids confusion about the production mechanism, which should be contrasted to that of e.g. \cite{savelainen2013}. The physical difference between NDI and DRI is that with standard model neutrinos (rather than sterile neutrinos or axions) NDI can only be produced if the modulus decays after neutrino decoupling, while the DR by assumption decoupled at very high temperatures and so DRI is produced by modulus decay at any temperature. The NDI and DRI do, however, produce the same CMB spectra, as can be seen from the equations of motion, which are not sensitive to the fermionic or bosonic character of radiation \cite{bertschinger1995}.}. In contrast to the effect of varying the axion contribution to DM in the axion CDI mode \cite{5yearWMAP}, here constraints to $\Delta N_{\rm eff}$ mean this factor can be determined, just like the ultra-light axion contribution can be determined to break a similar degeneracy as discussed further in Ref.~\cite{marsh2013} (see also the next subsection). 

The contribution to CMB power from correlations, $\hat{C}_\ell^{\rm cor.}$, can be calculated given the values of $\alpha_i$, $r_i$ and $r_{ij}$. It is given by
\begin{align}
\hat{C}_\ell^{\rm cor.}&= \frac{1}{A_s}(C_\ell^{CDI,\rm ad.}+C_\ell^{DRI,\rm ad.}+C_\ell^{CDI,DRI}) , \nonumber \\
&=r_{DM}\sqrt{\alpha_{DM}(1-\sum_i \alpha_i)} \hat{C}_\ell^{CDI,\rm ad.} \nonumber \\
&+r_{DR}\sqrt{\alpha_{DR}(1-\sum_i \alpha_i)} \hat{C}_\ell^{DRI,\rm ad.} \nonumber \\
&+r_{DM,DR}\sqrt{\alpha_{DM}\alpha_{DR}}\hat{C}_\ell^{CDI,DRI} \, .
\end{align}
Again we must be careful with DR versus neutrinos and note that
\begin{align}
\hat{C}_\ell^{DRI,\rm ad.}&=\frac{\Delta N_{\rm eff}}{N_{\rm eff}}\hat{C}_\ell^{NDI, \rm ad.} \, , \\
\hat{C}_\ell^{CDI,DRI}&=\frac{\Delta N_{\rm eff}}{N_{\rm eff}}\hat{C}_\ell^{CDI, NDI} \, .
\end{align}

We normalise the total scalar power, $A_s$, to its Planck central value of $A_s=2.2\times 10^{-9}$ \cite{Ade:2013zuv}. With fixed $H_I$ this normalisation fixes the slow roll parameter $\epsilon$ using Eq.~(\ref{eqn:find_epsilon}), discussed further in Appendix~\ref{appendix:norm}. We discuss the spectral indices in Appendix~\ref{appendix:indices}. The effect of varying $H_I$ is discussed in Section~\ref{sec:results}.

Finally, as our model is similar in spirit to the curvaton model, one would like to compare the two. In the curvaton model as first proposed, all the radiation is generated by decay of the inflaton and all of the curvature is generated by the curvaton conversion of isocurvature power. However, this need not be the case and we can define the parameter \cite{langlois2010}
\be
\lambda = \frac{P_{\zeta \zeta}(N_{\rm end})}{P_{\zeta \zeta}(0)}-1 \, ,
\ee
to measure how much of the final curvature was due to inflaton perturbations and how much due to modulus perturbations. The limit $\lambda \gg 1$ and $r_{DM}=1$ corresponds to the original curvaton model (of course with $\alpha_{DR}=\Delta N_{\rm eff}=0$). Using $\lambda$ we can rewrite the total auto-power scalar amplitude, $A_s$, as
\begin{align}
A_s=&\frac{1}{2\epsilon}\left( \frac{H_I}{2\pi M_{pl}} \right)^2 (1+\lambda) \nonumber \\
& \left[ 1+\frac{\alpha_{DM}+\alpha_{DR}-4(1+\lambda)\alpha_{DM}\alpha_{DR}}{1-(1+\lambda)(\alpha_{DM}+\alpha_{DR})}  \right] \, .
\end{align}

With a single modulus field we take two uncorrelated initial condition modes from the inflaton and the modulus, and project them onto three correlated CMB modes in curvature, CDI and DRI. Therefore there must exist a relationship between the CMB modes. In fact this implies the CDI and DRI modes are totally correlated, i.e. $r_{DM,DR}=1$ and $r_{DM}=r_{DR}$: the single source relations. To violate this relationship we must introduce a second isocurvature field. 

One natural option for this second field is a component of axion DM, which we discuss in the next subsection, or including the thermal component of WIMP DM, which carries no isocurvature. The other alternative is the more complicated option of a second decaying modulus. The equations of motion and initial conditions for this second modulus will be exact copies of those for the first modulus with different parameters specifying the mass, initial displacement, decay rate and branching ratios. We leave study of this second option to a future work.

\subsection{Axion Dark Matter}
\label{sec:axion_dm}

In our model presented so far, DM and DR isocurvature are generated simultaneously by the non-thermal decay of the modulus to WIMPs and relativistic axions. This implies the single source relations $r_{DM}=r_{DR}$ and $r_{DM,DR}=1$. It is entirely natural in this framework for the axion partner of the modulus to also be produced by the vacuum realignment mechanism and contribute to the DM density \cite{turner1986,berezhiani1992}, as well as to DR (e.g. \cite{Jeong:2013oza}). In this case the axion DM will carry its own isocurvature perturbations, completely uncorrelated to any other perturbations \cite{axenides1983}. These add to the total DM isocurvature and in particular will allow us to violate the single source relations, in a way which we now describe. 

For the isocurvature in this scenario we have
\be
S_{DM}=S_{DM}^{\rm mod}+S_{DM}^{\rm ax} \, .
\ee
Defining
\be
P_{CDI}^{\rm mod}=P_{ S_{DM}^{\rm mod}S_{DM}^{\rm mod}} \, ,
\ee
and
\be
P_{CDI}^{\rm ax} = P_{S_{DM}^{\rm ax}S_{DM}^{\rm ax}} =  (\delta_a^{\rm ax})^2 \, ,
\ee
where $\delta_a$ is the initial overdensity in the axion isocurvature mode. For an axion field $a=a_0+\delta a$, at early times
\be
\langle \delta_a^2 \rangle \approx \left\langle\left(\frac{\delta a}{a_0}\right)^2 \right \rangle \, .
\ee

The cross correlation is then given by
\be
r_{DM,DR} = (1+P_{CDI}^{\rm ax}/P_{CDI}^{\rm mod})^{-1/2} \, .
\ee
The axion perturbation power spectrum for $\delta a$ is the same as the modulus perturbation power spectrum $\delta \sigma$~\footnote{If the modulus evolves during inflation, both spectra can be affected, as discussed in \cite{kasuya2009}.}, in Eq.~(\ref{eqn:modulus_initial_ds}), so that the inflationary energy scale $H_I$ drops out of the cross correlation. 

The initial displacement of the homogeneous axion field, $a_0$, determines the relic density of axion DM. For a general axion-like particle (ALP) we have \cite{marsh2010}
\be
\Omega_a = \frac{1}{6}(9 \Omega_r)^{3/4}\left(\frac{m_a}{H_0}\right)^{1/2} \left(\frac{a_0}{M_{pl}} \right)^2 \, ,
\ee
where $\Omega_r$ is the density in radiation today, $H_0$ is the Hubble rate, and $m_a$ is the mass of the ALP. We have assumed it is heavy enough to behave entirely as DM \footnote{If it  does not, one must use the ultra-light axion isocurvature mode \cite{marsh2013}.}. One can therefore specify $r_{DM,DR}$ using the relic density of ALP DM, $\Omega_a$, the mass, $m_a$, and $P_{CDI}^{\rm mod}$, which our methods compute.

In the context of this extended model with ALP DM, constraints on the DM-DR cross correlation, such as \cite{moodley2004}, in turn constrain the combination of parameters $\Omega_a m_a^{1/2}$, and eliminates this combination from the overall normalisation in $\alpha_{DM}^{\rm ax}$ (e.g. \cite{5yearWMAP}). For the QCD axion, $\Omega_a m_a^{1/2}$ can be used to constrain the decay constant, $f_a$. We leave a full exploration of this partially correlated mode to future work.

\section{Results}
\label{sec:results}

In this Section we present the results of our computation of isocurvature observables as functions of the modulus parameters. We then go on to use these results to compute priors on the isocurvature observables informed by particle physics priors on SUSY and inflation. In our main investigations we hold $H_I=10^5$ TeV$=10 m_{\sigma,\rm max}$ fixed and normalise to $A_s=2.2\times 10^{-9}$ (we discuss our normalisation procedure in Appendix~\ref{appendix:norm}). We discuss the effect of varying $H_I$ towards the end of Section~\ref{sec:results_iso_single}. We fix $B_{DR}=0.21$ and we have checked that our results, except for the value of $\Delta N_{\rm eff}$, do not strongly depend on this choice. The reason being that DR is always, for sensible values of $\Delta N_{\rm eff}$, sub-dominant in the energy budget, while on super-horizon scales effects in the perturbations, such as increased damping, are irrelevant. The effective branching ratio to DM plays no dynamical role, and we set it to a small value which gives approximately the correct DM abundance. It is always possible in principle to find the exact value which produces the fixed point from annihilations, Eq.~(\ref{eqn:lsp_annihilation_fixedpoint}).

%The effective branching ratio to DM is fixed at $B_{DM}=10^{-15}$. This produces a reasonable DM abundance, and has no other affect. 

\subsection{Isocurvature From A Decaying Modulus}
\label{sec:results_iso_single}

\begin{figure*}[htbp!]
\centering
$\begin{array}{@{\hspace{-0.3in}}l@{\hspace{+0in}}l}
\includegraphics[scale=0.5]{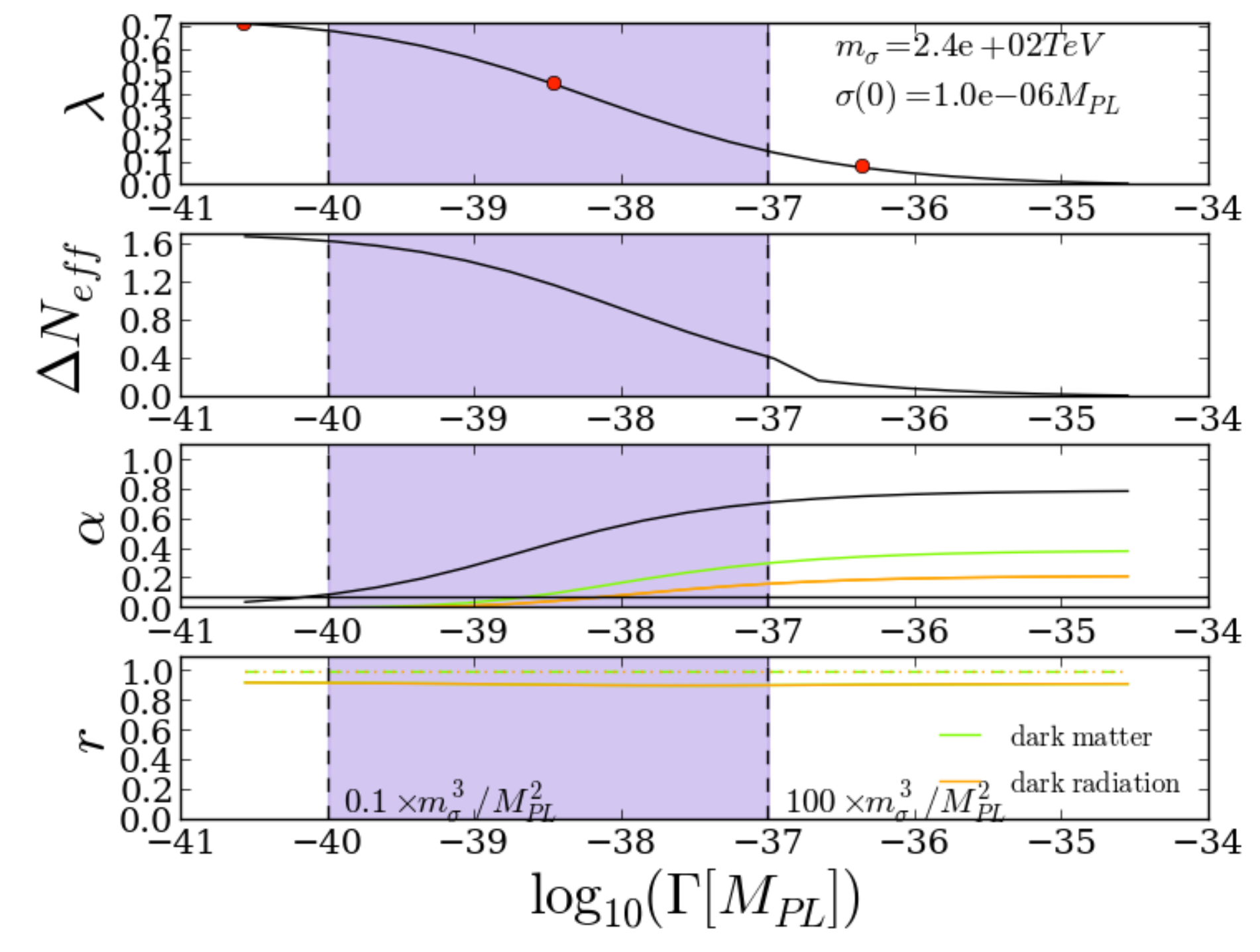}&
\includegraphics[scale=0.5]{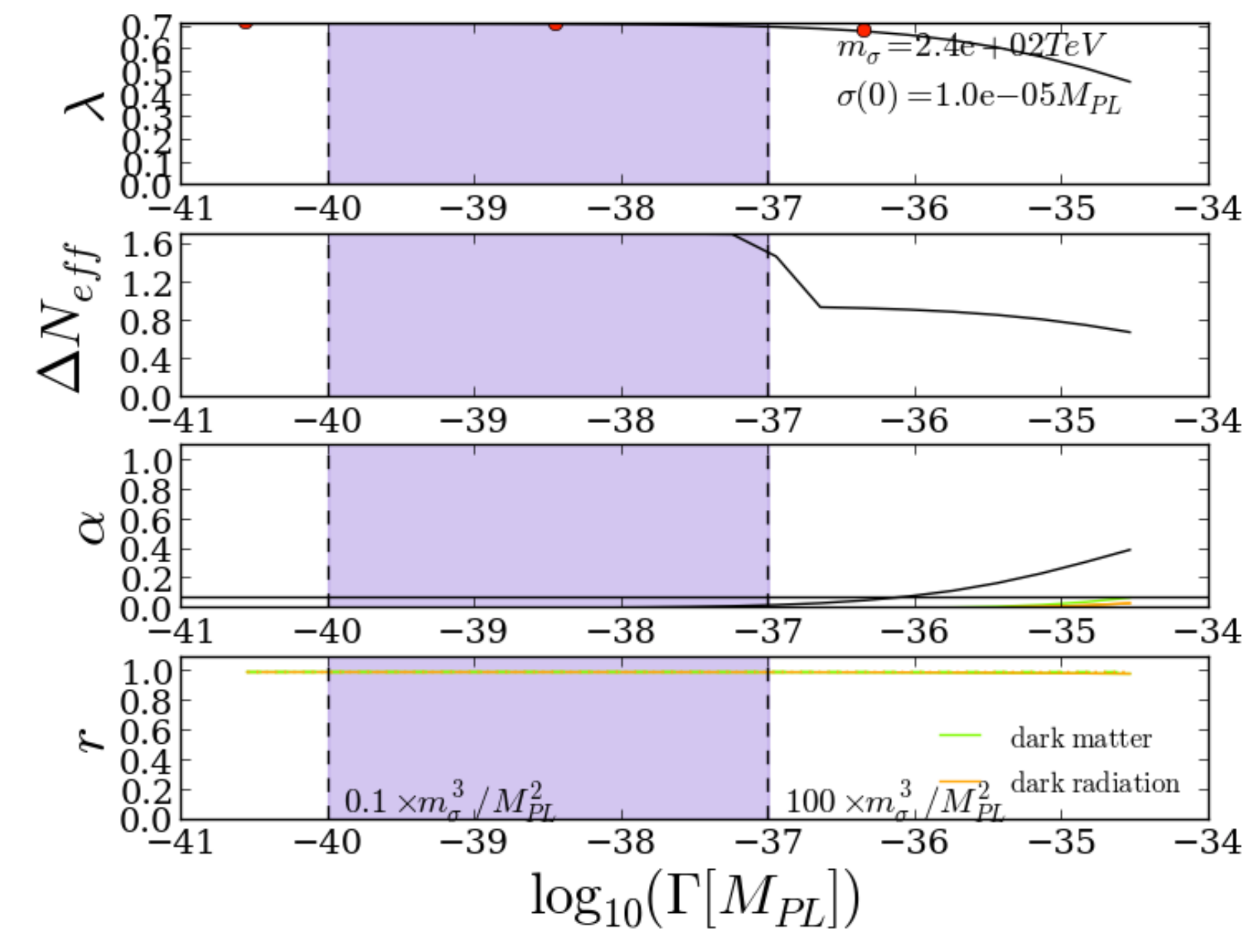}\\[0.0in] 
\end{array}$
\caption{Isocurvature quantities as a function of decay rate at fixed modulus mass and misalignment. The contribution of the modulus to the total curvature, $\lambda$, increases as the decay rate gets slower (smaller $\Gamma$). Isocurvature fractions for DM, DR, and total $f_{ISO}$ (black) increase when the decay is faster. Correlation parameters, $r$, obey the single source relation: the cross correlation $r_{DM,DR}=1$ (dot-dashed line, alternating colours). Decay rates allowed with $1/(4 \pi)\lesssim c_3\lesssim 100$ are shown in purple. \emph{Left Panel}: Small misalignment has more isocurvature and less dark radiation. \emph{Right Panel}: Larger misalignment has opposite behaviour.}
\label{fig:params_decayrate}
\end{figure*}

In Fig.~\ref{fig:params_decayrate} we show the dependence of isocurvature parameters on the decay rate of the modulus, $\Gamma_\sigma$. The modulus mass is fixed at $m_\sigma = 2.4 \times 10^2$ TeV. On each plot we mark the range of $\Gamma_\sigma$ allowed with $1/(4 \pi)\lesssim c_3\lesssim 100$, the typical decay rate expected in SUGRA. In both plots $f_{ISO}$ is seen to monotonically increase with $\Gamma_\sigma$, in accordance with our expectation that isocurvature is only produced if the modulus decays rapidly when it is sub-dominant in the energy density. Varying $c_3$ in its allowed range with all other parameters fixed can in this case cause large variation in the isocurvature fractions.

We observe that the relationship $\alpha_{DM}+\alpha_{DR}=1=f_{ISO}$ is obeyed when the modulus decays rapidly, so that it is sub-dominant and produces large amounts of isocurvature. The DM and DR isocurvature parameters in this regime satisfy $\alpha_{DM}/\alpha_{DR}\approx 2$. However, very little DR is produced in this regime, with $\Delta N_{\rm eff}\approx 0$, and so by Eq.~(\ref{eqn:dri_def}) the DRI mode will contribute little to the CMB spectrum.

Fig.~\ref{fig:params_decayrate} Left Panel illustrates the case of $\sigma_\star/M_{pl}=10^{-6}$, while Fig.~\ref{fig:params_decayrate} Right Panel takes $\sigma_\star/M_{pl}=10^{-5}$. In the range allowed for $c_3$ it is only the lower value of $\sigma_\star$ that leads to appreciable values of $\alpha_i$ for both isocurvature modes at this mass. Once again this is explained by the change to the energy density fraction contributed by the modulus at the time of decay, with smaller $\sigma_\star$ reducing $\rho_\sigma$, increasing the isocurvature fractions and reducing $\Delta N_{\rm eff}$.

The correlation parameters are observed to obey the relation $r_{DM}=r_{DR}$, $r_{DM,DR}=1$ expected from decay of a single modulus. The correlation parameters are not observed to vary with the decay rate, therefore we should expect that they will not depend on mass either. However we have found that this depends on the choice of $H_I$: with larger $H_I$ the dependence of the correlation parameters on $m_\sigma$ and $\Gamma_\sigma$ is enhanced.

\begin{figure*}
\begin{center}
$\begin{array}{@{\hspace{+0.1in}}l@{\hspace{+0.1in}}l}
\includegraphics[scale=0.34]{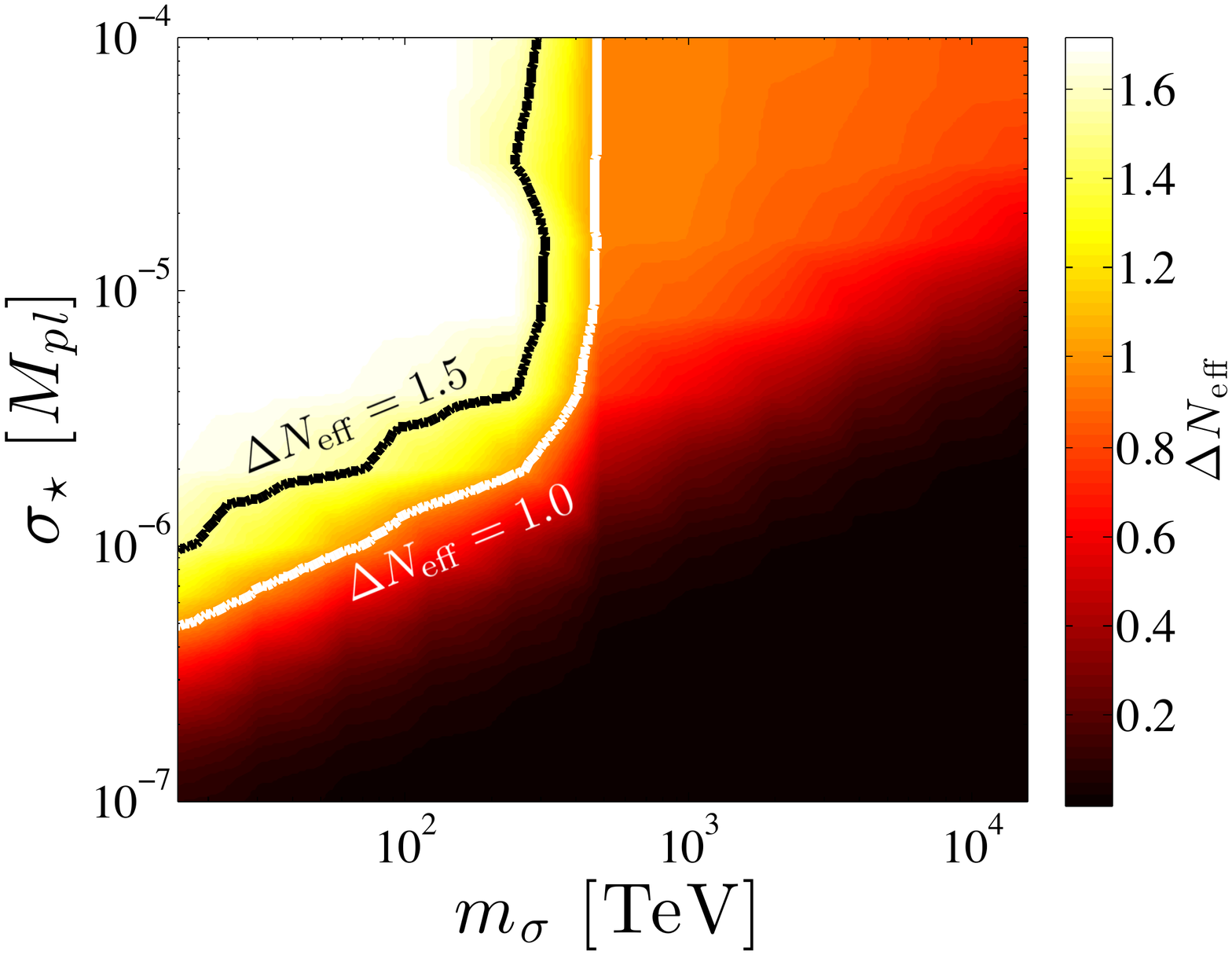}&
\includegraphics[scale=0.34]{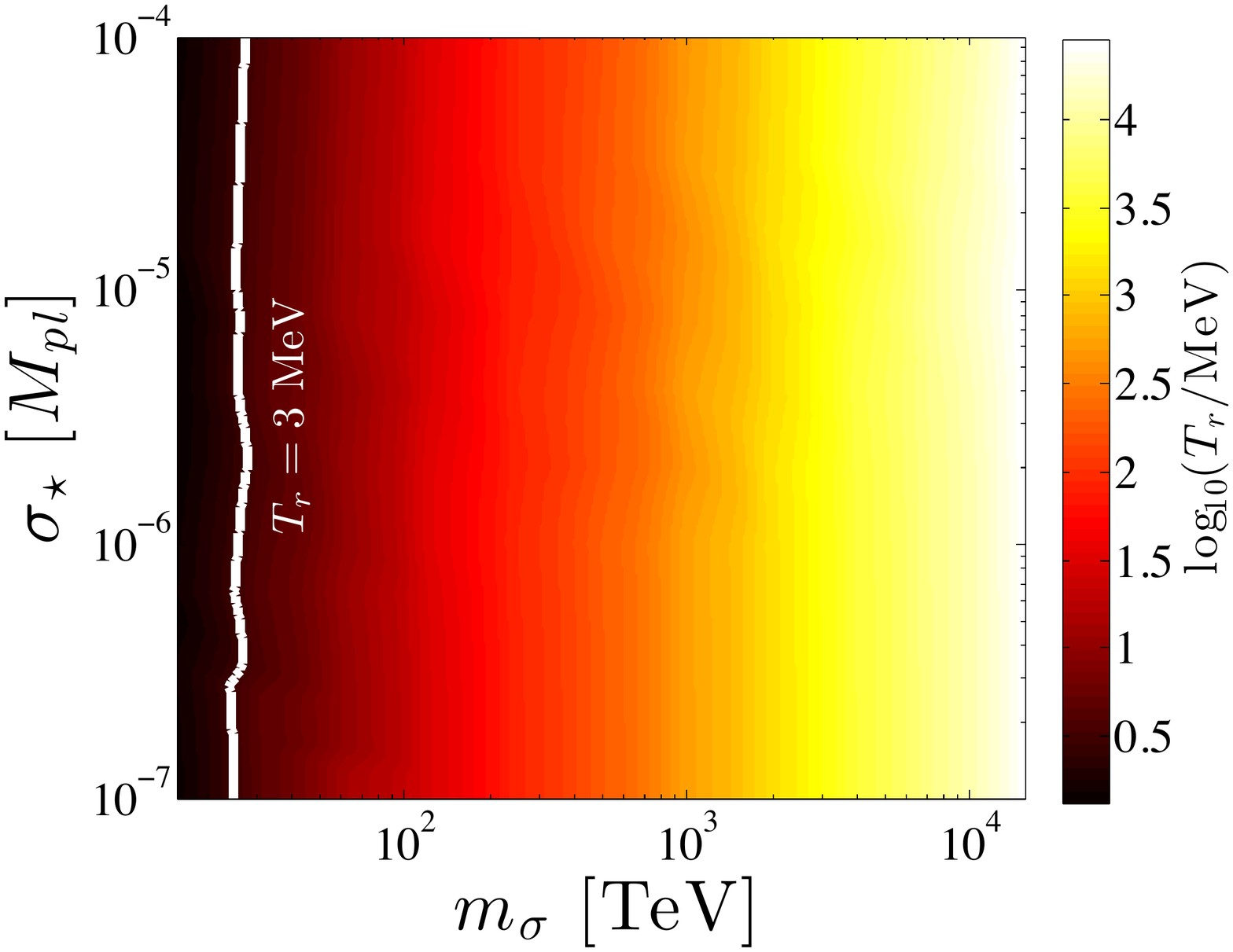}\\[0.0in]
\includegraphics[scale=0.34]{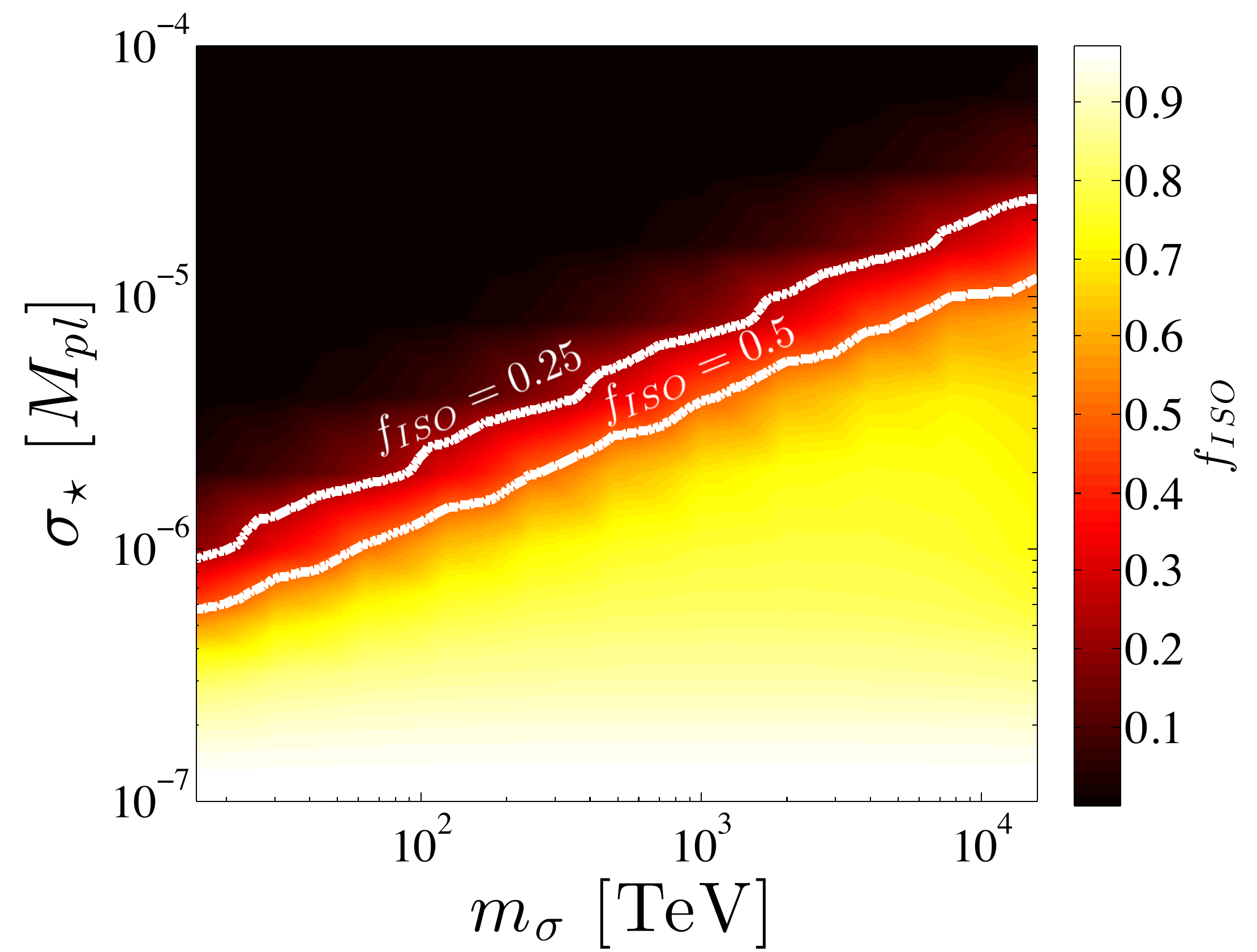}&
\includegraphics[scale=0.4]{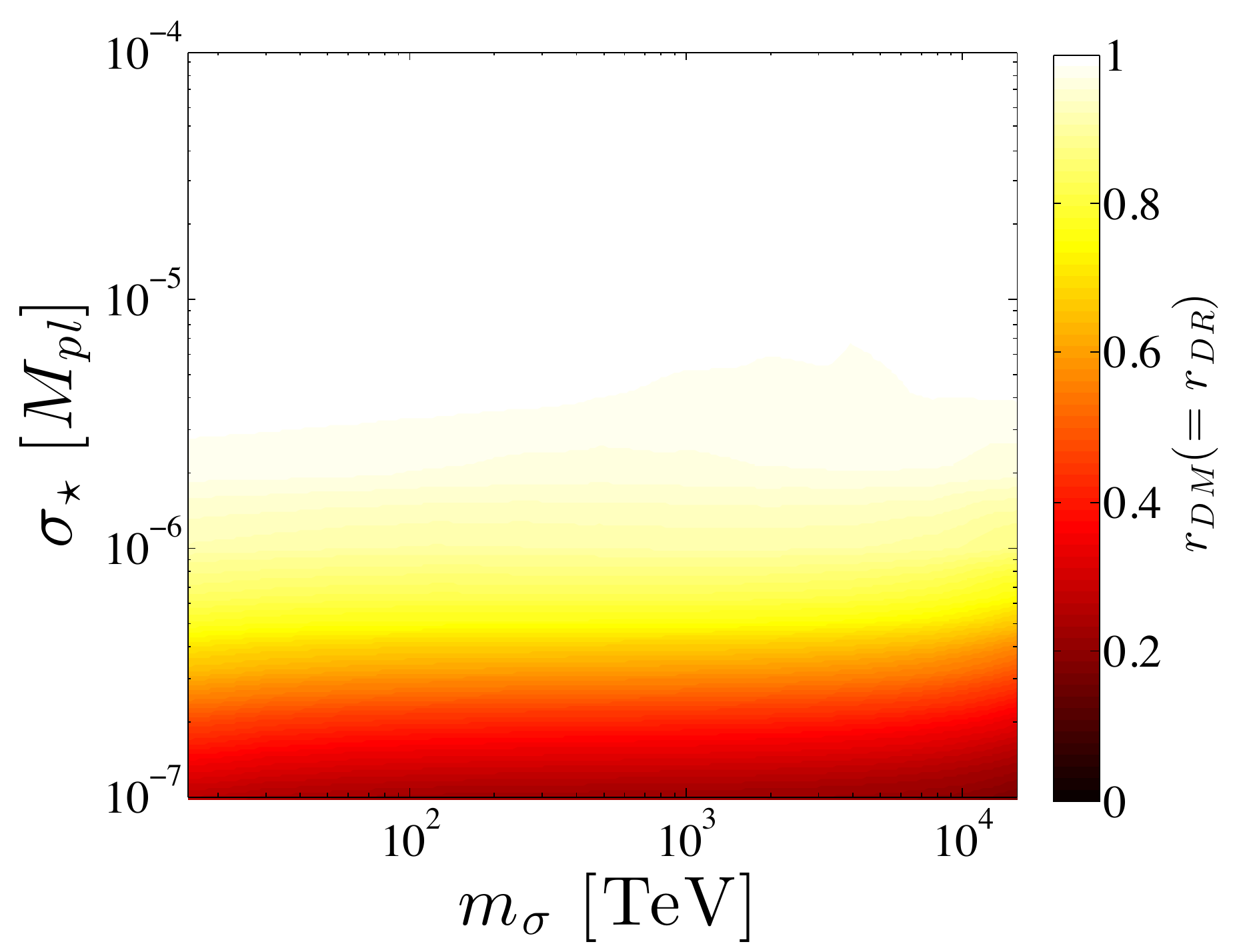}\\[0.0in]
\includegraphics[scale=0.4]{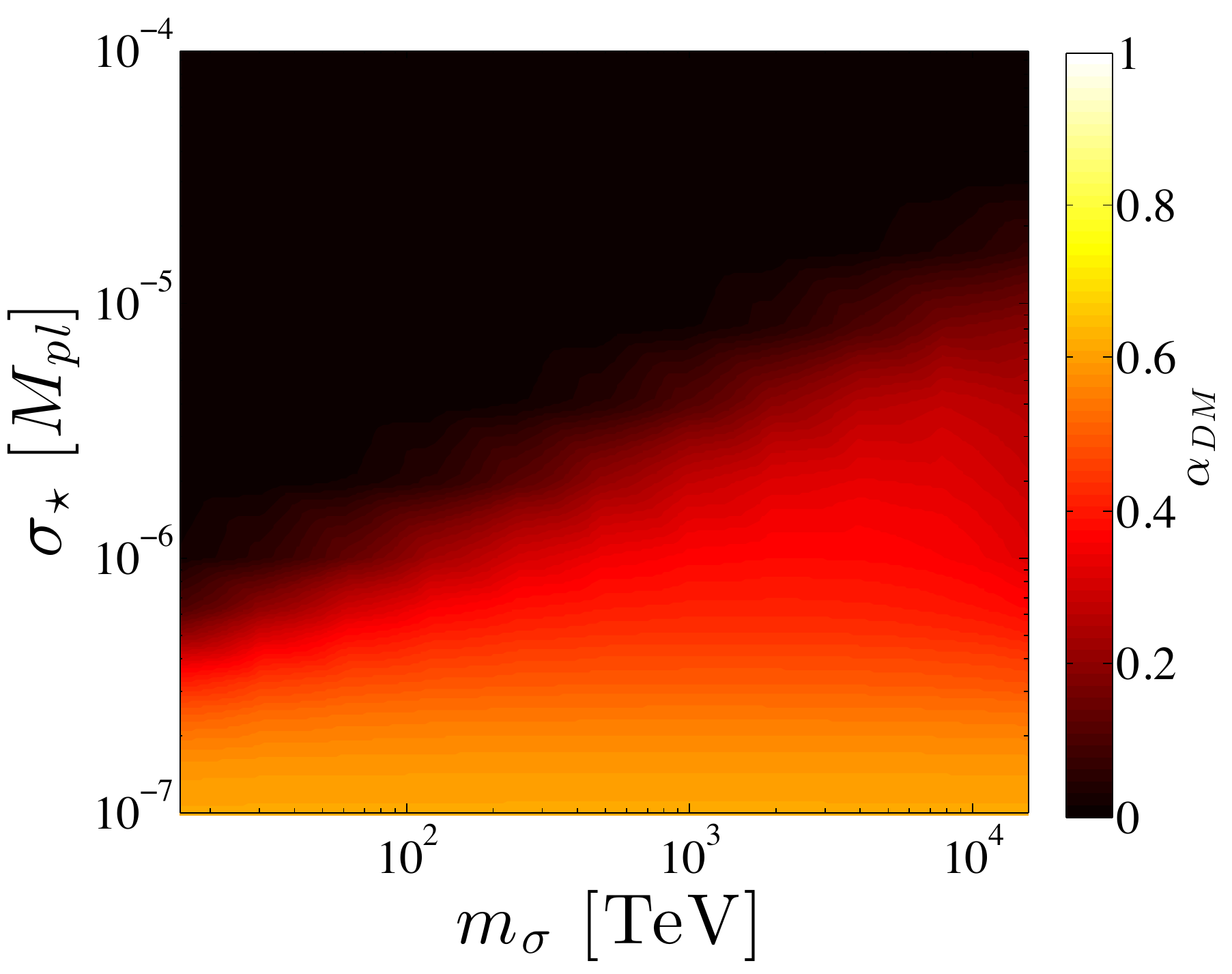}&
\includegraphics[scale=0.4]{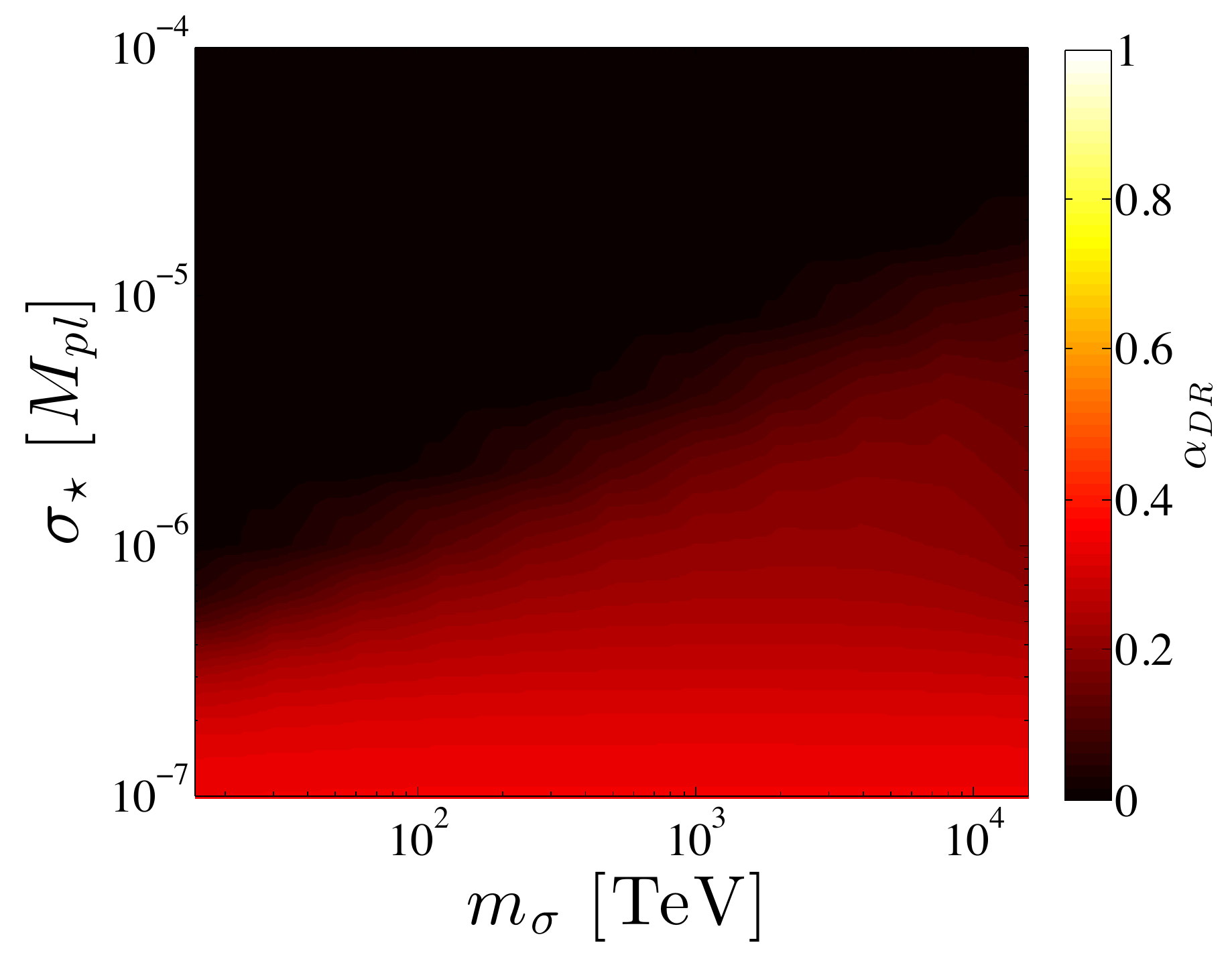}\\[0.0in] 
 \end{array}$
 \caption{Isocurvature observables in the $(m_\sigma,\sigma_\star)$ plane of modulus parameters, at fixed branching to DR, $B_{DR}$=0.21, fixed decay parameter, $c_3=100$, and fixed inflation scale, $H_I=10^5$ TeV, with overall scalar power normalised to its Planck value $A_s=2.2\times 10^{-9}$. The reheat temperature, $T_r$, depends only on the modulus mass, as it should. All isocurvature observables show significant variation over the range $\sigma_\star/M_{pl}\in [10^{-7},10^{-5}] $. Isocuravture depends on $\rho_R/\rho_\sigma\sim m_\sigma\sigma_\star^{-2}$ or $\sim m_\sigma^{3/4}\sigma_\star^{-8/3}$ if the modulus dominates at the time of decay (Eqs.~(\ref{eqn:iso_param_scaling1}) and~(\ref{eqn:iso_param_scaling2}) ), hence stronger scaling with $\sigma_\star$ is observed, in particular for $r_{DM}$ which hardly depends on $m_\sigma$. The amount of DR, measured by $\Delta N_{\rm eff}$, and the total amount of isocurvature, measured by $f_{ISO}$, show opposite behaviours in the plane. At fixed branching, DR constraints can be alleviated somewhat by reducing $\sigma_\star$, but only at the expense of introducing isocurvature. The correlations are fixed by the single source relation, and the value of $r_{DM}$.}\label{fig:contour_panels}
\end{center}
 \end{figure*}

In Fig.~\ref{fig:contour_panels} we show the dependence of isocurvature observables in the $(m_\sigma,\sigma_\star)$ plane, with $c_3=100$ held fixed. The larger $c_3$ means larger decay rates, and so maximal amounts of isocurvature. The mass and misalignment of the modulus determine the energy fraction that the modulus contributes during decay, between $N_{\rm osc}$ and $N_{\rm end}$. Heavier moduli decay more rapidly, and at fixed misalignment contribute less to the energy density. At fixed mass, reducing the misalignment reduces the energy density. 

With fixed $c_3$ the reheat temperature, $T_r$, depends only on the modulus mass, as expected (see Eq.~(\ref{reheattemp})).  We stress again that since the modulus can be sub-dominant, this is not strictly `reheating' in the usual sense, but the temperature at which decay is completed (see Fig.~\ref{fig:density_zoom}). As such it need not satisfy $T_r>T_{\rm BBN}\approx 3$ MeV (see Section~\ref{sec:background}). Because we allow the modulus to be sub-dominant at the time of decay $\Delta N_{\rm eff}$ depends on both $m_\sigma$ and $\sigma_\star$ at fixed $B_{DR}$ and $c_3$, with larger values and stronger dependence on the mass through $g_\star (T_r)$ where the modulus dominates the energy density. This leads to the rapid change in $\Delta N_{\rm eff}$ around $m_\sigma = 300$ TeV when $T_r$ goes through the QCD phase transition and $g_\star$ increases by a large factor. Using $B_{DR}$ one can compute the asymptotic value for dominant decay (Eq.~(\ref{eqn:delta_neff_dominant})), e.g. $\Delta N_{\rm eff}\approx 0.74$ when $g_\star (T_r)=106.75$, which is valid for large $m_\sigma$ when $T_r>m_t$, with $m_t$ the top quark mass. 

Isocurvature fractions $f_{ISO}$, $\alpha_{DM}$ and $\alpha_{DR}$ show the expected trend with $(m_\sigma,\sigma_\star)$ due to energy fraction as in the curvaton scenario. If the modulus dominates the energy density at the time of decay, there is no isocurvature, while if it is sub-dominant, isocurvature fractions are large (see Fig.~\ref{fig:density_zoom}). As in Fig.~\ref{fig:params_decayrate} we also observe that in regime of a sub dominant modulus we have $\alpha_{DM}+\alpha_{DR}=1=f_{ISO}$ with $\alpha_{DM}/\alpha_{DR}\approx 2$. We have not been able to explain this asymptotic ratio, but have found that it does not depend on $B_{DR}$. Once in the asymptotic region of a sub-dominant modulus it does not depend on $\Gamma_\sigma$, $m_\sigma$, $\sigma_\star$ or $H_I$ either, suggesting it is kinematical, possibly fixed by the equations of state of DM and DR.

For the reasons we have described, $\Delta N_{\rm eff}$ and $f_{ISO}$ have opposite behaviour in the $(m_\sigma,\sigma_\star)$ plane. At fixed branching fraction, the amount of DR produced can be reduced by decreasing the initial displacement of the modulus field, $\sigma_\star$. However, if this makes the modulus sub-dominant at the time of decay, it also predicts larger amounts of isocurvature. The simple requirement of cosmology to keep both the contribution to $\Delta N_{\rm eff}$ and $f_{ISO}$ relatively small requires a trade-off. In order not to overproduce isocurvature the misalignment must be large, but at fixed branching to DR it cannot be so large that it increases $\Delta N_{\rm eff}$ beyond acceptable levels. For example, the string model of Ref.~\cite{cicoli2012f}, shown in Ref.~\cite{angus2013} to be inconsistent with Planck constraints to $\Delta N_{\rm eff}$ even when including loop corrections to the decay rate, could be allowed if the misalignment is reduced and a small amount of isocurvature predicted. The authors of Ref.~\cite{evans2013} used small displacement of strongly-stabilised moduli to make the CMP a `non-problem'. However, this non-problem may predict large amounts of isocurvature inconsistent with current constraints, depending on the details of inflation in this model.

Demonstrating this explicitly we show contours for liberal and less liberal values of $f_{ISO}=0.25,0.5$ and $\Delta N_{\rm eff}=1.0,1.5$ that could be allowed when this model is constrained using Planck data.

\begin{figure}
\begin{center}
$\begin{array}{@{\hspace{-0.3in}}l}
\includegraphics[scale=0.4]{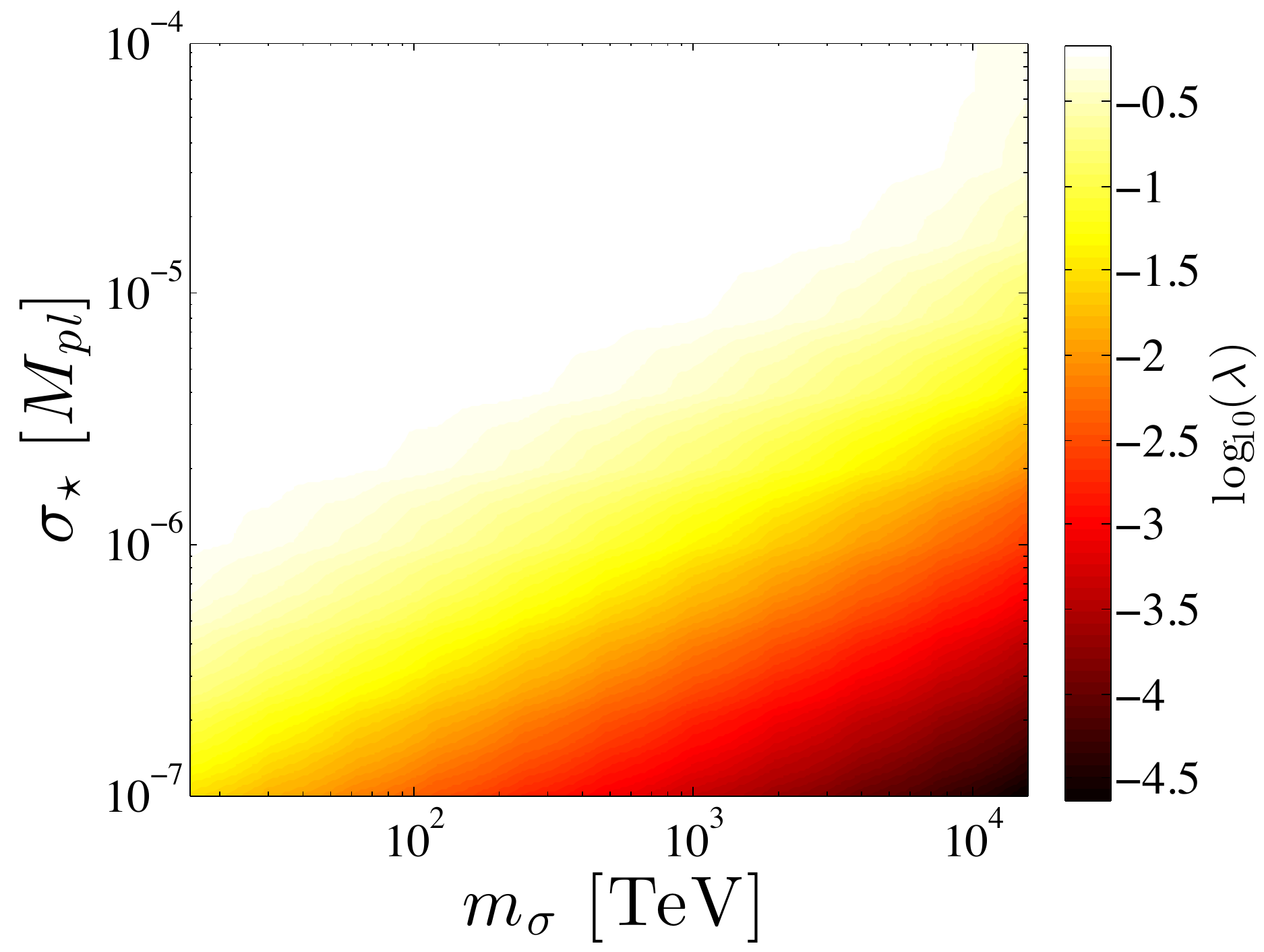} \\ [0.0cm]
 \end{array}$
 \caption{The parameter $\lambda$ measures how much of the total curvature is contributed by the modulus relative to the inflaton. We observe that $\lambda\sim \mathcal{O}(1)$ occurs as $f_{ISO}\rightarrow 0$. In the simplest curvaton scenario $\lambda\rightarrow \infty$ and so $r\rightarrow 1$ for all $f_{ISO}$.}\label{fig:mass_sigma_lambda}
\end{center}
 \end{figure}

The independent correlation parameter $r_{DM}$ varies across the $(m_\sigma,\sigma_\star)$ plane, with correlation vanishing as the total curvature vanishes and $f_{ISO}=1$, and total correlation when curvature dominates. This is due to the fact that when curvature dominates it is produced by the modulus decay, with $\lambda\sim \mathcal{O}(1)$ (see Fig.~\ref{fig:mass_sigma_lambda}).
 
In our model we see a stronger dependence of isocurvature observables on $\sigma_\star$ than on $m_\sigma$. The total amount of isocurvature depends on the ratio $\rho_R(N_{\rm end})/\rho_\sigma(N_{\rm end})$ (see Eq.~(\ref{eqn:sdmr_scott})). The transition from modulus dominance to sub-dominance is always approached while $\rho_R\sim a^{-4}$ (Eq.~(\ref{eqn:densities_source_solution}) and Fig.~\ref{fig:density_zoom}). Assuming that modulus oscillations begin in a radiation dominated universe we find that (setting $M_{pl}=1$ below for simplicity):
\begin{align}
m_\sigma^2 &\sim H_{\rm osc}^2\sim \rho_R(a_{\rm osc})\sim a_{\rm osc}^{-4} \, , \nonumber \\
\Rightarrow a_{\rm osc}&\sim m^{-1/2} \, , \nonumber \\
\rho_R(a)&= \rho_R(a_{\rm osc}) (a/a_{\rm osc})^{-4} \sim m_\sigma^2 (a/a_{\rm osc})^{-4} \, , \nonumber \\
\rho_\sigma (a) &\approx m_\sigma^2 \sigma_\star^2 (a/a_{\rm osc})^{-3} \, , \nonumber \\
\Rightarrow \frac{\rho_R (a)}{\rho_\sigma (a)} &\sim \sigma_\star^{-2} (a/a_{\rm osc})^{-1}\sim \sigma_\star^{-2}m^{-1/2}a^{-1} \, .
\end{align}
If decay occurs in the radiation dominated universe (Fig.~\ref{fig:density_zoom}, right panel) we have (e.g. \cite{book:lyth_and_liddle})
\begin{align}
\Gamma_\sigma \sim m_\sigma^3 \sim H_r &\sim (\rho_R)^{1/2}\sim a_r^{-2} \, , \nonumber  \\
\Rightarrow a_r & \sim m_\sigma^{-3/2} \, , \nonumber \\
\Rightarrow \frac{\rho_R(a_r)}{\rho_\sigma(a_r)}&\sim m_\sigma\sigma_\star^{-2} \, ,\label{eqn:iso_param_scaling1}
\end{align}
while if decay occurs in a modulus dominated universe (Fig.~\ref{fig:density_zoom}, left panel)
\begin{align}
\Gamma_\sigma \sim m_\sigma^3 \sim H_r \sim (\rho_\sigma& (a_r))^{1/2}\sim m_\sigma \sigma_\star (a_r/a_{\rm osc})^{-3/2} \, , \nonumber  \\
\Rightarrow a_r & \sim m_\sigma^{-5/4}\sigma_\star^{2/3} \, , \nonumber \\
\Rightarrow \frac{\rho_R(a_r)}{\rho_\sigma(a_r)}&\sim m_\sigma^{3/4}\sigma_\star^{-8/3} \, ,\label{eqn:iso_param_scaling2}
\end{align}
%\begin{align}
%m_\sigma^2 &\sim H_{\rm osc}^2\sim \rho_R/M_{pl}^2\sim M_{pl}^2 a_{\rm osc}^{-4} \, , \nonumber \\
%\Rightarrow a_{\rm osc}&\sim M_{pl}^{1/2} m_\sigma^{-1/2} \, , \nonumber \\
%\rho_R(a)&= \rho_R(a_{\rm osc}) (a/a_{\rm osc})^{-4} \sim M_{pl}^2m_\sigma^2 (a/a_{\rm osc})^{-4} \, , \nonumber \\
%\rho_\sigma (a) &\approx m_\sigma^2 \sigma_\star^2 (a/a_{\rm osc})^{-3} \, , \nonumber \\
%\Rightarrow \frac{\rho_R (a)}{\rho_\sigma (a)} &\sim M_{pl}^2 \sigma_\star^{-2} (a/a_{\rm osc})^{-1}\sim M_{pl}^{5/2}\sigma_\star^{-2}m_\sigma^{-1/2}a^{-1} \, .
%\end{align}
%If decay occurs in the radiation dominated universe (Fig.~\ref{fig:density_zoom}, right panel) we have (e.g. \cite{book:lyth_and_liddle})
%\begin{align}
%\Gamma_\sigma \sim m_\sigma^3/M_{pl}^2 \sim H_r &\sim (\rho_R/M_{pl})^{1/2}\sim M_{pl}a_r^{-2} \, , \nonumber  \\
%\Rightarrow a_r & \sim M_{pl}^{3/2}m_\sigma^{-3/2} \, , \nonumber \\
%\Rightarrow \frac{\rho_R(a_r)}{\rho_\sigma(a_r)}&\sim M_{pl}m_\sigma\sigma_\star^{-2} \, ,\label{eqn:iso_param_scaling1}
%\end{align}
%while if decay occurs in a modulus dominated universe (Fig.~\ref{fig:density_zoom}, left panel)
%\begin{align}
%\Gamma_\sigma \sim m_\sigma^3/M_{pl}^2 \sim H_r \sim (\rho_\sigma& (a_r)/M_{pl}^2)^{1/2}\sim m_\sigma \sigma_\star/M_{pl} (a_r/a_{\rm osc})^{-3/2} \, , \nonumber  \\
%\Rightarrow a_r & \sim M_{pl}^{7/12}m_\sigma^{-5/4}\sigma_\star^{2/3} \, , \nonumber \\
%\Rightarrow \frac{\rho_R(a_r)}{\rho_\sigma(a_r)}&\sim M_{pl}^{7/4}m_\sigma^{3/4}\sigma_\star^{-8/3} \, ,\label{eqn:iso_param_scaling2}
%\end{align}
hence there is always a stronger power law dependence of isocurvature observables on $\sigma_\star$. If SUSY is to make predictions for isocurvature it must also make contact with inflationary theory to set the scale of this initial displacement (Eq.~(\ref{potentialex})). In our model, all isocurvature observables display interesting transitionary behaviour with misalignments in the range $\sigma_\star/M_{pl}\in [10^{-7},10^{-5}]$. We will discuss in the next Section the possible significance of this energy range, in the `desert' of particle physics, for the Natural and Split-SUSY scenarios. 

This range of $\sigma_\star$ values is specific to the low-scale inflation with $H_I=10^5$ TeV that we have been considering. We show in Appendix~\ref{appendix:norm} how, as expected for a low-scale inflation model, the scalar normalisation to $A_s=2.2\times 10^{-9}$ then implies the extremely small slow-roll parameter $\epsilon\sim 10^{-11}$. Since the tensor-to-scalar ratio $r_h=P_h/P_{\zeta\zeta}^{\rm inf}(0)=16\epsilon$, then this implies unobservably small tensor modes in the CMB. Taken another way, observation of tensor modes would rule out Natural-SUSY with small, $\sigma_\star\sim 10^{12}$ GeV, initial modulus displacement.

\begin{figure}
\begin{center}
$\begin{array}{@{\hspace{-0.3in}}l}
\includegraphics[scale=0.45]{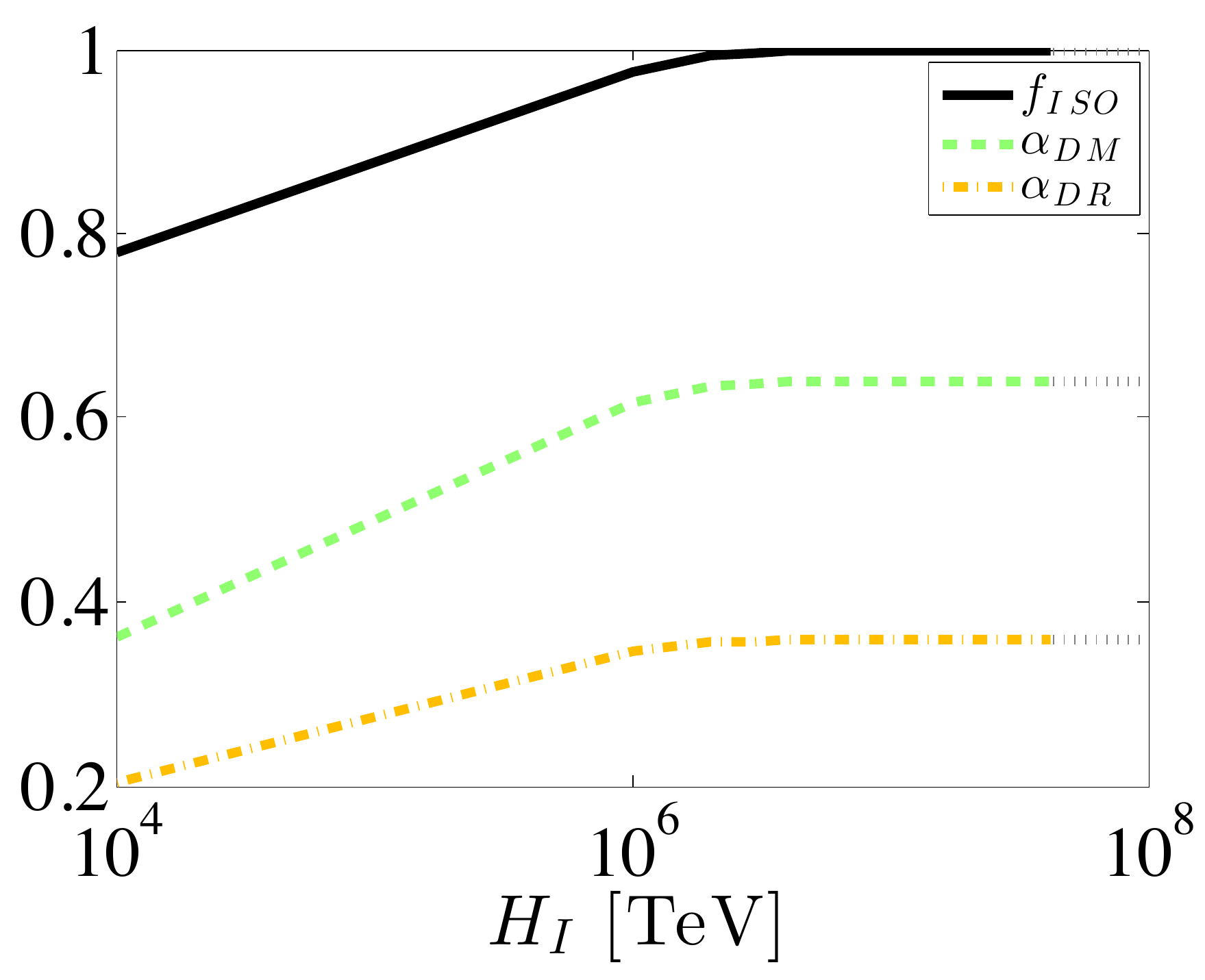} \\ [0.0cm]
 \end{array}$
 \caption{Isocurvature fractions as a function of $H_I$. The model shown corresponds to the largest decay rate in Fig.~\ref{fig:params_decayrate} (Left Panel), with $f_{ISO}=1$. Decreasing the energy scale of inflation decreases the amount of isocurvature. With large $H_I$ it is therefore necessary to have larger field displacements, $\sigma_\star$, to keep isocurvature small. Therefore with large $H_I$ small field displacement cannot be used to reduce DR contributions. With other parameters fixed there is a maximum $H_I$ beyond which the spectra cannot be normalised to $A_s=2.2\times 10^{-9}$ (shown as grey dotted, see Appendix~\ref{appendix:norm} for more details), but this occurs in the disallowed regime of isocurvature dominance.}\label{fig:fiso_hi}
\end{center}
 \end{figure}

Finally, in Fig.~\ref{fig:fiso_hi}, we show the effect of varying $H_I$ on the isocurvature fractions. For illustration we take a model with artificially high decay rate, with $\sigma_\star=10^{-6}M_{pl}$, $m_\sigma=240$ TeV and $f_{ISO}=1$ (largest decay rate in Fig.~\ref{fig:params_decayrate}, Left Panel).The isocurvature fractions decrease with $H_I$. With large $H_I$ in order to lower them again to acceptable levels one must increase the field displacement $\sigma_\star$. At large $H_I$ one shifts the transition regime in $\sigma_\star$ for isocurvature fractions away from the transition regime for $\Delta N_{\rm eff}$. Therefore with large $H_I$ one is not free to reduce $\sigma_\star$ far enough in order to reduce $\Delta N_{\rm eff}$ while still not overproducing isocurvature.

\subsection{Priors on Iscourvature from SUSY}

The mixed correlated isocurvature scenario with CDI and DRI modes, when compared to WMAP data, is poorly constrained and prior dependent \cite{moodley2004,bucher2004}. In order to confront this model with Planck data in a forthcoming work, we use our modulus model to derive theoretical priors. Isocurvature observables depend strongly on the high energy physics SUSY parameters $m_\sigma$ and $\sigma_\star$. In order for SUSY to be constrained by isocurvature it must therefore give isocurvature priors from UV considerations. We compute the priors for isocurvature observables given the priors $P(m_\sigma)$ and $P(\sigma_\star)$ holding other parameters fixed to $H_I=10^5$ TeV, $B_{DR}=0.21$ and $c_3=100$. We consider two scenarios which we loosely call `Natural-SUSY' and `Split-SUSY'. We use notation where $U(a,b)$ is the uniform distribution with minimum $a$ and maximum $b$.

\subsubsection{`Natural-SUSY'}

In this model we choose priors
\begin{align}
P(m_\sigma/\text{TeV}) &= U(10,100)\, , \\
  P(\sigma_\star/M_{pl}) &= \left\{ 
  \begin{array}{l l }
    U(0,10^{-5})\, ,\\
    U(0,5\times 10^{-6})\, , \\
    U(0,10^{-6}) \, .
  \end{array} \right.
\end{align} 
For the modulus mass, we consider `natural' as keeping $m_\sigma\lesssim 100$ TeV, where fine tuning will be around the level of $10^{-5}$ -- again we emphasize that the exact amount of tuning is an intricate question (see e.g. \cite{Baer:2012cf}). In order to avoid having been detected at LHC already, the mass must be larger than a few TeV. As noted in the previous section, there is interesting isocurvature phenomenology when $\sigma_\star/M_{pl}\in [10^{-7},10^{-5}]$. From Eq.~(\ref{eqn:sig_vev_inf}), the natural range of $\sigma_\star$ is specified by the mass scale $M$ (though we note the dependence on $n$, the dimension of the operator), which places $M$ in the desert of particle physics. In this range $M$ may be related to the axion decay constant ($f_a\approx 10^{12}$ GeV for DM constituted of a QCD axion with no fine tuning) or the geometric mean of the gravitino mass and the Planck scale, where SUSY is broken, $\Lambda_{\tiny SUSY}^2 = m_{3/2} M_{pl}$. We consider $\sigma_\star$ to be uniformly distributed over the range from zero to some value on the scale of $M$. Our Natural-SUSY priors specify three models, Nat1, Nat2, Nat3, given in Table~\ref{tab:natural_models}.
\begin{table}[dtp]
\begin{center}
\begin{tabular}{|c|c|c|}
\hline
Name &  $P(m_\sigma/\text{TeV})$ & $P(\sigma_\star/M_{pl})$ \\
\hline
\hline
Nat1 & $U(10,100)$  & $U(0,10^{-5})$ \\
Nat2 & $U(10,100)$  & $U(0,5\times 10^{-5})$ \\
Nat3 & $U(10,100)$  & $U(0,10^{-6})$ \\
\hline
         &$P(\log_{10}(m_\sigma/\text{TeV}))$ & $P(\log_{10}(\sigma_\star/M_{pl}))$ \\
\hline
Split & $U(1,4)$  & $U(-9,-2)$ \\
\hline
\end{tabular}\\
\end{center}
\caption{We consider four SUSY models, three Natural and one Split, with different priors on the modulus mass and misalignment.}
\label{tab:natural_models}
\end{table}

\subsubsection{`Split-SUSY'}

In this model we choose priors
\begin{align}
P(\log_{10} (m_\sigma/TeV)) &= U(1,4) \, , \\
P(\log_{10} (\sigma_\star/M_{pl})) &= U(-9,-2) \, ,
\end{align}
Foregoing naturalness, modulus parameters coming from SUSY need not be tied to the TeV scale. This scenario is what we will refer to as Split-SUSY. The log-flat prior is the scale invariant Jeffreys prior and represents the most conservative prior when there is no information about the scale of a parameter.

The log-flat prior can also be motivated physically. In UV complete models like string theory, the energy scales of low-energy physics, $f$, typically depend on geometric fields of the compact space, $\phi$ (such as moduli, or the dilaton), as $f\sim e^{-C \phi}$, where $C$ is some order one constant. If the geometric fields $\phi$ take uniform distributions at the Planck scale (or indeed some other high scale), this leads to uniform distributions in log space for the energy scales of low energy physics.

Our choice for the upper and lower bounds of the log-flat distributions for Split-SUSY are arbitrary, but span a phenomenologically interesting range. Our Split-SUSY priors on $\sigma_\star$ span a wider range than is plotted in Fig.~\ref{fig:contour_panels}.

\subsubsection{Isocurvature and Dark Radiation Priors}

\begin{figure*}
\begin{center}
$\begin{array}{@{\hspace{+0.1in}}l@{\hspace{+0.1in}}l}
\includegraphics[scale=0.38]{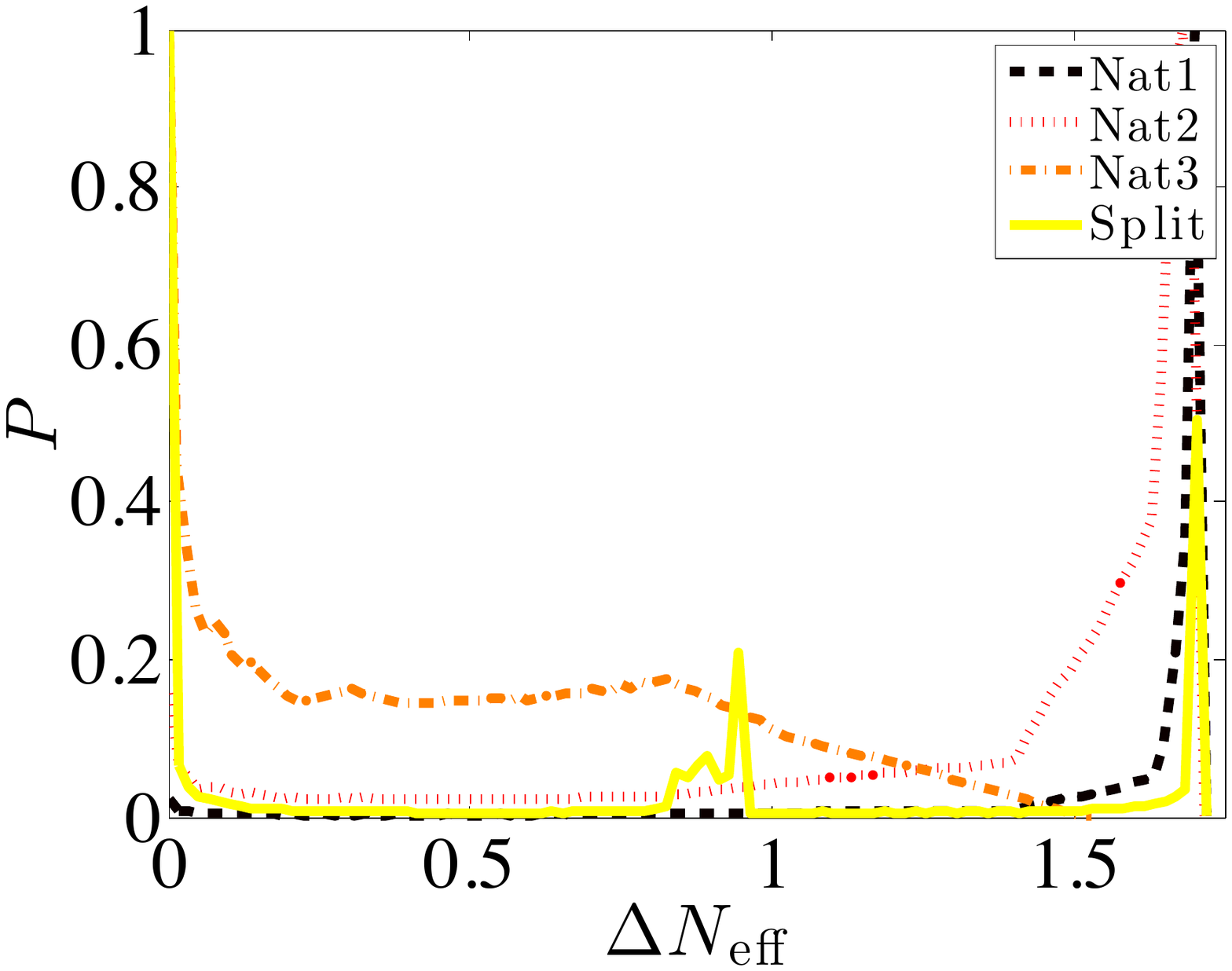}&
\includegraphics[scale=0.38]{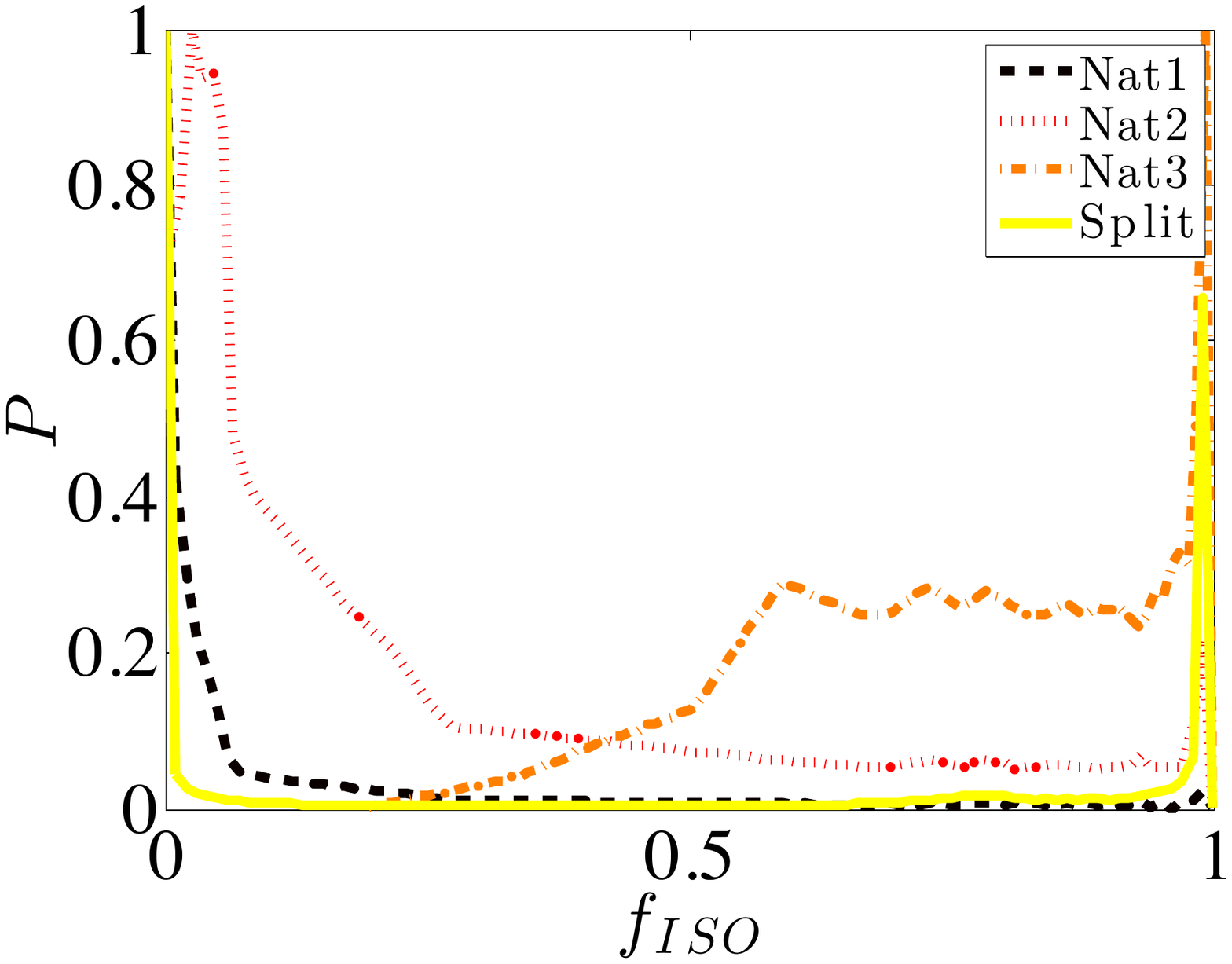}\\[0.0in]
\includegraphics[scale=0.4]{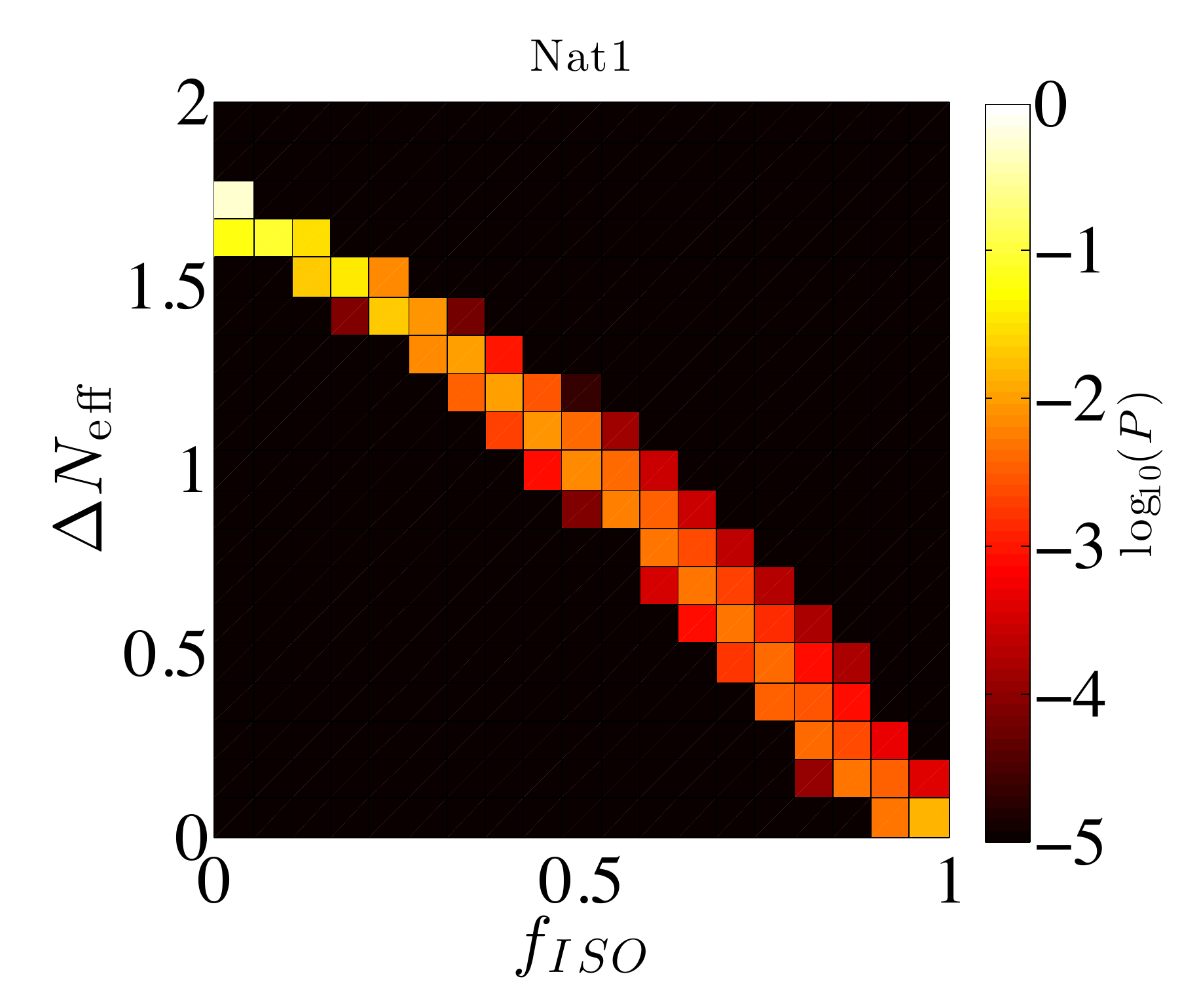}&
\includegraphics[scale=0.4]{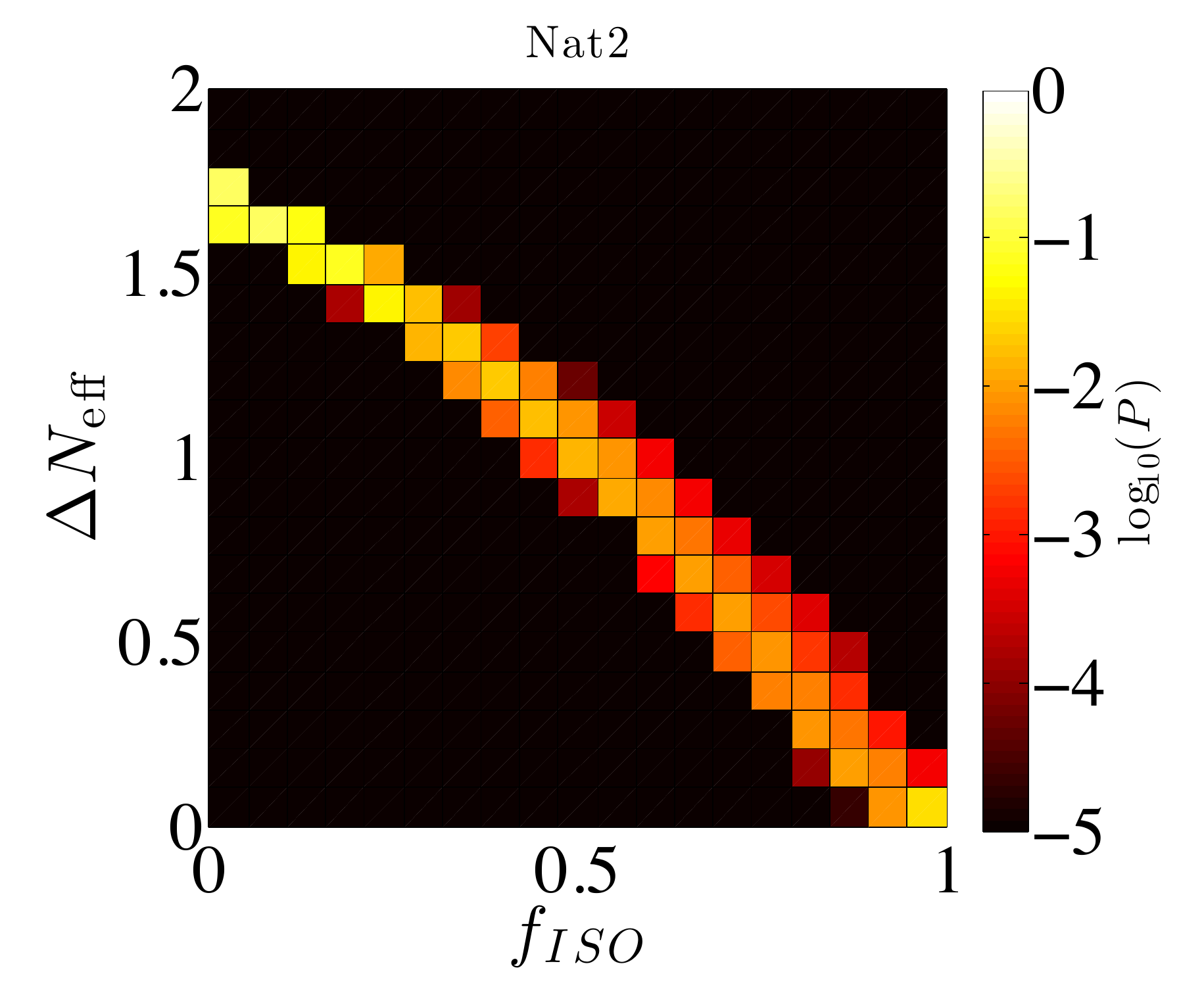}\\[0.0in]
\includegraphics[scale=0.4]{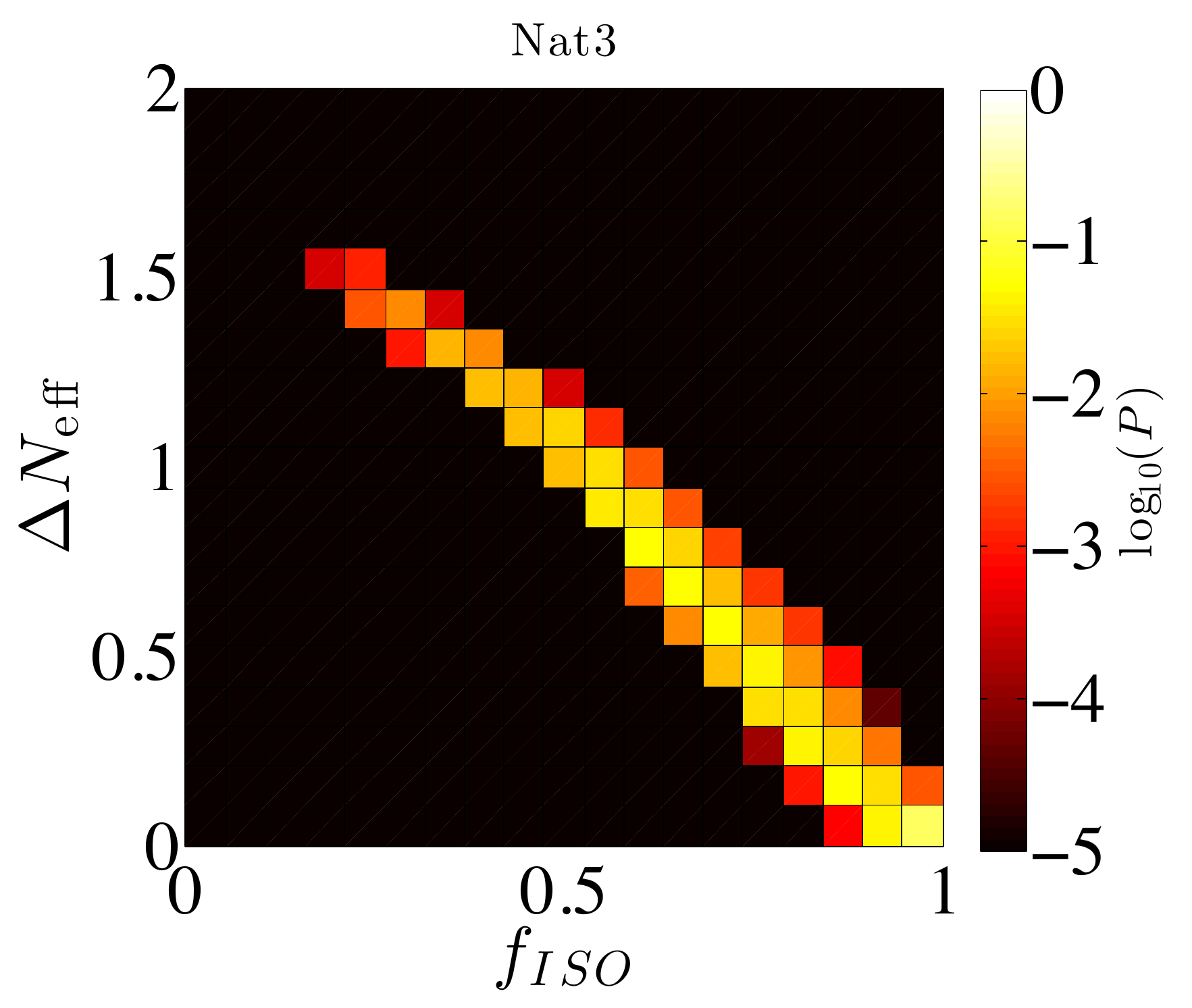}&
\includegraphics[scale=0.4]{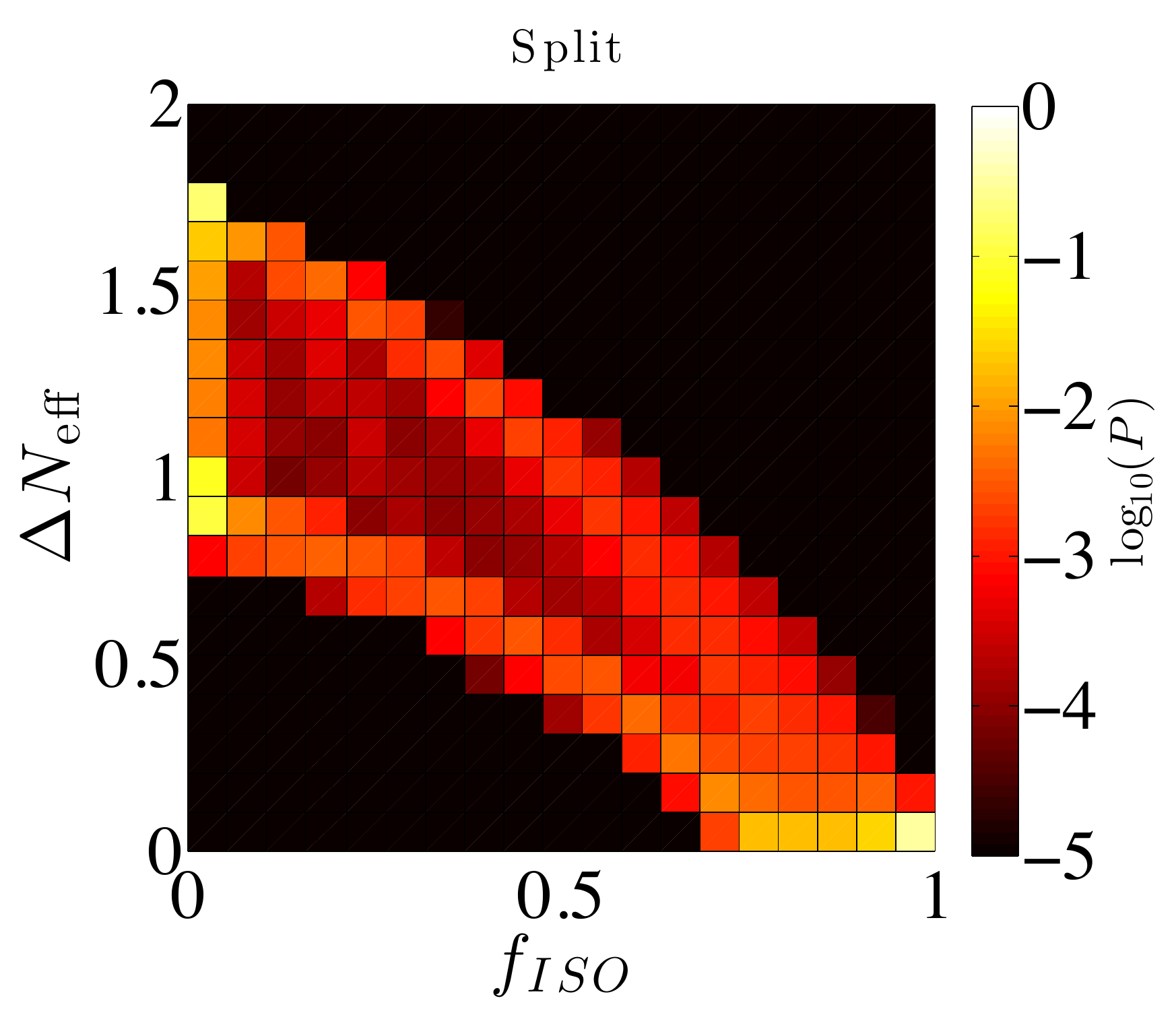}\\[0.0in] 
 \end{array}$
 \caption{Priors on isocurvature and DR in various SUSY models. In Natural-SUSY, the priors are driven by the prior on $\sigma_\star$, with larger priors favouring more DR and less isocurvature. There is a degeneracy between $f_{ISO}$ and $\Delta N_{\rm eff}$, so that reducing one increases the other. Split-SUSY favours a bi-modal distribution sampling both extremes. In Natural-SUSY as the prior range on $\sigma_\star$ is reduced the cosmological priors `migrate' between the two Split-SUSY modes.}\label{fig:prior_panels}
\end{center}
\end{figure*}

We show the isocurvature priors derived from our SUSY priors on $m_\sigma$ and $\sigma_\star$ in Fig.~\ref{fig:prior_panels}, sampling $9\times 10^4$ points from each distribution. We compute the priors by binning according to the observables, and normalise the probabilities by dividing by the total number of sample points, giving the probability in a bin and an unnormalised PDF. We show priors only on $f_{ISO}$ and $\Delta N_{\rm eff}$. It is obvious from Fig.~\ref{fig:contour_panels} that priors on $\alpha_{DM}$ and $\alpha_{DR}$ are simply scaled versions of the priors on $f_{ISO}$, while priors on $T_r$ and $r_{DM}$ reflect only the priors on $m_{\sigma}$ and $\sigma_\star$ respectively. For example, our Natural-SUSY priors favour low temperature reheating and partially correlated isocurvature.

Split-SUSY favours a bi-modal distribution for all parameters, with a mode at each extreme. In the limit that the prior range becomes infinite, the two modes become delta functions\footnote{This is not strictly true. At some extremely small value of $\sigma_\star$ the modulus energy density will drop below the DM energy density and all effects of the modulus will vanish, with no isocurvature or DR. We thank Cliff Burgess for discussion on this point.}. Therefore, Split-SUSY with log flat distributions favours either a totally isocurvature universe, which is certainly ruled out by CMB observations, or one with no isocurvature. For DR in the allowed adiabatic mode, it favours maximal DR given by the branching ratio. In this upper part of the $(m_\sigma, \sigma_\star)$ plane the reheat temperature is unconstrained. The reason for the Split-SUSY bi-modal distribution is clear from the results of Fig.~\ref{fig:contour_panels}: sampling the whole space includes observables covering the entire range of possible values. The space is sharply split into two with a small transition regime in log space, and this is reflected in the priors. Restricting the prior range on $\sigma_\star$ in Natural-SUSY priors causes the peak of the distribution to `migrate' between the two modes of Split-SUSY. 

We also show the degeneracy between $\Delta N_{\rm eff}$ and $f_{ISO}$. This demonstrates clearly that in the Natural-SUSY models small amounts of isocurvature favour larger values of $\Delta N_{\rm eff}$. This also implies that $\Delta N_{\rm eff}$ can be reduced by introducing small amounts of correlated isocurvature.

%%%%%%%%%%%  Conclusions

\section{Discussion and Conclusions}

In this paper we have considered the importance of isocurvature perturbations and dark radiation (DR) arising in cosmologies with a SUSY post-inflationary history in the presence of moduli. With guidance on mass scales from LHC, these models predict a non-thermal history where the decay of moduli provide additional sources of both dark matter (DM) and DR.  We've seen this implies the possibility of both correlated isocurvature perturbations and DR -- the latter providing an additional effective number of neutrino species, $\Delta N_{\rm eff}$ -- with both leading to new constraints on model building.

We find that the strongest constraints result when the branching ratio for moduli decay to DR is $\mathcal{O}(1)$ (as expected if it has an axion partner) and/or if the mass of the moduli during inflation is sub-Hubble -- the latter typically requiring that the physics responsible for lifting the flat-direction enters at sub-Planckian scales and therefore does not lead to too large an initial displacement (and so eventual amplitude of oscillation) of the field. Given models with these initial data, we find the possibility of production of DR isocurvature by direct decay, and established the necessity of including $\Delta N_{\rm eff}$ as a parameter along with considering this mode. In addition, in these models the modulus does not source all the curvature perturbation and as a consequence the isocurvature perturbations are not totally correlated. Generating two isocurvature modes also changes the specific relations between the amplitude of DM isocurvature and the fraction of energy density in the modulus/curvaton at the time of decay. Thus, we have found that SUSY non-thermal histories naturally provide a theoretical framework for motivating the past work of Refs.~\cite{bucher2004,moodley2004} where the importance of mixed isocurvature perturbations were explored in a phenomenological setting.

Stated another way, we've demonstrated that isocurvature priors can be computed using a high energy model. The priors in the simplest models are restrictive, in particular having only one independent correlation parameter, with the others fixed by a single source relation as discussed in Section \ref{sec:results}. This can be relaxed by introducing two components to the DM, as expected for example if the DM is a mixture of WIMPs and axions, or a mixture of thermal and non-thermal components. The correlation parameters then measure the fraction of non-thermal DM coming from modulus decay to the other component. The model also has all correlations strictly positive, with no anti-correlation possible. This relation can be broken only by introducing non-trivial interactions between the components. For example, a two field model with a repulsive interaction between the moduli could lead to two mutually anti-correlated isocurvature modes.

Again, we emphasize that the isocurvature and DR constraints on these models depend crucially on the initial displacement of the modulus field, $\sigma_\star$. What is the significance of the displacement in general for SUSY? Knowing that moduli must be heavier than some mass scale implies that displacement cannot be too small. Heavy moduli overproduce isocurvature if their displacement occurs on small energy scales, either by fine tuning or by hierarchy. As our knowledge of the scale of SUSY improves, and limits on moduli masses become stronger, so too does the interpretation via isocurvature constraints on the minimum scale of displacement. For example, with $m_\sigma\approx 10$ TeV isocurvature vanishes with $\sigma_\star\gtrsim 10^{-6} M_{pl}$, while if the moduli mass is instead  $m_\sigma\approx 10^4$ TeV then isocurvature vanishes for $\sigma_\star\gtrsim 10^{-5} M_{pl}$. Before LHC the moduli could have been as light as $m_\sigma \approx 10^{-1}$ TeV, but with much smaller displacements allowed by isocurvature bounds (about an order of magnitude below the $m_\sigma\approx 10$ TeV case we consider). At very low displacement these light moduli do not dominate the energy density of the universe, and so even though they decay after BBN, they do not suffer from the cosmological moduli problem (CMP).

We have found that the most interesting cases typically require a rather low scale for inflation with $H_I\sim 10^5$ TeV. This choice has been particularly interesting in the context of DR and isocurvature since with typical (gravitationally suppressed) decay rates for the moduli and $\mathcal{O}(1)$ branching to DR ($B_{DR}=0.21$) both $\Delta N_{\rm eff}$ and $f_{ISO}$ undergo transitions at the same value of the displacement $\sigma_\star\sim 10^{-6} M_{pl}\sim 10^{12}$ GeV in the Natural-SUSY mass range $10$ TeV $\leq m_\sigma \leq 100$ TeV. This means that with low-scale inflation one can reduce the impact of the `moduli induced axion problem' \cite{higaki2013} by reducing the inflationary minimum for the modulus (either with dynamical mechanisms involving higher-dimensional operators, or by fine-tuning) without at the same time introducing large amounts of isocurvature. If the inflationary energy scale is much higher than this, the transition in $f_{ISO}$ occurs at larger displacements than the transition in $\Delta N_{\rm eff}$, and so one cannot reduce the DR without introducing unacceptable amounts of isocurvature. One can restate this: if $m_\sigma$ is natural, and $\sigma_\star\sim 10^{12}$ GeV, an interesting energy scale in the desert and possibly related to axion physics, then in order not to be ruled out by isocurvature one requires low scale inflation with $H_I\lesssim 10^5$ TeV. Saturating this bound with the highest energy scale for inflation, the allowed isocurvature mode produced is a correlated DM-DR mode with interesting phenomenology yet to be explored with Planck data.

\begin{figure}
\begin{center}
$\begin{array}{@{\hspace{-0.2in}}l}
\includegraphics[scale=0.4]{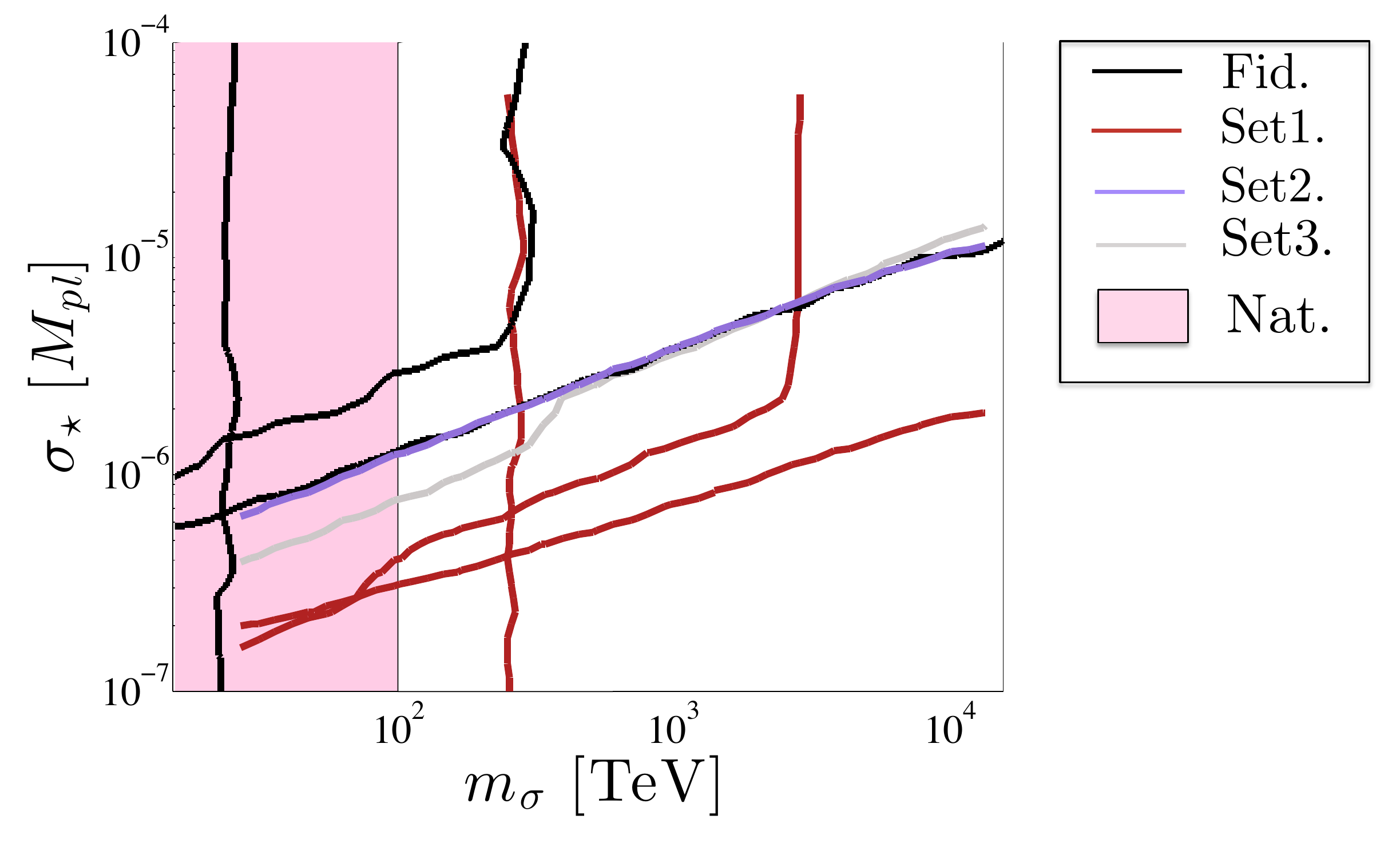} \\ [0.0cm]
 \end{array}$
 \caption{Summary of the model with fiducial parameters $H_I=10^5$ TeV, $B_{DR}=0.21$, $c_3=100$, and Sets 1-3 discussed in the text. Contours  show $T_r=3$ MeV (vertical), $f_{ISO}=0.5$ (diagonal), and $\Delta N_{\rm eff}=1.5$ (upper left corner). Allowed regions are above $f_{ISO}$, to the right of $T_r$, and below $\Delta N_{\rm eff}$. Natural-SUSY requires either suppressed branching to light particles in the dark sector, very low-scale inflation ($H_I\ll 10^4$ TeV), or both. Low $c_3$ worsens the CMP and rules out Natural-SUSY.}\label{fig:money_plot}
\end{center}
 \end{figure}

Our results are summarised in Fig.~\ref{fig:money_plot}. We show liberal contours for $\Delta N_{\rm eff}=1.5$ and $f_{ISO}=0.5$. For DR this corresponds to the edge of the $~3\sigma$ allowed region, while for $f_{ISO}$ this is the largest possible amount of isocurvature including all modes with free correlations. We take these liberal values since to the best of our knowledge the combined model with DR and isocurvature has never been constrained with CMB data and so we cannot take any constraints in this model as firm. In the allowed region of isocurvature, the modulus dominates the energy density at decay, and so we show the BBN bound of $T_r>3$ MeV. With our fiducial choices of parameters, in particular $H_I= 10^5$ TeV and $B_{DR}=0.21$, the Natural-SUSY models with $m_\sigma \in [10,100]$ TeV are put under severe pressure, leaving only a thin strip of parameter space in the allowed region. With less (but still extremely) liberal bounds on $f_{ISO}\lesssim 0.25$ and $\Delta N_{\rm eff}\lesssim 1$ then there is no allowed Natural-SUSY region with this branching and Hubble rate. 

Finally we briefly discuss the effects of varying our fiducial model with three additional parameter sets with parameters $(c_3,B_{DR},H_I/\text{TeV})$. {\bf Set 1:} $(1/4\pi,0.21,10^5)$, {\bf Set 2:} $(100,0.21,10^4)$, {\bf Set 3:} $(100,0.5,10^5)$. We show the contours for allowed values of $T_r$, $H_I$, and $\Delta N_{\rm eff}$ for these models also in Fig.~\ref{fig:money_plot}, only indicating contours when they change from the fiducial model. Lowering $c_3$ affects all contours, and leaves no Natural SUSY region allowed, forcing high modulus masses. Increasing $B_{DR}$ rules out the entire parameter space, leaving no region with $f_{ISO}<0.5$ and $\Delta N_{\rm eff}<1.5$. On the other hand, we found that lower $B_{DR}=0.1$ has the entire plane with allowed $\Delta N_{\rm eff}<0.8$, which is within $1\sigma$ allowed by Planck. Lowering $H_I$ has virtually no effect.

The power-law dependence of cosmological parameters on the underlying SUSY parameters generically predicts prior distributions steeply falling as a power law away from some extreme. In Split-SUSY models where masses can be raised significantly above the TeV scale, the derived priors on cosmological parameters have the generic feature of a bi-modal distribution. Parameters giving small amounts of isocurvature typically have the largest possible amount of DR, and vice versa.

We have seen that SUSY non-thermal histories provide a simple scenario to generate correlated DM-DR isocurvature from moduli decay to non-thermal DM and DR, and we have examined the resulting constraints from BBN and the CMB. Some immediately obvious extensions of our model are the computation of non-Gaussianities, and the inclusion of multiple decaying species \cite{Langlois:2011zz}, and using the super-horizon evolution of the spectral tilt and the full shape of the power spectra rather than just the scalar amplitude. Even considering liberal constraints from cosmological observables has led to strong and general constraints on Natural-SUSY, excluding models with large branching ratios to DR and/or high-scale inflation. A full analysis of the Planck CMB constraints on this model, incorporating the SUSY priors we have derived for it, will be forthcoming. We expect this analysis to place both stronger and more robust constraints on SUSY moduli.

\vspace{-0.25in}
% ------------------------ ACKNOWLEDGEMENTS ----------------------------------
\begin{acknowledgments}
\vspace{-0.1in}
We would like to thank Dan Hooper, Ogan Ozsoy, Kendrick Smith, David Mulryne, David Spergel, Joe Conlon, Moumita Aich, Michele Cicoli, Cliff Burgess, Will Kinney and John March-Russell for useful discussions, Ren\'{e}e Hlozek for providing some code, and Adrienne Erickcek for useful comments on the manuscript, and discussion.
The work of SW is supported in part by NASA Astrophysics Theory Grant NNH12ZDA001N, and DOE grant number DE-FG02-85ER40237.
SW would also like to thank the DAMTP, Cambridge University, for hospitality. Research at Perimeter Institute is supported by the Government of Canada through Industry Canada and by the Province of Ontario through the Ministry of Research and Innovation.
\end{acknowledgments}

% \renewcommand{\theequation}{A\arabic{equation}}
  % redefine the command that creates the equation no.
 % \setcounter{equation}{0} 
\appendix
\section{Computation of Isocurvature Observables}
\label{appendix:computation}

This Appendix contains details of the equations of motion for the modulus and its decay products, and how they are solved to compute the observables. The background equations of motion are given in Section~\ref{sec:iso_production}, while the initial conditions and equations of motion for the perturbations are given below. We briefly discuss the relation to the transport equations, and also why a fluid approximation was necessary to solve the modulus equations of motion with realistic SUSY decay rates, and demonstrate the accuracy of this
approximation. \\
\indent We solve the equations of motion for each initial condition mode separately, coupled to a background evolution with fixed parameters. The solution is obtained numerically using a fourth-order Runge-Kutta method implemented in Python \footnote{Interested readers can contact us for further information, or a copy of our code.}. The modules are highly adaptable as they can compute the flow for any number of modulus fields, species and modulus potential. They are scalable as they scale linearly in the number of fields and species.

\subsection{Cosmological Perturbations}
\label{appendix:cosmo_pert}

We work with scalar metric perturbations in the Newtonian gauge, where the metric is given by
\be
ds^2 = -(1+2 \Psi(t,\vec{x}))dt^2+a(t)^2(1-2\Phi(t,\vec{x}))\delta_{ij}dx^i dx^j \, . 
\ee
In the absence of anisotropic stress, we have that $\Psi=\Phi$. We decompose all perturbations $\xi(t,\vec{x})$ into Fourier modes as 
\be
\xi (t,\vec{x}) = \int d^3 k \, e^{i\vec{k}\cdot\vec{x}} \tilde{\xi}(t,\vec{k}) \, .
\ee
Being somewhat sloppy with notation we drop tildes and denote a function and its Fourier transform by the same symbol, where functional dependence clarifies the context. The dynamical Einstein equation we choose to work with is the first order equation
\be
\dot{\Psi}= -\frac{1}{6 M_{pl}^2}\frac{a}{H} \sum_i \delta \rho_i - \left( \frac{k^2}{3 a H}+aH \right)\Psi \, ,
\label{eqn:newtonian_potential}
\ee
where $\delta\rho_i$ is the density perturbation in species $i$: $\rho(t,\vec{x})=\rho(t)+\delta\rho(t,\vec{x})$.

In the presence of a source from modulus decays, the fluid overdensity, $\delta_i=\delta\rho_i/\rho_i$, and gradient of the fluid velocity, $\theta_i=i \vec{k}\cdot\vec{v}_i$, evolve as (see e.g. \cite{erickcek2011}):
\begin{align}
\dot{\delta}_i+(1+w_i)\frac{\theta_i}{a}-3(1+w_i)&\dot{\Psi}=B_i\Gamma_\sigma \frac{\rho_\sigma}{\rho_i}(\delta_\sigma-\delta_i+\Psi)\, , \label{eqn:fluid_delta_eom} \\
\dot{\theta}_i - \frac{k^2}{a}\left( \Psi+\frac{3}{4}w_i\delta_i \right) + (1-&3w_i)H\theta_i = \nonumber\\
 &\quad B_i \Gamma_\sigma \frac{\rho_\sigma}{\rho_i}\left( \frac{1}{1+w_i}\theta_\sigma -\theta_i\right)\, .&\label{eqn:fluid_theta_eom} 
\end{align}
where the $\Gamma_\sigma$ term is absent prior to $t_{\rm osc}$.

We perturb the modulus field as $\sigma(t,\vec{x})=\sigma(t)+\delta\sigma(t,\vec{x})$. After decay begins in the background field, the modulus perturbations evolve as
\be
\delta\ddot{\sigma}+(3H +\Gamma_\sigma)\delta\dot{\sigma}+\left( \frac{k^2}{a^2}+m^2_\sigma \right)\delta \sigma = -2 m_\sigma^2 \sigma \Psi+4 \dot{\sigma}\dot{\Psi} \, .
\label{eqn:pert_kg}
\ee
The density, pressure and $\theta$ perturbations arising from the modulus $\sigma$ are given by
\begin{align}
\delta\rho_\sigma &= \dot{\sigma}\delta\dot{\sigma}-\Psi \dot{\sigma}^2+m_\sigma^2\sigma\delta\sigma  \, , \\
\delta P_\sigma &= \dot{\sigma}\delta\dot{\sigma}-\Psi \dot{\sigma}^2-m_\sigma^2\sigma\delta\sigma \, , \\
(\rho_\sigma+P_\sigma)\theta_\sigma &=i k^2 \dot{\sigma}\delta\sigma \, .
\end{align}

As with the background, for computational simplicity once oscillations have begun we in fact follow an effective fluid description of the modulus perturbations. On super-horizon scales the time averaged sound speed in these perturbations, $c_s^2=\langle\delta P/\delta\rho\rangle$ goes to zero and the equations of motion for $\delta_\sigma$ and $\theta_\sigma$ are given by
\begin{align}
\dot{\delta}_\sigma+\frac{\theta_\sigma}{a}-3 \dot{\Psi}&=-\Gamma_\sigma \Psi \, ,\label{eqn:modulus_delta_eom} \\
\dot{\theta}_\sigma-\frac{k^2}{a}\Psi+H\theta_\sigma &= 0 \, .\label{eqn:modulus_theta_eom} 
\end{align}

To summarise, prior to $t_{\rm osc}$ we evolve Eqs.~(\ref{eqn:fluid_delta_eom}), (\ref{eqn:fluid_theta_eom}) and (\ref{eqn:pert_kg}) with $\Gamma_\sigma=0$. After $t_{\rm osc}$ we continue to evolve the R, DM and DR fluids  with Eqs.~(\ref{eqn:fluid_delta_eom}) and (\ref{eqn:fluid_theta_eom}) with $\Gamma_\sigma\neq 0$ and we treat the modulus field as an effective fluid obeying Eqs.~(\ref{eqn:modulus_delta_eom}) and (\ref{eqn:modulus_theta_eom}). The Newtonian potential is solved for at all times using Eq.~(\ref{eqn:newtonian_potential}).

\subsection{Initial Condition Modes}
\label{appendix:initial_modes}

We follow the evolution of a single super-horizon wavenumber, $k_0$, corresponding to a common CMB pivot scale for adiabatic and isocurvature modes of $k_0=0.002$~$h^{-1}$~Mpc. We construct the spectral indices in Appendix~\ref{appendix:indices}. The spectral indices are assumed not to evolve, although in principal we can easily compute any such evolution by simply following more $k$ values.

All species inherit adiabatic curvature perturbations from the decay of the inflaton. This adiabatic mode is given by
\begin{align}
\delta_R&=\delta_{DR}=-2\Psi \, , \\
\delta_{DM}&=\frac{3}{4}\delta_R=-\frac{3}{2}\Psi \, , \\
\theta_R&=\theta_{DR}=\theta_{DM}=\frac{1}{2} (k^2 \tau) \Psi \, , \\
\delta \sigma &= \delta\dot{\sigma} = 0 \, .
\end{align}
The conformal time is $\tau$ which we define initially, when $t=0$, $N=0$, $a=1$, as $\tau=1/aH$, and we use this only to set the initial conditions in terms of the initial Hubble rate. 

The isocurvature mode associated to curvaton perturbations is given by
\be
\sqrt{\langle\delta \sigma^2\rangle}(k_0) = \frac{H_I}{2\pi} \, .
\label{eqn:modulus_initial_ds}
\ee
with all other perturbations set to zero\footnote{Higher order corrections to this mode can be computed, as in the axion isocurvature mode of \cite{marsh_inprep}, but these are irrelevant at very early times when modes are extremely super-horizon.}.

When combining the two modes we distinguish variables in the inflaton sourced adiabatic mode with superscript `inf', and those arising from modulus isocurvature by superscript `mod'. The normalisation of the adiabatic mode is set by the inflationary spectrum for $\Psi^{\rm inf}$. We have that
\be
\zeta^{\rm inf}=-\frac{3}{2}\Psi^{\rm inf} \, .
\ee
The spectrum of curvature perturbations is
\be
P_{\zeta\zeta}^{\rm inf}(k_0) = \frac{1}{2\epsilon}\left( \frac{H_I/M_{pl}}{2\pi} \right)^2 \, ,
\label{eqn:zeta_power}
\ee
where $\epsilon=|\dot{H}|/H^2$ is the inflationary slow roll parameter. We normalise $\Psi^{\rm inf}=1$ in the inf mode, and $\delta\sigma^{\rm mod}=1$ in the mod mode. The relative normalisation of the two modes at the pivot scale is given by the ratio of the initial conditions
\be
\frac{\delta\sigma^{\rm mod}}{\Psi^{\rm inf}} = -3 M_{pl}\sqrt{\epsilon} \, .
\label{eqn:relative_norm}
\ee
We discuss the normalisation of the total scalar power in more detail in Section~\ref{sec:observables} and in Appendix~\ref{appendix:norm}.

Since the modulus and inflaton are not coupled, the adiabatic and isocurvature modes defined in this way are completely uncorrelated. In addition, since we are not interested in computing non-Gaussianities it is sufficient to follow linear evolution, which does not couple these modes. These two simplifications allow us to evolve each mode (inflaton adiabatic or modulus isocurvature) independently with a common background evolution, as independent universes. To construct observables, however, we must project these uncorrelated modes onto the observable modes of the CMB, combining the two universes. This projection, as in the case of a curvaton, allows the modulus isocurvature mode to source the adiabatic curvature perturbations in the CMB. It also generates DM and DR isocurvature perturbations, which now have specific correlations to the curvature and thus the CMB adiabatic mode.

\subsection{Relation to the Transport Equations}
\label{appendix:transport_eqn}

While it is clear how one obtains the equation of motion for the perturbations of all the species, evolving the correlation functions between these perturbations, in order to obtain the power spectra in CMB, is more troublesome. By following e.g. Refs.~\cite{mulryne2009,mulryne2013}, one can evolve  the curvature correlation matrix $\langle \zeta_i \zeta_j \rangle$. Performing such a numerical evolution proves problematic -- the problem scales as the square of the total number of species and the evolution of such matrices is prone to numerical errors. \\
\indent On the other hand, we notice that in the period between inflation and the start of curvaton oscillations, the correlation functions between perturbations of different species are given by the correlations in the quantum fluctuations of the fields from the inflationary period which decayed into these species. Since after inflation, we assume that all the species have decayed from the inflaton fields, we consider that the perturbations in radiation, dark matter and dark radiation are fully correlated, $\langle{\delta_i(\mathbf k) \delta_j(\mathbf k')}\rangle =(2\pi)^3 \delta^3(\mathbf k - \mathbf k') \delta_i(\mathbf k) \delta_j(\mathbf k')$. If we consider that the modulus field(fields) is independent from the inflation field \footnote{Which means that $[a_{\mathbf k}, b_{\mathbf k'}] = 0$ where $a_{\mathbf k}$ and $b_{\mathbf k'}$ are the creation/annihilator operators for the curvaton and inflation field}, then after inflation $\langle{\delta \sigma(\mathbf{k}) \delta_i(\mathbf k')}\rangle = 0$ and $\langle{\delta \sigma(\mathbf{k}) \delta \sigma(\mathbf{k'})} \rangle= (2\pi)^3 \delta^3(\mathbf k - \mathbf k') |\delta \sigma(\mathbf k)|^2$. Thus note that we can define two sets of fully independent ``cosmologies'', as we did in the previous subsection, where we have defined an adiabatic and an isocurvature mode \footnote{Or a greater number of fully independent cosmological perturbations for multiple-modulus models.}, each having initially the perturbations in all the species fully correlated.\\
\indent As we work with a linear ODE system, we can independently evolve these two sets of cosmological perturbations.  When evolving the two-point functions $\langle{\delta_i(\mathbf k) \delta_j(\mathbf k')}\rangle$ for modes where the perturbations are initially fully correlated, they will remain fully-correlated throughout the evolution; this means that the two-point functions will depend only on the evolution of the amplitudes of the perturbations which is given by the equations of Appendix~\ref{appendix:cosmo_pert}. For the two-point functions we are interested in we can thus avoid the whole mechanism of evolving correlation functions according to the transport equations, by independently evolving the perturbation amplitudes for the adiabatic and isocurvature modes and by super-posing the two cosmologies with the appropriate normalization factors. Thus, the correlation matrix for the perturbations is given by, $\langle{\delta_i(\mathbf k) \delta_j(\mathbf k')}\rangle = (2\pi)^3 \delta^3(\mathbf k - \mathbf k') (\delta_i^{\rm inf} \delta_j^{\rm inf} +  \delta_i^{\rm mod} \delta_j^{\rm mod})$ and consequently the correlation matrix for the curvature perturbations is given by, $\langle{\zeta_i(\mathbf k) \zeta_j(\mathbf k')}\rangle = (2\pi)^3 \delta^3(\mathbf k - \mathbf k') (\zeta_i^{\rm inf}\zeta_j^{\rm inf} +  \zeta_i^{\rm mod} \zeta_j^{\rm mod})$.

\subsection{Why The Fluid Approximation?}
\label{appendix:accuracy_approx}
In both the Klein-Gordon equations for the background and field perturbations there are two time scales -- the time scale of oscillations ($\sim 1/m_{\sigma}$) and the time scale of decay ($\sim 1/\Gamma_{\sigma}$).

It is practically impossible to follow both time scales as we would have to follow a great number of oscillations ($\sim M_{pl}^2/m_\sigma^2 \sim 10^{26} $ if $m_\sigma=100$TeV), in order to be able to simulate the evolution with physical values of $\Gamma_{\sigma}$ we need to eliminate the time-scale of oscillations. In order to do this once the oscillation of the field starts we approximate the evolution of the scalar field as a fluid with the equation of state $w=0$ and sound speed $\langle c_s^2 \rangle=0$, both of which can be derived via solution of the Klein-Gordon equation under a WKB approximation. According to the Averaging Theorem
and using a redefinition of the modulus field, the approximation will be in a neighbourhood of size $O(\Gamma_{\sigma})$ of the solution of our initial system for a time of order $O(1/\Gamma_{\sigma})$, in other words, until our modulus decays.

\begin{figure}
\begin{center}
$\begin{array}{@{\hspace{-0.1in}}l}
\includegraphics[scale=0.45]{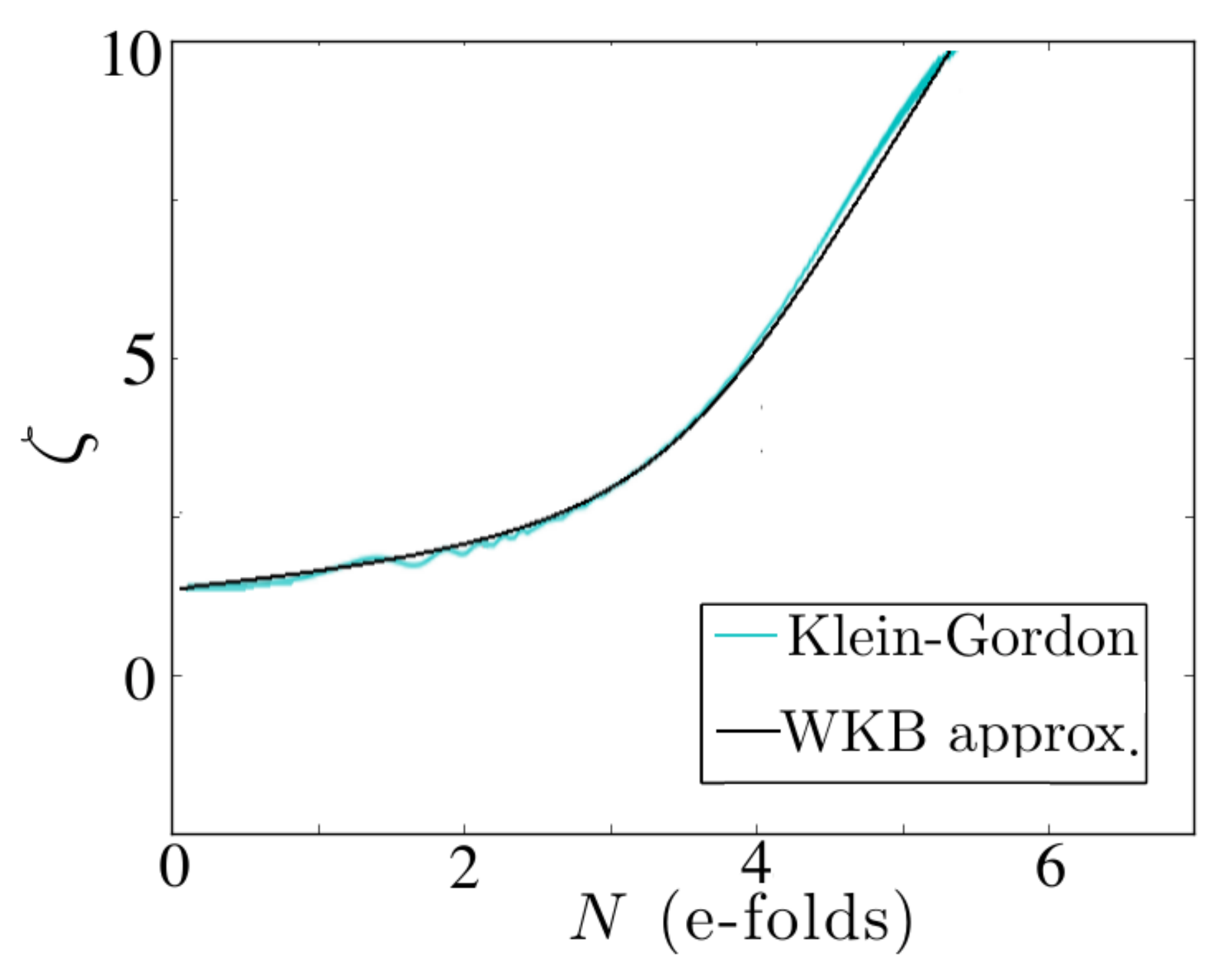} \\ [0.0cm]
 \end{array}$
 \caption{Evolution of the total curvature perturbation, $\zeta$, compared between the full solution of the perturbed Klein-Gordon equation and the fluid approximation. We have chosen artificially large decay rate so that the Klein-Gordon equation can be solved numerically over the range of time scales. The two methods are in good agreement. The accuracy of the fluid approximation improves as the time scales are further separated, i.e. as the decay rate is lowered towards the expected SUGRA values.}\label{fig:fluid_compare}
\end{center}
 \end{figure}

We find that the solution to the approximated differential equation system numerically agrees with the exact system. We show this in Fig.~\ref{fig:fluid_compare}, where we compare the evolution of the total curvature perturbation between the full KG equation solution and the fluid approximation. We have used large values for the decay rate such that we can numerically evolve the oscillating system over the same range of time scales as the fluid system. The accuracy of the fluid approximation for describing the decay, and hence isocurvature amplitudes and correlations, increases the further this time scale is separated from the oscillation time scale, and thus improves further for the realistic decay rates with $\Gamma_\sigma\sim m_\sigma^3/M_{pl}^2$.

\section{Power Spectra}
\label{appendix:spectra}

In this Appendix we discuss the normalisation of the power spectra using inflationary parameters and the scalar amplitude. We then discuss how to construct the spectral indices, assuming no $k-$dependence of the super-horizon evolution.

\subsection{Normalisation}
\label{appendix:norm}

In order to be able to physically interpret the cosmological perturbations one must normalize the evolution such that after modulus decay, the normalisation of the CMB $C_\ell$ (Eqs.~(\ref{eqn:c_ell_def}) and (\ref{eqn:as_def})) is $A_s  = 2.2 \times 10^{-9}$. Choosing the scalar amplitude such that we obtain the properly normalised CMB spectra is not the only scaling freedom that we encounter in our problem -- one can also choose the ratio between the power spectra of the adiabatic and iso-curvature by fixing the inflationary slow-roll parameter $\epsilon$, or the overall normalisation by the energy scale of inflation $H_I$. More precisely, because after inflation the curvature associated to all species in the `inf' mode will be equal $\zeta^{\rm inf} = \zeta_r = \zeta_{DM} = -3 \Psi/2$ and the the perturbation in the modulus field will be identical to that in the inflaton, $\delta \phi_{I} = \delta \sigma = {H_I/2k^3} $, we have Eq.~(\ref{eqn:relative_norm}) for the ratio between the amplitude of the modulus perturbation and the Newtonian potential. The overall normalisation by $H_I$ then enters through Eq.~(\ref{eqn:zeta_power}).

Let us denote $P_T^i=P_{\zeta\zeta}^i+\sum_j P_{S_jS_j}^i$, the total auto power in `inf' or `mod' universes, such that $A_s=\sum_j P_T^j(N_{\rm end})$. Our normalisation in the numerical computation sets $\Psi^{\rm inf}(0)=\delta\sigma^{\rm mod}(0)=1$, so that restoring the physical values we compute the power spectra at reaheating $N=N_{\rm end}$ as: $\tilde{P}_T^{\rm mod}=P_T^{\rm mod}(N_{\rm end})/\delta\sigma^{\rm mod}(0)^2$ and $\tilde{P}_T^{\rm inf}=P_T^{\rm inf}(N_{\rm end})/\Psi^{\rm inf}(0)^2$. This gives
\begin{equation}
A_s =\frac{1}{2\epsilon} \left( \frac{H_I}{2\pi M_{pl}} \right)^2 \left(\frac{4}{9}\tilde{P}_T^{\rm inf}+4\epsilon \tilde{P}_T^{\rm mod}\right) \, .
\label{eqn:total_norm}
\end{equation}
We choose $H_I$, and fix $\epsilon$ by normalisation. Rearranging
\be
\epsilon = \left( \frac{H_I}{M_{pl}} \right)^2 \frac{\tilde{P}_T^{\rm inf}/18 \pi^2}{A_s-(H_I/M_{pl})^2 \tilde{P}_T^{\rm mod}/2 \pi^2} \, .
\label{eqn:find_epsilon}
\ee
Thus, when iterating over models with different values of $m_{\sigma}$, $\sigma_\star$  and $\Gamma_\sigma$ we can input the energy scale of inflation $H_I$ for all such models and determine the slow-roll parameter $\epsilon$ that gives the correct $A_s$.

Note that if there is too much power generated by the modulus, then Eq.~(\ref{eqn:find_epsilon}) will only possess negative solutions. Since $\epsilon$ must be positive these solutions are not physical, and the corresponding cosmology cannot yield the correct normalisation for $A_s$. We encounter such cosmologies when $H_I$ is large.

\begin{figure}
\begin{center}
$\begin{array}{@{\hspace{-0.1in}}l}
\includegraphics[scale=0.45]{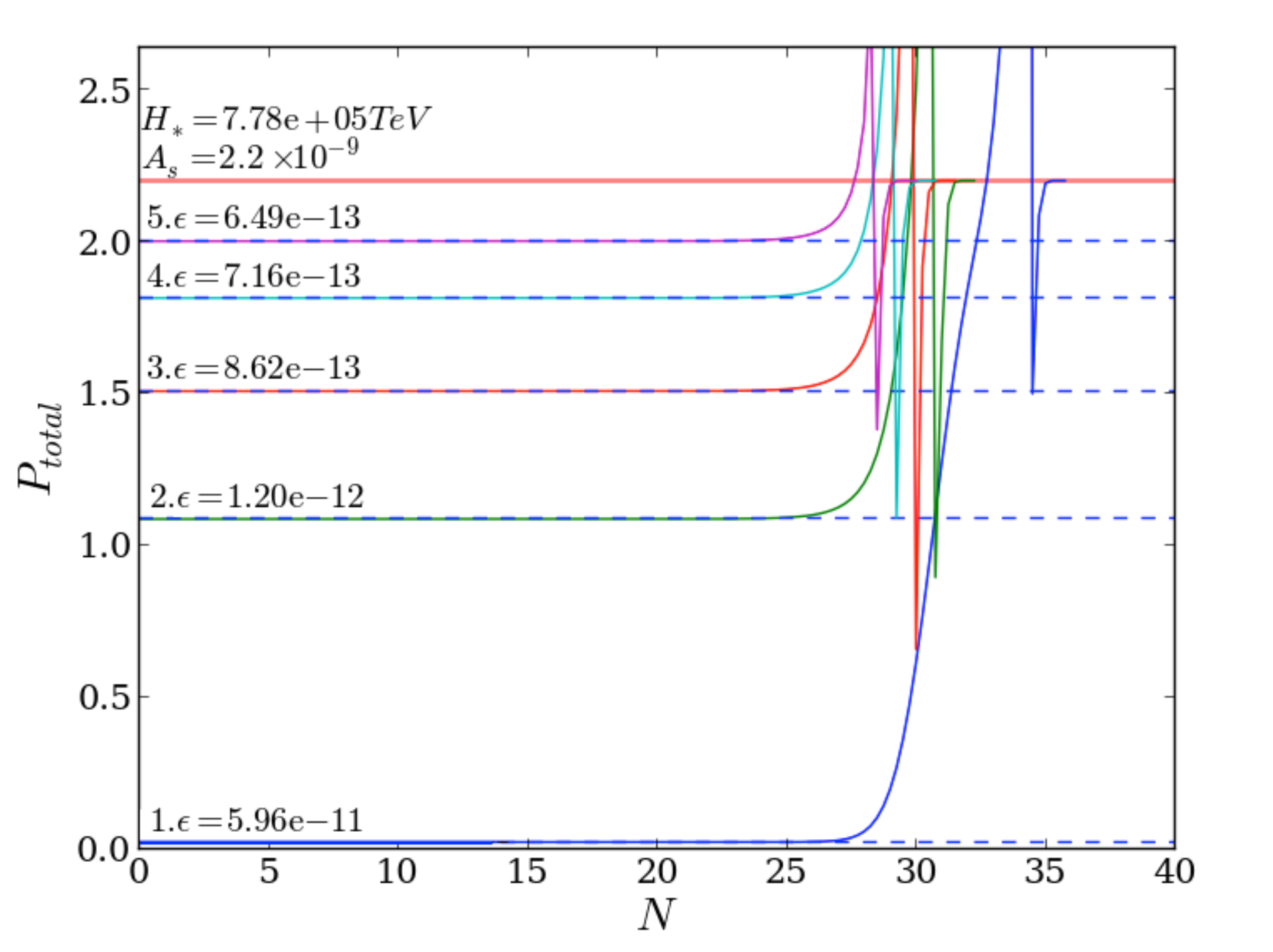} \\ [0.0cm]
 \end{array}$
 \caption{Normalisation to $A_s=2.2\times 10^{-9}$ for different choices of $\Gamma_\sigma$ at fixed $m_\sigma$, $\sigma_\star$ and $H_I$. The choice of $H_I$ sets the initial value of all the perturbations, up to a factor of $\epsilon$. This factor is used to fix the final value of the scalar power by shifting the relative normalisation of initial inflaton to modulus perturbations. Notice that the total power stops evolving at $N_{\rm end}$ when the modulus has completely decayed, and the expansion is radiation dominated and adiabatic and therefore $\dot{\zeta}=0$ (conservation of super-horizon curvature). Smaller values of $\Gamma_\sigma$ lead to larger $N_{\rm end}$, since the modulus takes longer to decay.}\label{fig:curvature_normalisation}
\end{center}
 \end{figure}

\begin{figure}
\begin{center}
$\begin{array}{@{\hspace{-0.1in}}l}
\includegraphics[scale=0.45]{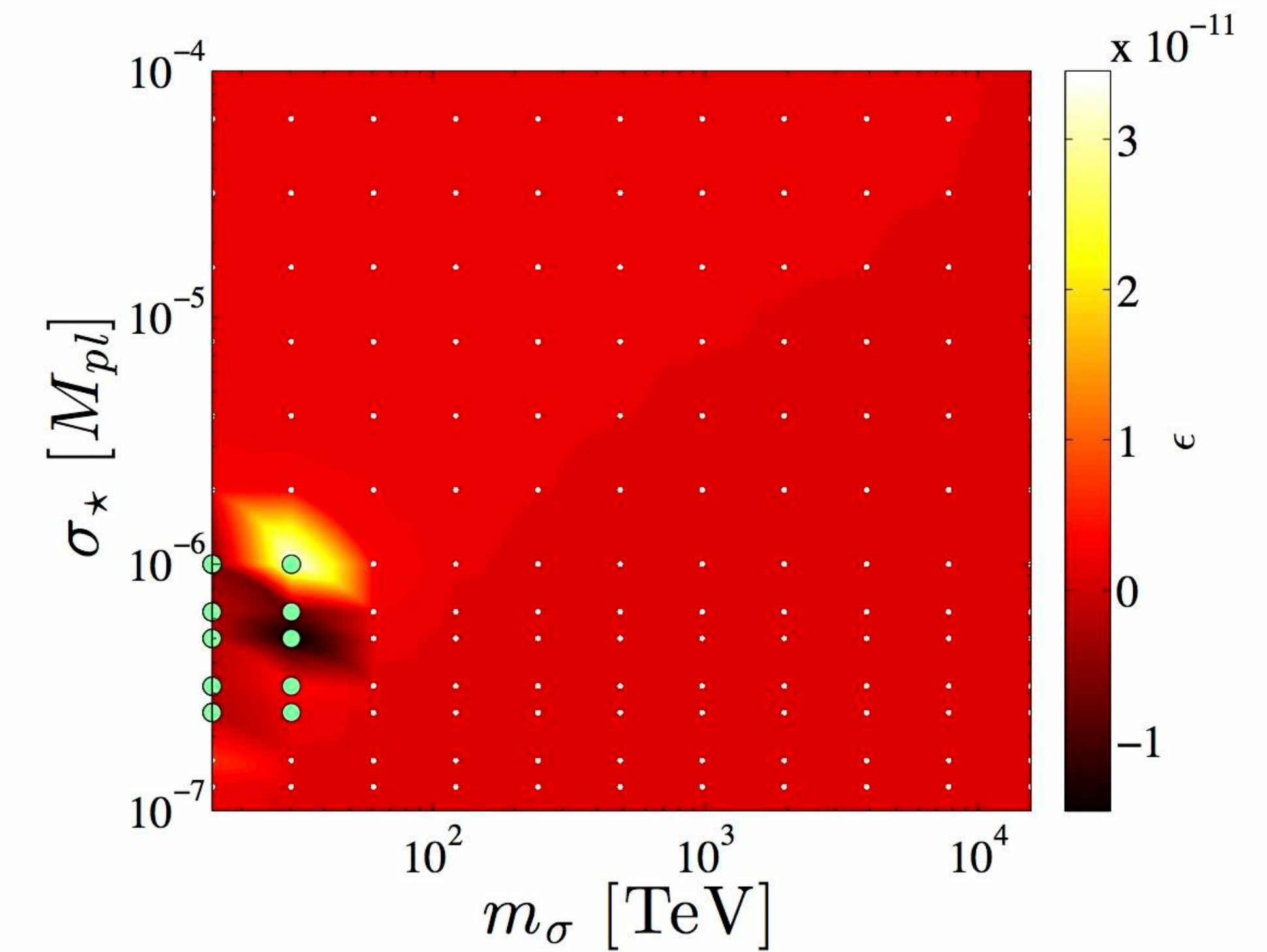} \\ [0.0cm]
 \end{array}$
 \caption{We take the parameters of Section~\ref{sec:results} and increase $H_I$ to $10^6$ TeV, plotting $\epsilon (m_\sigma,\sigma_\star)$ required to satisfy normalisation of $A_s$ (Eq.~(\ref{eqn:find_epsilon})). This leads to some sample points failing, by requiring unphysical $\epsilon<0$ (dark shaded). The original sample points are marked as points, while the sample points with $\epsilon<0$ are marked as large filled circles.}\label{fig:H1e6_epsilon_cont}
\end{center}
 \end{figure}

We illustrate our normalisation procedure in Fig.~\ref{fig:curvature_normalisation}, where we fix $H_I$ and compute $\epsilon$ with varying $\Gamma_\sigma$ (c.f. Fig.~\ref{fig:params_decayrate}). For our low-scale inflation with sub-dominant isocurvature we find as expected that $\epsilon$ must be extremely small, $\epsilon\sim \mathcal{O}(10^{-12})$, implying unobservably small primordial tensor modes, since $r_h=P_h/A_s=16\epsilon$. In the main text in Fig.~\ref{fig:fiso_hi} we also showed the dependence of isocurvature fractions on the choice of $H_I$ with all other parameters fixed. With these parameters, there was a maximum value of $H_I$ beyond which the spectrum could not be normalised. We demonstrate this effect again for the parameters used in Section~\ref{sec:results} in Fig.~\ref{fig:H1e6_epsilon_cont}, where we increase $H_I$ from $10^5$ to $10^6$ TeV. With $H_I=10^6$ TeV there are certain locations in the $(m_\sigma,\sigma_\star)$ plane where our normalisation procedure demands $\epsilon<0$, with the sample points that lead to this marked as large filled circles. Increasing $H_I$ increases the amount of isocurvature at larger values of $\sigma_\star$, so that the prior range on $\sigma_\star$ must be increased as $H_I$ increases in order not to overproduce isocurvature. The sample points with $\epsilon<0$ in Fig.~\ref{fig:H1e6_epsilon_cont} are deep in the disallowed large isocurvature regime, and therefore do not pose a problem.

\subsection{Spectral Indices}
\label{appendix:indices}

Inflaton seeded perturbations have spectral index
\be
n_{\rm inf} = 1+2\eta-6\epsilon \, ,
\ee
where $\epsilon$ and $\eta$ are the standard slow-roll inflation parameters, while the modulus seeded perturbations have spectral index \cite{lyth2002}
\be
n_{\rm mod}=1-2\epsilon+\frac{2}{3}\left( \frac{m_\sigma}{H_I} \right)^2 \, .
\ee
Unlike the curvaton model we do not assume that the modulus seeds all of the power, therefore the total power spectrum for $\langle \zeta_i \zeta_j \rangle$ takes the form of a broken power law. For example, the total curvature power spectrum is
\be
P_{\zeta\zeta} = \zeta^2 \left(\gamma^{\rm mod}_{\zeta\zeta} \left( \frac{k}{k_0}\right)^{1-n_{\rm mod}}+ \gamma^{\rm inf}_{\zeta\zeta} \left( \frac{k}{k_0}\right)^{1-n_{\rm inf}} \right) \, ,
\ee 
where $\zeta^2=(\zeta^{\rm mod})^2+(\zeta^{\rm inf})^2$ and $\gamma_{XY}^i=X_i Y_i/\sum_j X_j Y_j$. In this paper, for the sake of brevity, we do not compute the $\gamma^i_{XY}$, although they are simple to extract from our computations and can be used to compute the spectrum for any $\langle XY \rangle$. To the extent that $S^{\rm inf}_i=0$ the $\gamma^i_{XY}$ can be extracted from the fractional powers and correlations we define in Section~\ref{sec:observables}.  

As we saw in Appendix~\ref{appendix:norm}, our choice of $H_I\sim 10^5$ TeV typically leads to small values of $\epsilon$ when normalising $A_s$. We also saw in Fig.~\ref{fig:mass_sigma_lambda} that $\lambda \sim \mathcal{O}(1)$ across a wide range of models, which like $\gamma$ is a measure of how much modulus versus inflaton seeded perturbations contribute to the spectrum. In this regime the spectrum is a broken power law at the pivot scale and cannot be naively compared to the standard constraint of $n_s =0.96$, and will mimic a small positive running of the index \cite{Ade:2013zuv}. In order to reproduce a red tilt at $k>k_0$ with $n_{\rm mod}\approx 1$ one requires $\gamma^{\rm mod}_{\zeta\zeta}<1$ and then the freedom of $\eta$ in $n_{\rm inf}$ can be used to fix $n_s=0.96$ over some range of scales. For $\gamma^{\rm mod}_{\zeta\zeta}>1$ one requires a substantially higher $H_I$ to increase $\epsilon$ such that  $n_{\rm mod}< 1$. A complete analysis of this model with broken power law primordial spectra and free inflationary parameters $\{H_I,\epsilon,\eta \}$ is forthcoming.

%\section{Computation of Priors}
%\label{appendix:priors}

%The prior, $P(O)$, on an observable $O(\vec{x})$ that is a function of $D$ variables $\vec{x}=(x_i,\ldots,x_D)\in X$ with priors $P_i(x_i)$ is computed by a formal change of variables
%\be
%P(O)=\int d^Dx \Pi_{i=1,\ldots,D} P_i(x_i) \delta (O'(\vec{x})-O) \, ,
%\ee
%where $\delta (x)$ is the Dirac delta. Even when $O(\vec{x})$ is known this integral can be intractable analytically due to the inversion necessary to deal with the delta function. Fortunately, the procedure is much easier numerically. 

%We sample each variable $x_i$ according to its prior, $P_i(x_i)$, and compute the value of $O(\vec{x})$. We then bin $O(\vec{x})$ and count the number of points in each bin. Dividing by the total number of points sampled in $X$ we obtain the normalised probability in that bin. \tkDM{Luca to check} For each scenario we sample a total of \tkDM{$XX$} points.

% ---------------------- BIBLIOGRAPHY -----------------------------------------

\bibliographystyle{h-physrev3}
\bibliography{doddyoxford,scott_bib_1}
\end{document}